\documentclass[review,authoryear,11pt]{elsarticle}

\usepackage[margin=3cm]{geometry}
\usepackage{soul}
\usepackage{amsfonts}
\usepackage{amssymb}
\usepackage{amsbsy} % Per a \boldsymbol
\usepackage{amsmath}
\usepackage{subfigure}
\usepackage[T1]{fontenc}
\usepackage[latin1]{inputenc}
\usepackage{color}
\usepackage{array}
\usepackage{array,graphicx}
\usepackage{algorithm}
\usepackage{algpseudocode}
\usepackage{pifont}
\usepackage{lineno}
\usepackage{soul}
\usepackage{gensymb}
\usepackage[table,xcdraw,pdftex,divpsnames]{xcolor}  % Coloured text etc.

\usepackage{xargs}                      % Use more than one optional parameter in a new commands
\usepackage[colorinlistoftodos,prependcaption,textsize=tiny]{todonotes}
\usepackage{multirow}
\newcommand{\f}{{\mathbf f}}
\newcommand{\x}{{\mathbf x}}

\newcommand{\y}{{\mathbf y}}

\newcommand{\K}{{\mathbf K}}
\newcommand{\bK}{{\mathbf K}}
\newcommand{\bk}{{\mathbf k}}

\newcommand{\bby}{\bar{\y}}
\newcommand{\sK}{\boldsymbol{\mathcal{K}}}

\newcommand{\R}{{\mathbb R}}

\newcommand{\I}{{\mathbf I}}

\newcommand{\B}{{\mathbf B}}

\newcommand{\IG}{\includegraphics}
\newcommand{\Real}{\mathbb R}
\newcommand{\Normal}[0]{\mathcal{N}}

\newcommand{\Sigmabf}{\mathbf{\Sigma}}

\newcommand{\Deltabf}{\boldsymbol{\Delta}}

\newcommand{\black}[1]{\textcolor[rgb]{0,0,0}{#1}}          
        
\newcommand{\blue}[1]{\textcolor[rgb]{0,0,0.5}{#1}}

\usepackage{hyperref}
\hypersetup{
    unicode=false,          % non-Latin characters in Acrobat’s bookmarks
    pdftoolbar=true,        % show Acrobat’s toolbar?
    pdfmenubar=true,        % show Acrobat’s menu?
    pdffitwindow=false,     % window fit to page when opened
    pdfstartview={FitH},    % fits the width of the page to the window
    pdftitle={},    % title
    pdfauthor={Verrelst},     % author
    pdfsubject={GPR-BAT},   % subject of the document
    pdfcreator={Verrelst},   % creator of the document
    pdfproducer={Verrelst}, % producer of the document
    pdfkeywords={keyword1} {key2} {key3}, % list of keywords
    pdfnewwindow=true,      % links in new window
    colorlinks=true,       % false: boxed links; true: colored links
    linkcolor=blue,          % color of internal links (change box color with linkbordercolor)
    citecolor=blue,        % color of links to bibliography
    filecolor=blue,      % color of file links
    urlcolor=blue           % color of external links
}

\journal{Remote Sensing of Environment}

\begin{document}
%\linenumbers

\begin{frontmatter}

\title{Fusing Optical and SAR time series for LAI gap filling\\ with multioutput Gaussian processes
\footnote{Paper published at Remote Sensing of Environment Volume 235, 15 December 2019, 111452, doi: https://doi.org/10.1016/j.rse.2019.111452}} % GCV

% Authors (Add full first names)
\author[uv]{Luca Pipia \corref{cor1}}
\author[uv]{Jordi Mu\~{n}oz-Mar\'i}
\author[uv]{Eatidal Amin}
\author[uv]{Santiago Belda}
\author[uv]{Gustau Camps-Valls}
\author[uv]{Jochem Verrelst}
\cortext[cor1]{Corresponding author}
\address[uv]{Image Processing Laboratory (IPL), Parc Cient\'ific, Universitat de Val\`encia, 46980 Paterna, Val\`encia, Spain.}
\fntext[phones]{Tel: +34-96-354-40-67; Fax: +34-96-354-32-61.}

\begin{abstract}
The availability of satellite optical information \black{is often hampered by the natural presence of clouds, which can be problematic for many applications. Persistent clouds over agricultural fields can mask key stages of crop growth, leading to unreliable yield predictions. Synthetic Aperture Radar (SAR) provides all-weather imagery which can potentially overcome this limitation, but given its high and distinct sensitivity to different surface properties, the fusion of SAR and optical data still remains an open challenge}. In this work, we propose the use of Multi-Output Gaussian Process (MOGP) regression, \black{a machine learning technique that learns automatically the statistical relationships among multisensor time series, to detect vegetated areas over which the synergy between SAR-optical imageries is profitable. For this purpose, we use the Sentinel-1 Radar Vegetation Index (RVI) and Sentinel-2 Leaf Area Index (LAI) time series over a study area in \black{north west of the} Iberian peninsula. Through a physical interpretation of \black{MOGP trained models}, we show its ability to provide estimations of LAI even over cloudy periods using the information shared with RVI, which guarantees the solution keeps always tied to real measurements. Results demonstrate the advantage of MOGP especially for long data gaps, where optical-based methods notoriously fail. The leave-one-image-out assessment technique applied to the whole vegetation cover shows MOGP predictions improve standard GP estimations over short-time gaps (R$^2$ of 74\% vs 68\%, RMSE of 0.4 vs 0.44 $[m^2m^{-2}]$) and especially over long-time gaps (R$^2$ of 33\% vs 12\%, RMSE of 0.5 vs 1.09 $[m^2m^{-2}]$).} 
\black{A second assessment is focused on crop-specific regions, clustering pixels fulfilling specific model conditions \black{where} the synergy is profitable. Results reveal the MOGP performance is crop type and crop stage dependent. For long time gaps, best R$^2$ are obtained over maize, ranging from 0.1 (tillering) to 0.36 (development) up to 0.81 (maturity); for moderate time gap, R$^2$ = 0.93 (maturity) is obtained. Crops such as wheat, oats, rye and barley, can profit from the LAI-RVI synergy, with R$^2$ varying between 0.4 and 0.6\black{. For} beet or potatoes, MOGP provides poorer results, but alternative descriptors to RVI should be tested for these specific crops in the future before discarding synergy real benefits. In conclusion, active-passive sensor fusion with MOGP represents a novel and promising approach to cope with crop monitoring over \black{cloud-dominated} areas.}

\end{abstract}

\end{frontmatter}
% \linenumbers

\section{Introduction}
Monitoring the dynamics of Earth's surface properties constitutes the ultimate purpose of remote sensing. Among the objectives of the multiple Earth observation (EO) missions launched in the last decades, primary importance has been given to \black{natural and managed vegetation land covers}
\citep{chevrel1981,huete2002,zhang2003,boles2004,dierckx2014}. The strong relationship between the response of vegetation in the visible and near-infrared spectrum and its biophysical activities led the preference towards optical satellite sensors, \black{even if} on average clouds prevent measuring more than 70\% of Earth surface at any time over the globe \citep{chelton2005}. \black{Besides}, only to some extent the revisit time was considered a demanding constraint for these satellite missions, since its increase goes at the expense of the imagery's spatial resolution. This trade-off \black{narrowed down} the analysis to global vegetation dynamics, with limitations over areas with persistent cloud covers. 

A major breakthrough was achieved with the Sentinel space missions of the Copernicus Space programme \citep{Berger2012}. Covering a range of technologies, such as radar and multispectral imaging instruments, each mission comprises a constellation of two satellites, thereby allowing land, ocean and atmospheric monitoring with an unprecedented spatial-time-spectral sampling \citep{Malenovsky12}. With the Sentinel2A/B (S2) sensors, launched in June 2015 and April 2017 respectively, multispectral images \black{have become} freely available with a maximum revisit of 5 days \citep{drusch2012}. New challenges in monitoring the phenological evolution of vegetation land covers are then posed. 1) The high-resolution of S2 imagery allows extending global analysis to field-scale homogeneous areas and identifying different growing stages in terms of timing as well as abundance. 2) Simple vegetation indices such as the normalized difference vegetation index (NDVI) \citep{tucker1979} or more complex descriptors such as the green leaf area index (LAI) \citep{Jonckheere2004} can be estimated at high temporal sampling and a level of detail suitable for local studies. Nonetheless, the limitations due to the presence of clouds are reduced but not eliminated. The need for cloud-free acquisitions becomes even more critical in the case of crop monitoring, especially during the key development stages between seedling and flowering. Precise and timely delivered EO data provide farmers with the opportunity to intervene when necessary to improve crop yields or water-use efficiency \citep{baret2010,duchemin2015}. Yet, cloud contamination may drastically reduce the actual number of samples at disposal over a specific area of interest, \black{t}hereby limiting the reliability of farming recommendations.

Multiple methods have been proposed to deal with the presence of gaps in the temporal series of EO-based optical vegetation descriptors. These methods either consider the time series as a whole \citep{golyandina2007,verger2013,bacour2006}, or define a moving time window for the analysis \citep{chen2004,verger2011,jonsson2002}. The methods belonging to both approaches are compared in \citet{kandasamy2013}. \black{With the aim of generating finer and timely LAI products at potentially global scale, multisensors approach using spatio-temporal enhancement methods have been also pursued. MODIS and Landsat information have been merged using canopy spectral invariants theory \citep{ganguly2012} and spatio-temporal adaptive models, first at reflectance level \citep{gao2006,gao2012,wang2014} and later using biophysical parameters such as LAI \citep{houborg2016} or evapotranspiration \citep{cammalleri2014}.} The main limitation all these methods share is for long data gaps (>2 months), common on regions with dominant cloudiness, \black{vegetation changes are mostly ignored and retrievals cannot be trusted}.

An alternative solution is represented by synthetic aperture radar (SAR) imagery. On the one hand, clouds are not an issue for these sensors. On the other hand, SAR backscattering is related to observation geometry, but also to physical properties of the surface such as soil roughness, soil moisture and, particularly interesting in this context, vegetation cover \citep{ulaby1981,ferrazzoli1992}. The presence of vegetation generates the so-called volumetric scattering, which causes the redistribution of energy in the cross-polar band and evinces the sensitivity of SAR response to vegetation thickness \citep{bousbih2017}. In case of crop monitoring, the backscattering from the ground dominates the crop's early and late phenological stages but the vegetation contribution becomes higher in between \citep{mattia2003,cloude2010}. Thus, SAR sensors represent a suitable solution to create gap-free data collections for continuous vegetation mapping. Several studies exist on the description of crop phenology evolution with SAR imagery alone \citep{bousbih2017}, using both single-pol \citep{satalino2014} and full-pol \citep{wiseman2014,lopez2013} information. The sensitivity of SAR backscattering to vegetation growth also led to the definition of a Radar Vegetation Index (RVI), when dealing with fully polarimetric data \citep{kim2009,kim2014,veloso2017}, and a simplified version of it when only dual-pol data are available \citep{kim2001}. In general, SAR co-polar and cross-polar combinations are more related to optical vegetation descriptors as NDVI and LAI than single bands \citep{kim2012,fieuzal2013}, but the degree of this relationship is crop- as well as wavelength-dependent \citep{veloso2017}. Additionally, shorter wavelengths, such as C-($\sim$5cm) or X ($\sim$ 3cm) bands, are more sensitive to vegetation small-scale properties than L-band ($\sim$ 23cm), and their interaction with the soil may be dominant in case of sparse vegetation layers such as croplands in early stages \citep{ulaby1981}. 
From this point of view, the Sentinel1A/B (S1) sensors, launched in April 2014 and 2016 respectively, are suited for monitoring croplands: they provide SAR images at C-band of all over the world with a maximum revisit time of 6 days \citep{torres2012}. Successful use of S1 collections include crop classification \citep{torbick2017,xu2019} and monitoring (via classification of different vegetative stages \citep{mandal2018}) but, due to SAR sensitivity to many surface parameters, these approaches are not yet robust enough for operational monitoring services or crop yield estimation. At present, filling the temporal gaps of optical data over vegetation-covered surfaces with SAR observations is still an open challenge that requires finding a proper way to exploit their synergy: these two data modalities contain both complementary as well as common information that should be disentangled for their synergistic exploitation. 

In the last years, several efforts have been devoted to blend radar and optical information. Many applications benefit from the fused information, including building structure detection \citep{tupin2003}, urban classification \citep{haas2017}, heterogeneous land cover and crop classification \citep{camps2008,amarsaikhan2007,dusseux2014}, yield prediction \citep{baup2015}, soil moisture mapping \citep{gao2017}, LAI estimation \citep{dusseux2014} and environmental hazards detection \citep{errico2015}, among others. Nonetheless, despite the variety of fusion techniques used during the years, a concrete rationale on their applicability is still missing \citep{joshi2016}. First results on the generation of synthetic RGB images using deep neural network (NN) architectures trained on S1 and S2 collections have been recently presented in \citet{he2018,schmitt2018}, yet their performances are still insufficient for any operational purpose. The complementarity of active and passive data for NDVI reconstruction has been shown in \citet{scarpa2018} using convolutional NNs trained over S1-S2 short time series. They provide an assessment based on similarity of the spatial patterns but not a quantitative estimation of absolute errors at pixel level, nor a physically interpretable model of the information merging or a characterization of the estimation uncertainty. Altogether, a robust method able to merge active-passive data over vegetation \black{and} improve the quantitative retrieval from a primary source (i.e., optical) by exploiting only the amount of information shared with a secondary source (i.e., SAR) is still to be defined.

In this work, we propose a novel fusion approach for exploiting the information of S1 and S2 time series data collections at pixel level, and tackle the problem of cloud-induced data gaps over vegetated areas. We explore the use of S1-based RVI time series to fill the missing S2-based LAI information with Multi-Output Gaussian Processes (MOGPs). MOGP is a powerful statistical method able to capture the dependencies among any kind of multiple, yet related, data collections \citep{alvarez2012}. The use of this technique for remote sensing applications is absolutely new. To the best of our knowledge, only a simplified version of MOGP has been used to provide gap-free soil moisture time series by merging multi-frequency soil moisture products \citep{piles2018}. The approach pursued here is more challenging because it investigates the potentials of MOGP to learn the relationship \black{between} two totally different sensors, build a cross-domain kernel function able to transfer information across the time series, and do predictions (and provide associated gapfilled intervals) on regions where no optical data are available. MOGP can be trained at any scale, e.g., per-pixel or averaged per land cover. For each multisensor time series, it creates a specific model providing a prediction of the vegetation descriptors \black{at any date} along with an estimation of its uncertainty. Of interest hereby is that the parameters of each model are shown to describe directly the correlation between optical and SAR time series. As such, the MOGP method provides a quantifiable measure on how well two distinct EO products sources are expected to complement each other in time series gap filling. 

After providing a theoretical background of standard GP and MOGP (Section \ref{TheoBG}), LAI and RVI collections over a study area in north west of Iberian peninsula (Section \ref{preprocessing}) are used to illustrate the proposed MOGP and the active-passive synergy. The results (Sections \ref{Results}) on LAI gap-filling over a specific crop field are presented for propaedeutic purposes. Their analysis allows presenting a clear interpretation of the key role played by MOGP model's parameters. This order facilitates understanding the findings obtained when MOGP is extended to the spatial domain for cloud-free LAI map retrievals, and paves the path to discuss the potentials and limitations of the proposed method (Section \ref{Discussions}).
\label{sec:intro}

\section{Methodology}\label{TheoBG}

Standard Gaussian Process Regression (GP) models are state-of-the-art statistical methods for regression and function approximation. In recent years, we have successfully applied GPs for the retrieval of biophysical parameters from optical imagery, see~\citep[e.g.][]{Verrelst2012,verrelst2013,Verrelst13a,campos2016,amin2018,Verrelst2015, camps2015,CampsValls16grsm,CampsValls19nsr}. 
GP models yield not only predictions of the phenomenon to be characterized by means of a non-parametric modelling, but also an estimation of their uncertainty. A general introduction to GP can be found in \cite{Rasmussen06,CampsValls16grsm}. In the following we briefly review the standard GP and MOGP formulations, adapted to the general needs of this study.

\subsection{GP modeling of single output variable}\label{SOGPR_theo}

Let $\mathcal{D}=\{t_i,y_i\}_{i=1}^N$ be a set of $N$ pairs of a generic parameter $y_i$ extracted from data acquired at times $t_i$. We use these pairs to learn a function $f$ able to predict the parameter estimates at new times. Instead of assuming a parametric form for $f$, we start by assuming an additive noise model: 
\begin{equation}\label{eq:GLR}
    y_{i} = f(t_i) + e_i,~~~e_i \sim\Normal(0,\sigma_n^2),
\end{equation}
where $t\in\Real$, $\sigma_n^2$ is the noise variance and $f(t)$ is the unknown (and nonparametric) latent function to be found. Defining $\mathbf{t}=[t_{1}, \ldots ,t_{N}]^\intercal$, the GP model assumes that $f(\mathbf{t})$ is a Gaussian-distributed random vector with zero-mean and covariance matrix ${\bf K(\mathbf{t},\mathbf{t})}$, i.e. $f(\mathbf{t})\sim\mathcal{N}(\boldsymbol{0},\bK)$. The elements $ij$ of the covariance matrix are calculated by means of a kernel function $k(t_i,t_j)$ encoding the similarity between input time points $t_i$ and $t_j$. The square exponential (SE) or the Matern 3/2 kernel functions are often used \black{for this purpose} \citep{Rasmussen06}. 

The Bayesian framework allows us to estimate the distribution of $f_\ast$ at the test point $t_\ast$ conditioned on the training data, $p(f_\ast|\mathcal{D},t_\ast)$. According to the GP formulation, $f(t_\ast)$ is normally distributed with mean and variance given by:%}\todo{i clarified a bit the notation}
\begin{align} \label{eq:gppred}
  \begin{aligned}
   f(t_*)=\mu_{\text{GP}} (t_\ast) &= \bk_{*}^\intercal (\bK + \sigma_n^2\I_N)^{-1}\y \\
   \mathbf{\sigma}^2_f(t_*)=\sigma^2_{\text{GP}} (t_\ast) &= c_\ast - \bk_{*}^\intercal (\bK + \sigma_n^2\I_N)^{-1} \bk_{*},
  \end{aligned}
\end{align}
where $\bk_{*} = [k(t_*,t_1), \ldots, k(t_*,t_N)]^\intercal$ is an $N\times 1$ vector, $\y = [y_1,..,y_N]^\intercal$ and $c_\ast = k(t_*,t_\ast) + \sigma_n^2$. The model hyperparameters are obtained by maximizing the marginal likelihood of the model \citep{rasmussen2004}. 

\subsection{GP modeling of multiple output variables}

One limitation of the previous GP model formulation is that it applies only to scalar functions, i.e. we can predict only one target variable at a time. A straightforward strategy to deal with several target variables is to develop as many individual GP models as target variables. While generally good performance is attained in practice, the approach has a clear shortcoming: the obtained models are independent and do not take into account the relationships among outputs.

In order to handle this problem, we propose a Multi-output Gaussian Process (MOGP) model based on the \emph{linear model of coregionalization} (LMC)~\citep{alvarez2012}, also known as \emph{co-kriging} in the field of geostatistics~\citep{Journel78}. \black{Here, we describe the rationale of the MOGP approach; see \citet{alvarez2012} for details of the mathematical formulation}. 

Being $D$ the number of outputs of interests, i.e. the number of correlated parameters we want to link in the prediction model, each output function $f_d(t)$ can be expressed as a linear combination of groups of $Rq$ latent functions~\citep{Journel78} sampled from $Q$ independent GPs: 
\begin{equation} \label{eq:lmc}
    f_d(t) = \sum_{q=1}^{Q} \sum_{r=1}^{R_q} a^r_{d,q}u^r_q(t),~~~~~~d=1,\ldots,D
\end{equation}
where $a^r_{d,q}$ are scalar coefficients, and $u^r_q(\x)$ are latent functions sampled from the GP $q$ with zero mean and covariance ${\mathbf{K}_{q}}$. \black{So as} to simplify the equations that follows, we assume that all outputs have the same number of training samples, $N$. The formulation where each output has a different number of sources can be straightforwardly obtained.

Being ${\mathbf{T}_{d}}=[t^{d}_1,..,t^{d}_{N}]^\intercal$ the vector containing the time samples of the output $d$ and $\mathbf{T} = [\mathbf{T}_{1}^\intercal,...,\mathbf{T}_{D}^\intercal]^\intercal$ the one containing all of them, it can be shown~\citep{alvarez2012} that the full covariance (matrix) of the LMC model can be expressed as: 
\begin{equation} \label{eq:sos-lmc}
    \sK(\mathbf{T},\mathbf{T}) = \sum_{q=1}^Q \B_q \otimes \mathbf{K}_q(\mathbf{T},\mathbf{T}),
\end{equation}
where $\otimes$ is the Kronecker product and $\B_q\in R^{D\times D}$ are rank-$R_q$ positive definite matrices known as \emph{coregionalization} matrices. These matrices are the real core of the method because they encode the relationships among \black{the} multiple outputs. 
\iffalse
$\sK(\mathbf{T},\mathbf{T})$ is made up of $D\times D$ blocks $\K(\mathbf{T}_d,\mathbf{T}_{d'} )$ of $N\times N$ for $d,d'=1,..,D$. When the training kernel matrix $\sK(\mathbf{T},\mathbf{T})$ is block diagonal, that is, $\K(\mathbf{T}_d,\mathbf{T}_{d'}) = \boldsymbol{0}$ for all $d\neq d'$, then each output is independent of the others, and we have individual SOGP models. In other terms, the non-diagonal matrices are in charge of establishing relationships among the outputs. 
\fi

Rearranging the scalar coefficients of each output $q$ in (\ref{eq:lmc}) as $\mathbf{a}^i_q=[a^1_{1,q},...,a^1_{D,q}]^T$ with $i=1,...,R_q$, and stacking them in the matrix $\mathbf{A}_q=[\mathbf{a}^1_q,...,\mathbf{a}^{R_q}_q]$, each \emph{coregionalization} matrix is related to the coefficients of LMC model as follows: 
\begin{equation}  \label{eq:coreg_m}
   \B_q=\mathbf{A}_q\mathbf{A}_q^T.
\end{equation}
By grouping the output functions as $\f(t)=[f_1(t),...,f_D(t)]^\intercal$, the multi-output prediction and uncertainty at a new input value $t_*$ is given by: 
\begin{equation}\label{eq:2OPred}
  \begin{aligned}
    \mathbf{f}(t_*) = \mu_{\text{MOGP}}(t_*)= \sK_{t_*}^\top(\sK(\mathbf{T},\mathbf{T}) + \Sigma \otimes \I_ N)^{-1}\bby,\\
    \boldsymbol{\sigma}^2_f(t_*) = \sigma^2_{MOGP}(t_*)= \sK_{**}-\sK_{t_*}^\top(\sK(\mathbf{T},\mathbf{T}) + \Sigma \otimes \I_ N)^{-1} \sK_{t_*}, 
  \end{aligned}
\end{equation}
where $\Sigma\in\R^{D\times D}$ is a diagonal matrix containing the variance of the noise of each output, $\bby$ is the vector obtained by concatenating all the samples of the $D$ outputs, $\sK_{t*}\in\R^{D\times ND}$ is composed by blocks $\K_q(t_*,\mathbf{T}_d)$ for $d=1,..,D$, and $q=1,..,Q$  \citep{alvarez2012}.

\subsection{MOGP model simplification}\label{modelsimplification}
\black{In order to} adapt MOGP formulation to model the active-passive synergy, we assume only one parameter time series is available from each imagery ($D=2$). Referring to them as $P_{S1}$ and $P_{S2}$, \black{ the collections $\mathcal{D}_{P_{S1}}=\{t_i,P_{S1,i}\}_{i=1}^{N_{P_{S1}}}$ and $\mathcal{D}_{P_{S2}}=\{t_i,P_{S2,i}\}_{i=1}^{N_{P_{S2}}}$ represent the information to be used for training purposes.} Moreover, we posit that the two outputs are obtained as a linear combination of latent functions from two independent GPs ($Q=2$) sampled only once ($R_q=1$). This simplification, also known as \emph{semiparametric latent factor model} (SLFM) \citep{alvarez2012} leads to the definition of the reference model as follows: 
\begin{equation}\label{eq:mogp_system}
    \mathbf{f}(t) =\begin{bmatrix}
      f_{P_{S1}}(t) \\ f_{P_{S2}}(t)
    \end{bmatrix}=
    \begin{bmatrix}
      a^1_{1,1}u^1_1(t)+a^1_{1,2}u^1_2(t) \\ a^1_{2,1}u^1_1(t)+a^1_{2,2}u^1_2(t)
    \end{bmatrix}=\mathbf{a}^1_{1}u^1_1(t)+\mathbf{a}^1_{2}u^1_2(t).
\end{equation}
Accordingly, the two \emph{coregionalization} matrices are rank-1 and related to the model's coefficients as: 
\begin{eqnarray}\label{eq:B12_matrices}
[B_{1}] = \mathbf{a}^1_{1}{\mathbf{a}^1_{1}}^T
%\begin{bmatrix}a^{1}_{1,1} \\a^{1}_{2,1}  \end{bmatrix} \begin{bmatrix} a^{1}_{1,1} & a^{1}_{2,1} \\\end{bmatrix}
\enspace\enspace\enspace
[B_{2}] = \mathbf{a}^1_{2}{\mathbf{a}^1_{2}}^T.
%\begin{bmatrix}a^{1}_{1,2} \\a^{1}_{2,2}  \end{bmatrix} \begin{bmatrix}  a^{1}_{1,2} & a^{1}_{2,2} \\\end{bmatrix}.
\label{Eqn:TGP1}
\end{eqnarray}
\begin{figure}[t]
\centering
\footnotesize
\resizebox{1.0\textwidth}{!}{ 
\begin{tabular}{ll}
\IG[width=0.8\textwidth,trim={5cm 8cm 6cm 4cm},clip]{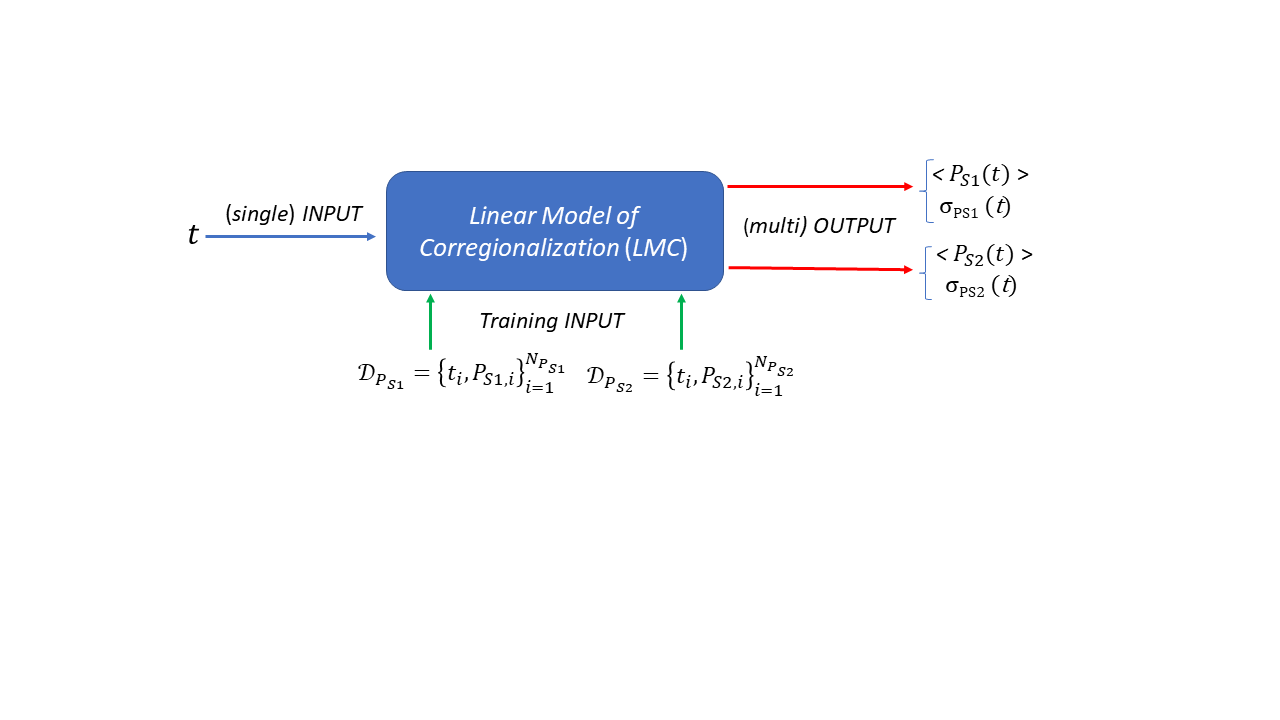}
\end{tabular}}
\vspace{-0.5cm}
\caption{Sketch of MOGP modelling based on the Linear Corregionalization Model (LCM). The time series of the two input parameters to be linked by MOGP,$\mathcal{D}_{P_{S1}}$ and $\mathcal{D}_{P_{S1}}$, are used to train the model. \black{The trained MOGP model provides a prediction of $P_{S1}$ and $P_{S1}$ along with their uncertainty for each input time $t$}.}
\label{Fig:MOGP_sketch}
\end{figure}
The expressions in (\ref{eq:mogp_system}) and (\ref{eq:B12_matrices}) constitute the basic structure of the multi-output model used for our analysis. On the one hand, the two independent GPs will be able to account for details of the two time series with a magnitude and at a time scale defined by the hyperparameters of each $\mathbf{K}_q$ after the training step. On the other hand, whether or not these details are shared between the two \black{outputs} along time will be determined by the components of $\mathbf{a_1}$ and $\mathbf{a_2}$. Note that the ideal case is a trained model \black{with} a dominant GP with all-ones vector $\mathbf{a}$, i.e. completely shared by the two outputs, and a second possibly negligible GP \black{with} a null-vector $\mathbf{a}$. This corresponds to the case of two perfectly correlated parameters $P_{S1}$ and $P_{S2}$. \black{Finally, it is worth stressing that the only input to the trained MOGP model is the time $t$ at which the parameters must estimated, as described visually in the sketch of Figure \ref{Fig:MOGP_sketch}}.

\subsection{Model set-up and training} \label{ModeSetupTraining}

\black{For the implementation of the MOGP, we use the GPy python library freely distributed by the ML Group of the Sheffield University \citep{gpy2014}. From each stack of the two coregistered collections, $\mathcal{D}_{P_{S1}}$ and $\mathcal{D}_{P_{S2}}$, a time series is extracted at pixel level. Afterwards, the Covariance kernel is set to Matern3/2, an optimum trade-off between preserving the time variability of training information and avoiding excessive smoothing effects \citep{Rasmussen06}. The \black{maximum} degrees of freedom of the model described in Section \ref{modelsimplification} is 10: 3 for each GP Covariance parameters (variance, lengthscale, noise) and 2 for each rank-1 matrix $[B]$. \black{By fixing the kernel's variance to 1, we reduce them to 8 (6 model parameters +2 noise standard deviations) and let the matrix $B$ account directly for the data dispersion}. 
The two collections of training samples are normalized with respect to their mean value and variance, and stacked to generate the vectors $\mathbf{T}$ in \eqref{eq:sos-lmc} and $\bby,$ in \eqref{eq:2OPred}. The result is passed to GPy for the hyperparameter optimization through maximization of the marginal likelihood. Finally, the covariance matrices $k_q(\mathbf{T},\mathbf{T})$, with $q=1,2$ are calculated \citep{alvarez2012}. The computational time for the training step at pixel level increases with the total number of training samples and is proportional to the cost of inverting the expression in (\ref{eq:2OPred}), i.e. $O(n^3)$ for GPy library \citep{gpy2014}. Once the MOGP model is trained, the only input it expects is time: for any value, it provides the corresponding estimation of the two outputs, $\mathcal{P}_{S1}$ and $\mathcal{P}_{S2}$, along with \black{their} uncertainty. The new version of \black{optical-based} $\mathcal{P}_{S2}$ is essentially a smoothed variant of the training input over period where samples were available, but over \black{cloud-induced gaps the solution is obtained from SAR-based $\mathcal{P}_{S1}$ according to the linking rule MOGP has learned during the training.} The key benefit of MOGP is that the estimation of each output at $t_*$ is \black{always} obtained by keeping into account the relationship of the two outputs.}

\section{Study area and time series preprocessing}\label{preprocessing} 

\subsection{Study area}\label{Testsite}
\begin{figure}[t]
\centering
\footnotesize
\resizebox{1.0\textwidth}{!}{ 
\begin{tabular}{ll}
\IG[width=0.8\textwidth,trim={0cm 0cm 0cm 0cm},clip]{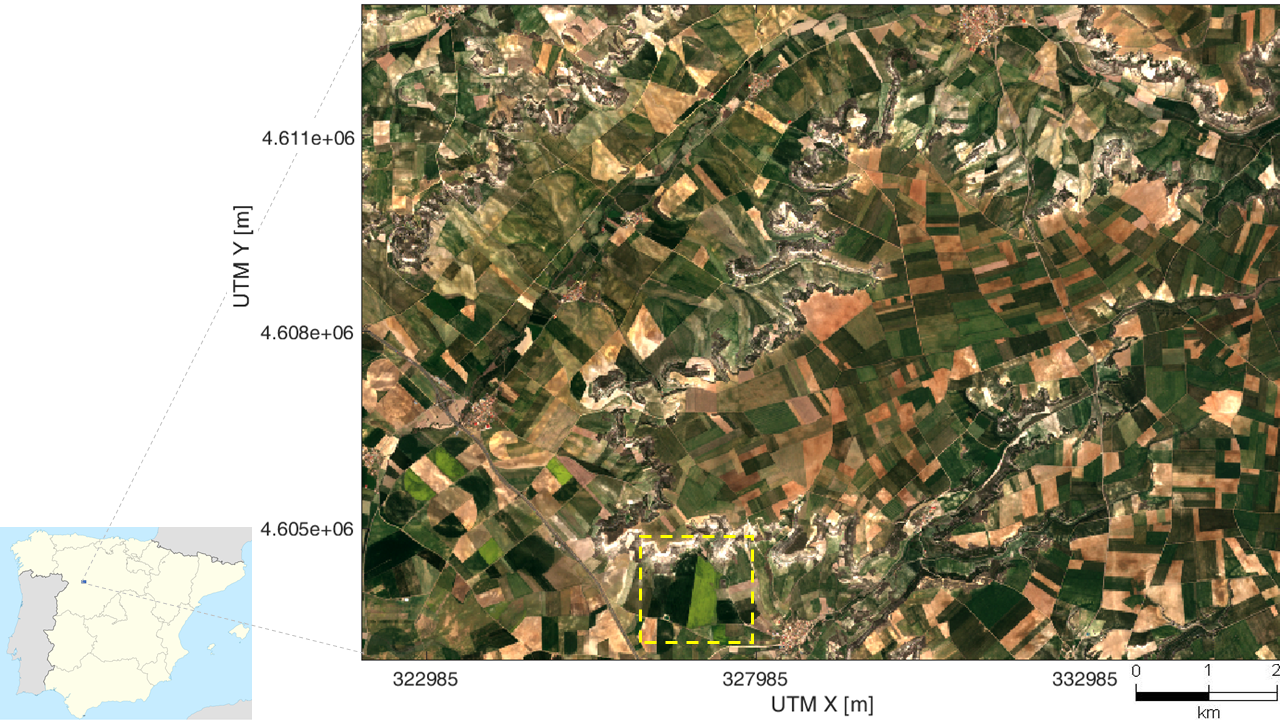}
\end{tabular}}
\vspace{-0.5cm}
\caption{RGB image of the crop AOI in Castile and Leon region, Northwest Iberian peninsula, from Sentinel2 capture of 2017, April 12$^{th}$.The yellow dotted rectangle delimits the subset analyzed in Section \ref{moderategap}.}
\label{Fig:RGB_AOI}
\end{figure}

The Area of Interest (AOI) selected for the study is \black{ a crop region in Castile and Le\'{o}n, in north west of \black{the} Iberian peninsula, of approximately 140 km$^2$ (Figure~\ref{Fig:RGB_AOI})}. The AOI belongs to a wider validation region of the H2020 Sensagri Project \citep{amin2018}. A land cover map is generated yearly since 2013 using a random forest classifier, trained on in-situ data collected by the ITACYL \citep{gomez2018}. The area selected for this work mainly corresponds to a dryland farming area with winter crops including cereals, wheat, barley and forage. A part of the arable land is also irrigated in summer with water stored in reservoirs. The main irrigated crops are maize, barley, wheat, sugar beet, alfalfa and potato. 

\subsection{Preprocessing of Optical time series}\label{Timeseries_genS2}

The collection of S2 images over the AOI was gathered from Theia datahub \citep{hagolle2010}, where multispectral reflectance images are available at tile level. A total amount of 97 cloud-free or at least partially-cloudy S2 images were downloaded, unevenly spaced between November 2015 and October 2018. 
From atmospherically corrected S2 data, two collections of products were generated. 1) NDVI time series at 10 m was obtained from the S2 topography-corrected reflectance collection. 2) The standard GP regression model presented in \cite{amin2018} was applied to generate green LAI time series at 10m. \black{The model, trained with LAI in-situ measurements collected in Barrax,\black{S}pain \citep{verrelst2013}, during different campaigns and both real and synthetic atmospherically corrected S2 data, provide\black{d} a $R^2=0.89$}. It accepts as input the S2 bands at 10m and 20 m and generates as output the prediction of LAI at pixel level along with its uncertainty.

\begin{table}[t]
\label{Table:spatialCorr}
\resizebox{1.0\textwidth}{!}{
\renewcommand{\arraystretch}{0.5}
\begin{tabular}{@{}l|ccccccccccc@{}}
\hline
\multicolumn{1}{c|}{\textbf{\%$\rho_s$>0.95}} & \multicolumn{1}{l}{\textbf{2/20}} & \multicolumn{1}{l}{\textbf{3/16}} & \multicolumn{1}{l}{\textbf{4/15}}  & \multicolumn{1}{l}{\textbf{5/15}}  & \multicolumn{1}{l}{\textbf{6/20}} & \multicolumn{1}{l}{\textbf{7/14}}  & \multicolumn{1}{l}{\textbf{8/19}}  & \multicolumn{1}{l}{\textbf{9/06}} & \multicolumn{1}{l}{\textbf{10/12}} & \multicolumn{1}{l}{\textbf{11/11}} & \multicolumn{1}{l}{\textbf{12/23}} \\ \hline
\textbf{VV}    & 96.6      & 97.2      & 97.1    & 97.7    & 97.5    & 97.3    & 97.2    & 97.3    & 97.4    & 97.2    & 96.5    \\
\textbf{VH}    & 97.1      & 97.6      & 97.0    & 97.2    & 96.4    & 95.8    & 95.5    & 96.4    & 96.1    & 96.7    & 96.8    \\
\textbf{VH/VV} & 96.6      & 97.2      & 97.1    & 97.7    & 97.5    & 97.4    & 97.2    & 97.3    & 97.4    & 97.2    & 96.5    \\
\textbf{RVI}   & {\color[HTML]{00009B} \textbf{97.7}} & {\color[HTML]{00009B} \textbf{98.1}} & {\color[HTML]{00009B} \textbf{98.3}} & {\color[HTML]{00009B} \textbf{98.6}} & {\color[HTML]{00009B} \textbf{98.1}} & {\color[HTML]{00009B} \textbf{97.5}} & {\color[HTML]{00009B} \textbf{97.0}} & {\color[HTML]{00009B} \textbf{97.4}} & {\color[HTML]{00009B} \textbf{97.4}} & {\color[HTML]{00009B} \textbf{97.8}} & {\color[HTML]{00009B} \textbf{97.4}} \\ \hline
\end{tabular}}
\caption{Percentage of vegetation pixels with spatial correlation $\rho_s>0.95$ between ascending and descending daily acquisitions, throughout 2017.}
\label{Table:ASCDES_corr}
\end{table}

\subsection{Preprocessing of SAR time series}\label{Timeseries_genS2}

The active nature of S1 imagery makes both ascendant and descendant orbits useful to monitor the AOI: orbits 74 and 154 were identified for this purpose. As no interferometric study was envisaged, radar amplitude data were downloaded from ESA's Scientific Hub in the Ground Range Detected format. A comprehensive collection of 246 and 267 images was created for the two observation geometries, covering a time span of almost 4 years, from January 2015 up to October 2018. As for S2 time series, the time sampling is not uniform as the second satellite (S1B) became available in October 2016. From that date on, two daily S1 passes are available every 6 days over the AOI, approximately at 6 am and 6 pm local time. 
The preprocessing applied to S1 collection backed up on the Graph Processing Tool distributed by ESA, and consists of four main steps: thermal noise removal, radiometric calibration, spatial speckle filtering (7$\times$7 Lee-Sigma) and terrain correction. Sudden peaks in the temporal profile were smoothed using an additional Gaussian filtering step at pixel level. Finally, data were projected onto UTM reference to achieve a subpixel coregistration of the two collections. For each S1 acquisition, four \black{products} were initially generated: 1)VH and 2) VV radar reflectivities, 3) VH to VV ratio (VH/VV) and 4) dual-pol Radar Vegetation Index (RVI) calculated as \citep{kim2012}:
\begin{eqnarray}
\text{RVI}= \dfrac{4\text{VH}}{\text{VH+VV}}.
\label{RVI}
\end{eqnarray}

To trim down the \black{time series length} and further reduce speckle residual effects at once, the \black{spatial correlation $\rho_s$ between S1 ascending and descending SAR daily products was estimated using a 7$\times$7 boxcar filter}. \black{The cumulative distribution of $\rho_s$ calculated on 11 pairs of ascending-descending images throughout 2017 (Table \ref{Table:ASCDES_corr})} confirms that more than 97\% of pixels have a correlation above 0.95 for VH and VV bands . The percentage is higher for VH/VV ratio, but the highest value is always obtained for RVI. This outcome agrees with \citet{veloso2017}, stating that for shallow incidence angles (>35-40\degree) like in S1 observations the ascending and descending acquisitions over vegetation provide close responses in all bands \citep{veloso2017}. Accordingly, S1 data acquired the same day were averaged.

\subsection{Selection of active-passive vegetation descriptors}\label{Vegparmsel}

\begin{figure}[]
\begin{center}
    \IG[width=11cm,trim={2cm 0cm 2cm 0cm},clip]{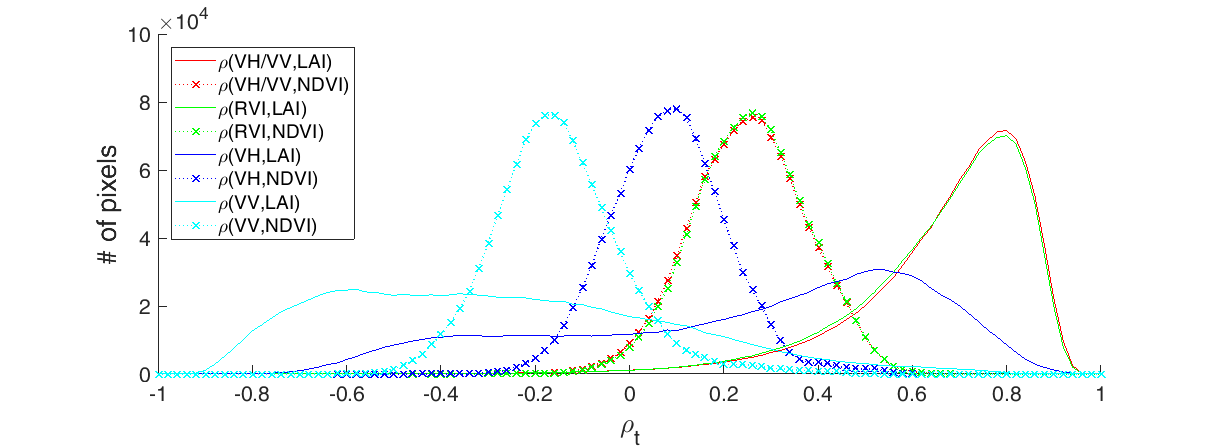} \\
\caption{Histograms of Pearson's correlation coefficient $\rho_t$ through time calculated per-pixel using all possible pairs of S2(NDVI,LAI) and S1(VV,VH,VH/V\black{V},RVI) parameter time series over the AOI.}
 \label{Fig:OpticSARCorrHistograms}
\end{center}
\end{figure}

To demonstrate the potential of MOGP for optical-SAR gap filling, the pair of active-passive vegetation descriptors providing the highest correlation in time \black{must be identified}. The closeness of S1 and S2 information can be quantified using the Pearson's correlation coefficient $\rho_t$ among all possible combination of optical (NDVI and LAI) and SAR (VV, VH, VH/VV, RVI) time series. The histograms in Figure~\ref{Fig:OpticSARCorrHistograms} show \black{that} the highest \black{values} are obtained for VH/VV and LAI and for RVI and LAI, in line with the results in \citet{veloso2017}.

\begin{figure}
\begin{center}
\setlength{\tabcolsep}{0pt}
\begin{tabular}{cc}
    \IG[width=8cm,trim={5cm 1.8cm 2cm 0cm},clip]{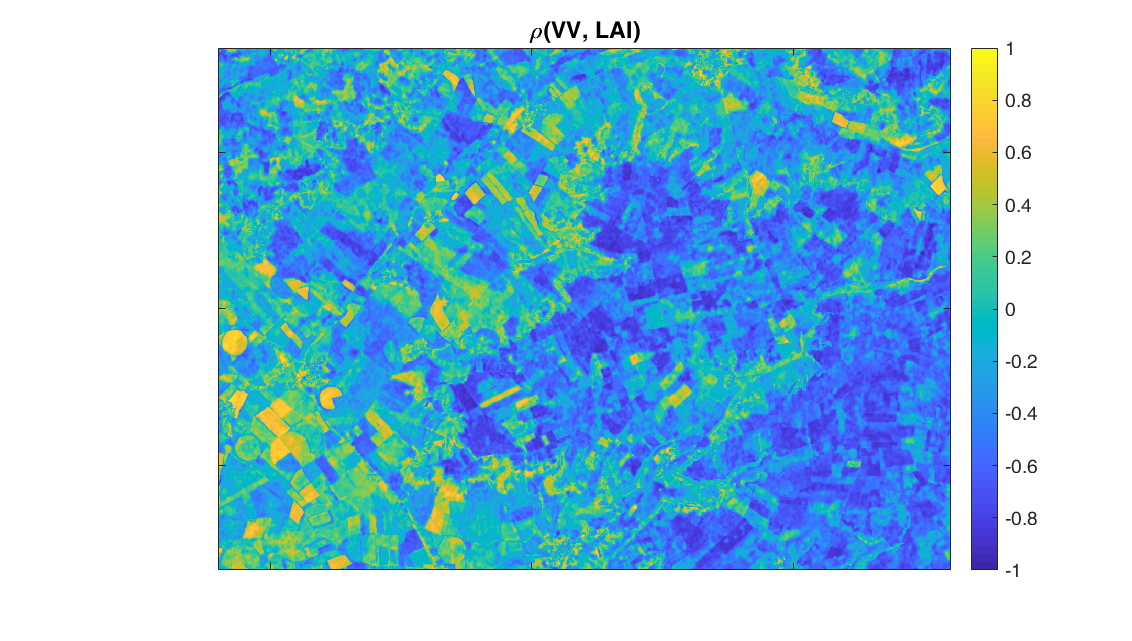} &
    \IG[width=8cm,trim={5cm 1.8cm 2cm 0cm},clip]{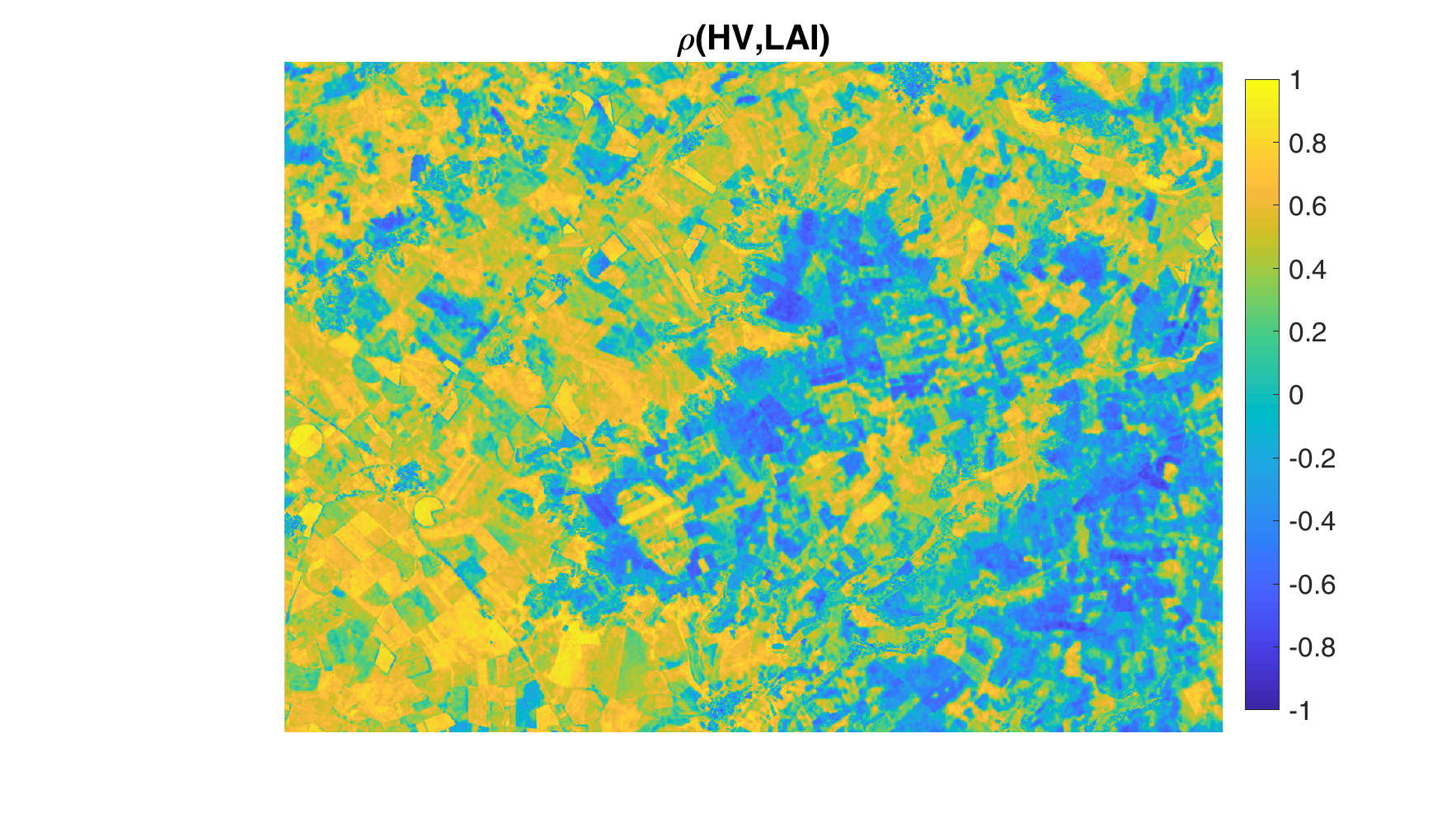} \\
    (a) & (b) \\
    \IG[width=8cm,trim={5cm 1.8cm 2cm 0cm},clip]{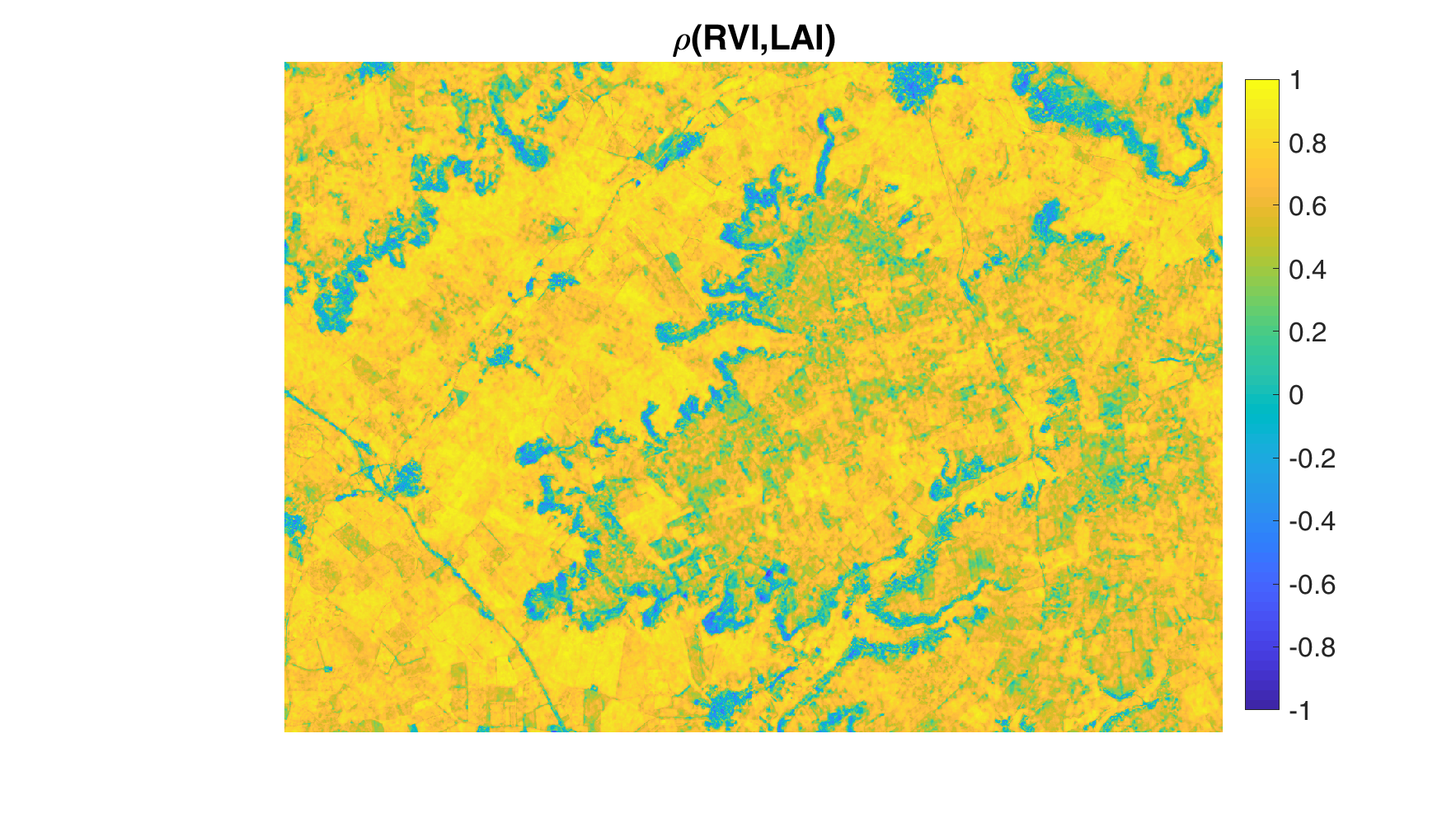} &
    \IG[width=8.6cm,trim={4.6cm 1.8cm 0cm 0cm},clip]{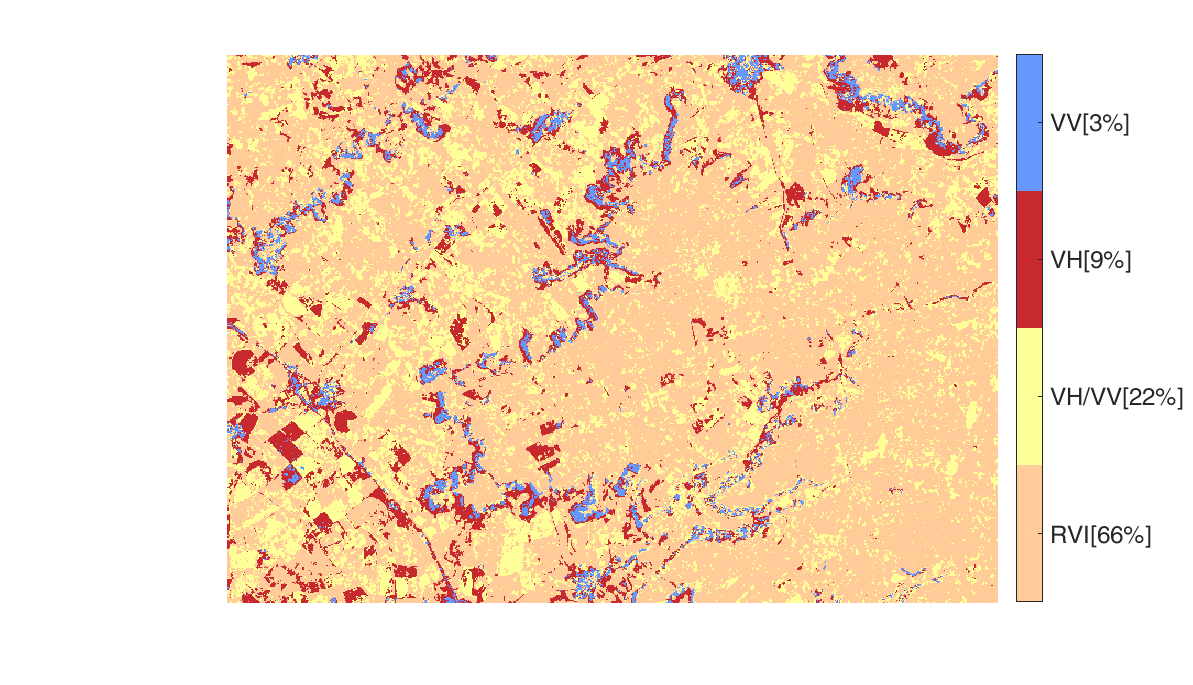} \\
    (c) & (d)
\end{tabular}
\end{center}
\vspace{-0.45cm}
\caption{Spatial distribution of temporal correlation $\rho_t$ between LAI and SAR vegetation descriptor time series - VV(a), VH(b), RVI (c) - and map of descriptor with the highest $\rho_t$ (d) over the AOI.}
\label{Fig:Pearson2DImgs}
\end{figure}

The spatial distribution of $\rho_t$ between LAI and VV,VH and RVI are shown in Figure \ref{Fig:Pearson2DImgs}. We omitted VH/VV because changes with respect to RVI are negligible. \black{RVI exhibits the highest $\rho_t$ in most of the study area}. The blueish strips in Figure \ref{Fig:Pearson2DImgs}c correspond to areas of low correlation, mainly pastureland. Here, volumetric scattering does not play an important role at C-band and surface backscattering dominates the SAR response, \black{hence explaining the higher correlation with VV band (see Figure \ref{Fig:Pearson2DImgs}a) \citep{cloude2010}. VV presents} its highest values of $\rho_t$ over some scattered parcels in the Southwest part of the AOI\black{, which are mainly labeled as sunflower,} but even there it is VH that outperforms the other three descriptors. 
Except for pasture, VV polarization tends to be less adequate for describing the crop stage evolution. \black{All this information is summarized} in Fig.~\ref{Fig:Pearson2DImgs}d, which shows the product with the highest \black{time} correlation score at each pixel: RVI outperforms the other three descriptors over more than 65\% of pixels, followed by VH/VV ratio in 23\%, VH in 9\% and finally VV in 3\% of pixels. Based on these results, RVI and LAI time series are chosen as the optimal pair of crop phenology descriptors to be input to the proposed MOGP model for optical-SAR fusion.

\section{Results}\label{Results}

\subsection{Temporal gap filling}\label{Results1D}

To understand the functioning of MOGP, we first consider the case of a single parcel and analyze the features of its active-passive time series, interpret its trained model and assess its predictions. According to the land cover map product described in \citet{gomez2018}, it corresponded to wheat in 2016, beet in 2017 and again wheat in 2018. 

\subsubsection{Training time series}\label{Trainingtimeseries}
\begin{figure}[!t]
    \centering
    	\footnotesize
    	\resizebox{1.1\textwidth}{!}{ 
    	\begin{tabular}{c}
         %\IG[width={1.0\textwidth},trim={7cm 0cm 10 0cm},clip]{figs/S2LAI_vs_S1ASC_DES_new_axis.png}  
         \IG[width={1.0\textwidth},trim={6cm 0cm 10 0cm},clip]{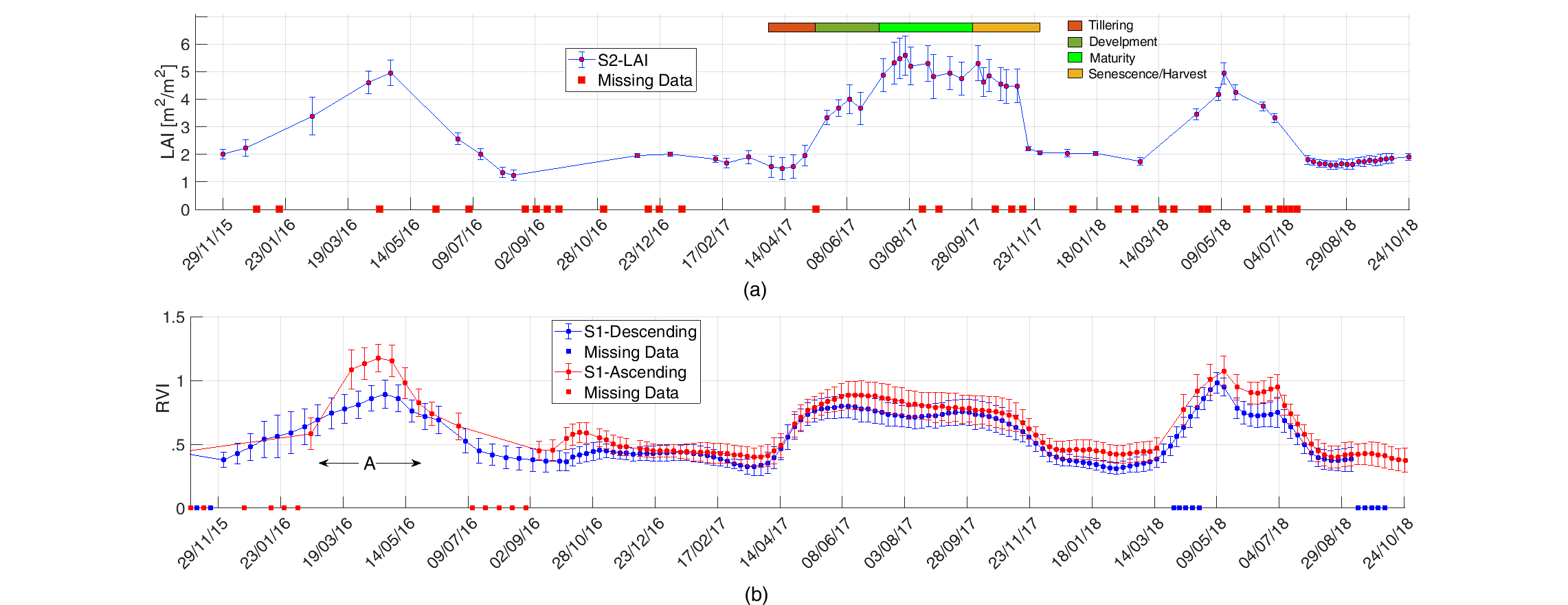}  
        \end{tabular}}
    \caption{Statistical behavior (mean value and standard deviation) of LAI(a) and ascending(red) and descending(blue) RVI(b) time series over a selected parcel from November 2015 to end of October 2018. Inteval A in (b) indicates soil contribution to RVI imparing SAR co-/cross-polar balance between the two geometries. The main phenological stages of a crop are qualitatively sketch along the second season in (a), as \black{qualitative} reference.}
    \label{Fig:InputTimeSeries1D}
\end{figure}

The images in Fig.~\ref{Fig:InputTimeSeries1D} show the time evolution of LAI (a) and ascending/descending RVI (b) statistics (mean value and standard deviation) over the selected parcel. Many data are missing in S2 collection due to clouds (orange squares along the time axis of the upper image). A much denser sampling is provided by S1 along the whole observation period, as expected.
For the second season, which offers the highest number of LAI samples, we sketched an example of the duration of the main crop phenological states (tillering, development, maturity, senescence/harvest) based on the evolution of LAI typically expected for each of them.
In general, a high resemblance characterizes the active-passive collections. Concerning RVI, slight differences are observable between ascending (red) and descending (blue) profiles, mainly during 2016 wintertime (interval A). \black{Soil anisotropic features, such as a specific plowing orientation, \black{can induce} backscattering differences under different observation geometries when \black{the illuminating microwave interacts with the ground after propagating through the vegetation layer}. Soil contributions impair the balance of energy between SAR co-polar and cross-polar bands expected over dense vegetation, explaining also RVI values higher than 1 \citep{szigarski2018}. Nonetheless, this effect is unusual and non-representative of the general behavior, as demonstrated in Table \ref{Table:spatialCorr}. Accordingly, the RVI collections from ascending and descending orbits are merged and averaged when two daily samples are available. \black{Finally}, LAI and daily-averaged RVI \black{time series} are used for training}.

\subsubsection{Physical Interpretation of MOGP model}\label{UndstrMOGPmodel} 
\black{T}he LAI/RVI collections of Figure \ref{Fig:InputTimeSeries1D} are used to optimize the free parameters \black{of the MOGP model} defined in Section \ref{ModeSetupTraining}. Their value is reported in Table \ref{Table:MOGPParams}. A specific lengthscale parameter $\ell$ is obtained for each independent GP, one being significantly higher than the other. Recalling that $\ell$ defines the smoothing properties of the covariance kernel \citep{rasmussen2004}, we refer to the GP with lower $\ell$ as LFGP (Low-Frequency), and to the other one as HFGP (High-Frequency). Note that the GP \black{suffix} has been omitted in Table \ref{Table:MOGPParams} for ease of notation. According to the formulation in (\ref{Eqn:TGP1}), \black{each semi-definite rank-1 $[B]$ matrices is expressed as the outer product of the corresponding vector $\mathbf{a}$}.

\begin{table}[t!]
\begin{center}
    \begin{tabular}{lcclcc}
    \hline
    \multicolumn{6}{c}{\textbf{MOGP Optimized Parameters}} \\ 
    \hline
    \textbf{$\ell_{LF}$}    & \multicolumn{2}{l}{\blue{76.44}}   & \textbf{$\ell_{HF}$}    & \multicolumn{2}{l}{\blue{13.43}} \\
    $[\mathbf{B}_{LF}]=\mathbf{a}_{LF}^{ }\mathbf{a}_{LF}^\intercal$    & \multicolumn{2}{l}{ $\begin{bmatrix}  0.709 & \blue{0.912}   \\  \blue{0.912} & 1.173   \\   \end{bmatrix}$ }      & 
    $[\mathbf{B}_{HF}]=\mathbf{a}_{HF}^{ }\mathbf{a}_{HF}^\intercal$   & \multicolumn{2}{l}{ $\begin{bmatrix}  0.180 & \blue{-0.013}   \\ \blue{-0.013} & 0.001   \\ \end{bmatrix}$ }      \\
    $\mathbf{a}_{LF}^\intercal=[\mathbf{a}^{LF}_{1} \mathbf{a}^{LF}_{2}]$  & \multicolumn{2}{l}{ $\begin{bmatrix}  0.8420 &  1.0831  \\ \end{bmatrix}$ }   & 
    $\mathbf{a}_{HF}^\intercal=[\mathbf{a}^{HF}_{1} \mathbf{a}^{HF}_{2}]$  & \multicolumn{2}{l}{ $\begin{bmatrix}  0.4243 & -0.036   \\ \end{bmatrix}$ }  \\
     \hline
\end{tabular}
\caption{Main hyperparameters of the MOGP model trained \black{with the LAI and RVI time series in Figure \ref{Fig:InputTimeSeries1D}}.}
\label{Table:MOGPParams}
\end{center}
\end{table}

\black{Each} $[B]$ contains all the rules linking the corresponding GP to the two outputs, but \black{its} off-diagonal term $b_{12}$ \black{directly describes their degree of correlation with respect to that GP}. As the off-diagonal element of $[B]_{LF}$ is close to 1, LFGP accounts for the common trends of LAI and RVI evolution. Moreover, being $\ell_{LF}$ significantly higher than $\ell_{HF}$ points out that only LFGP is able to describe the long-time changes of vegetation phenology. Conversely, $\ell_{HF}$ is low and the off-diagonal element of $[B]_{HF}$ close to zero, meaning HFGP fulfills essentially the task to absorb the short-time uncorrelated changes of the two outputs. 
The role of each GP can also be interpreted by examining the coefficients entering the linear combination in (\ref{eq:mogp_system}), which can be now rewritten as: 
\begin{eqnarray}
\begin{array}{cc}
LAI(t)= a^{LF}_{1}{u}_{LF}(t)+a^{HF}_{1}{u}_{HF}(t)\\
RVI(t)= a^{LF}_{2}{u}_{LF}(t)+a^{HF}_{2}{u}_{HF}(t),
\end{array}
\label{Eqn:LAI_RVI_system}
\end{eqnarray}
where $u_{LF}(t)$ and $u_{HF}(t)$ are the latent samples from LFGP and HFGP, respectively.
Table~\ref{Table:MOGPParams} shows that $a_1^{LF}>2a_1^{HF}$, implying the contribution of HFGP to LAI is of minor relevance. Because $a_2^{LF}>>a_2^{HF}$, it becomes negligible for RVI. Then, LAI and RVI outputs generated by MOGP are essentially independent with respect to HFGP, and the prediction of LAI over cloudy dates mimics the \black{local} trend shown by the RVI output. 

\iffalse
\begin{figure}[!t]
 \centering
	\footnotesize
	\resizebox{1.1\textwidth}{!}{ 
	\begin{tabular}{l}
    %\IG[width={1.0\textwidth},trim={7cm 0cm 10 0cm},clip]{figs/MOGP_RVI_LAI_new_axis.png}  
    \IG[width={1.0\textwidth},trim={6cm 0cm 10 0cm},clip]{figs/MOGP_RVI_LAI_new_axis_real_sigma.png}  
\end{tabular}}
\caption{MOGP predictions (circles) of LAI (a) and RVI (b) on S2 acquisition dates provided by the MOGP model trained on input time series (asterisks). The uncertainty of $\pm$ 2$\sigma$ is represented by the boundary shade (grey) at any estimate. In (a), interval B indicates the more realistic end of tillering dynamics obtained by fusing RVI information; interval C the higher uncertainty of LAI retrieval due to local differences between training samples of LAI (C) and of RVI (E) over the same period. In (b), intervals D and F indicate local inconsistencies between ascending and descending RVI samples.}
 \label{Fig:MOGPRetrieval}
\end{figure}
\fi

\begin{figure}[!t]
 \centering
	\footnotesize
	\resizebox{1.1\textwidth}{!}{ 
	\begin{tabular}{l}
    %\IG[width={1.0\textwidth},trim={7cm 0cm 10 0cm},clip]{figs/MOGP_RVI_LAI_new_axis.png}  
    \IG[width={1.0\textwidth},trim={6cm 0cm 10 0cm},clip]{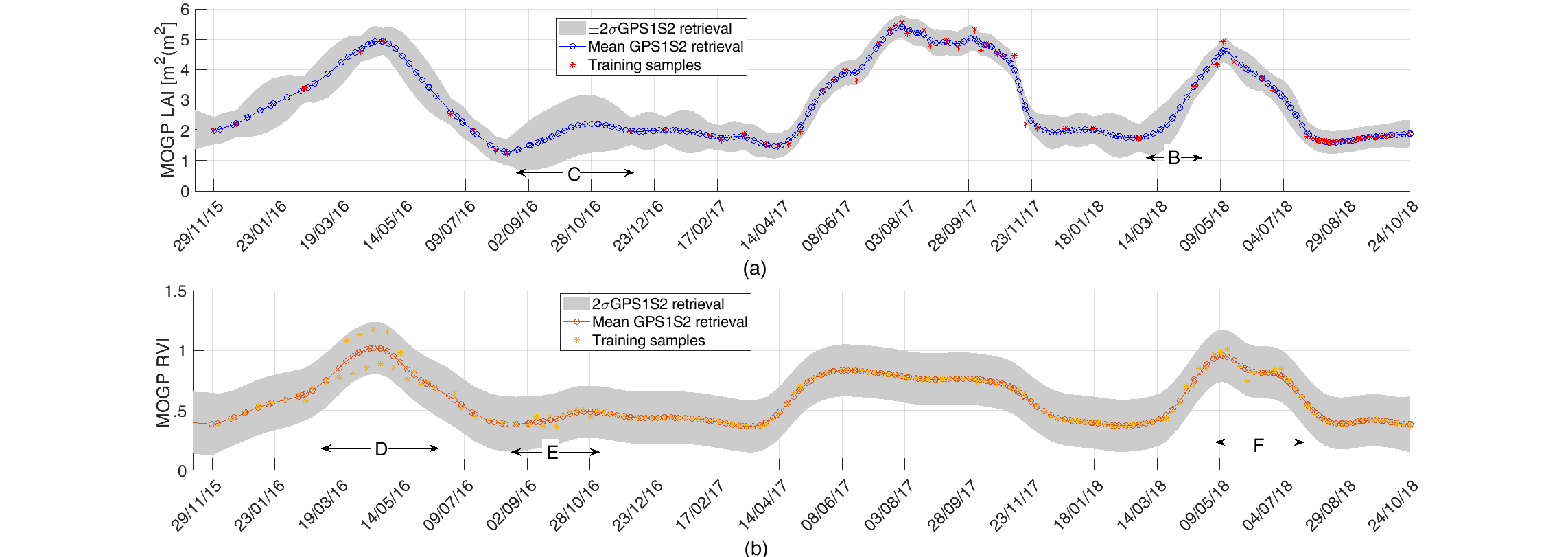}  
\end{tabular}}
\caption{MOGP predictions (circles) of LAI (a) and RVI (b) on \black{the union of S1 and S2 acquisition dates} provided by the MOGP model trained on input time series (asterisks). The uncertainty of $\pm$ 2$\sigma$ is represented by the boundary shade (grey) at any estimate. In (a), interval B indicates the realistic end of tillering dynamics obtained by fusing RVI information; interval C the higher uncertainty of LAI retrieval due to local differences between training samples of LAI (C) and of RVI (E) over the same period. In (b), intervals D and F indicate local inconsistencies between ascending and descending RVI samples \black{likely due to anisotropic soil contributions}.}
 \label{Fig:MOGPRetrieval2}
\end{figure}

\subsubsection{Prediction analysis of temporal profile} \label{PredictionAnalysis}

\black{Once the MOGP model has been trained, it can be applied to predict \black{LAI at any time}. It is worth recalling from the sketch in Figure~\ref{Fig:MOGP_sketch}, that time is the only input to the model}, and it is completely arbitrary. \black{For instance, the use of the time vector corresponding to the union of the S2 (i.e. including the cloudy days) and S1 allows showing the potential of MOGP in blending the information of the two time series. This high prediction sampling makes it easier to appreciate the capability of the technique to infer LAI missing information backing up on RVI time dynamics.}
%\black{To show the potential of MOGP to blend the information of the two time series, we have chosen a time vector corresponding to the union of the S2 nominal (i.e. including the cloudy days) and S1 capture dates. The high prediction sampling makes it easier to appreciate the capability of the technique to infer LAI missing information backing up on RVI time dynamics.}
\black{ Figure~\ref{Fig:MOGPRetrieval2} shows the LAI (a) and RVI (b) predictions of the model examined in Section \ref{UndstrMOGPmodel}, along with their uncertainty. The two quantities are given by \eqref{eq:coreg_m}, with two different blocks of $\sK$ obtained by the training samples and the optimized model's hyperparameters. The training samples are labeled by asterisks, the prediction mean values by circles, and their $\pm$ 2$\sigma$ uncertainty by the boundary shade (grey).}

\black{LAI predictions generally match LAI training values, but the covariance filtering leads to a smoother time evolution over periods where training information is noisier, as it happens during the summer season in 2017. Note that smoother prediction profiles do not necessarily imply higher uncertainties if the sampling density \black{allows discriminating noisy information}. \black{Over periods of densely sampled optical information}, LAI training information dominates \black{LAI} prediction whereas RVI contributions are almost irrelevant. \black{It is over periods with scarce optical samples}, such as in 2016, that the advantage of the multioutput becomes clear. \black{Instead of using only the LAI training samples, MOGP retrieves LAI using also the statistical correlation with RVI established during the training. MOGP is then able to reconstruct the transitional \black{LAI} behaviors that have been undersampled by S2}. For instance, the rise of LAI in March 2018 (interval B), which corresponds to the end of the tillering period and the so-called Stard-Of-Season (SOS) \citep{jonsson2004}, is reconstructed in a realistic way because \black{this fast} evolution \black{was} captured by RVI.}

Concerning the uncertainty of the LAI prediction at a specific date, \black{both the availability of samples and the local agreement between the two training collections play a role}. The uncertainty is low when there is good match, but increases when only one parameter is available and some inconsistency is detected between them at the temporal extreme of the gap to be filled. The period labeled as interval C, between August and December 2016, shows a clear example. \black{In relative terms, the increase of LAI value required to tie the two extremes of the gap (>13\%) is higher than RVI change (<6\%) around E. Then, the trend sensed by RVI must be distorted for S2 sample fitting, generating a higher uncertainty (>1 $m^2/m^2$) with respect to other filled intervals (~0.4 $m^2/m^2$)}. 
With respect to uncertainty of RVI, the higher sampling density makes the prediction mean values match the training samples over most of the time series. The effects of noisy inputs is reduced by the covariance smoothing and the \emph{coregionalization} link to LAI, forcing the output profile not to follow inconsistent patterns. \black{Note for instance the samples in winter 2016 (interval D) following the trend of LAI and uncertainty limits defined by the inconsistency between ascending and descending orbits, or in fall 2016 and winter 2018 (intervals E and F, respectively). As for LAI output, RVI output is also influenced by the S2 training samples, although their effect is less evident}.

\black{A final remark on the information provided by MOGP is in order. The Standard GPR is a probabilistic nonparametric approach to regression and allows to compute uncertainties (predictive variance) of the inferred function. The gap filled time series provided by MOGP are obtained by transferring across domain/data sources not only the function predictions but also the uncertainties, allowing error quantification and uncertainty propagation by embedding the models in a higher level parameterization \citep{johnson2019}.}

\subsubsection{Assessment of temporal profile}\label{Assessment1D}

\begin{figure}[!t]
 \centering
	\footnotesize{ 
	\begin{tabular}{r}
    %\IG[width={1.0\textwidth},trim={7cm 0cm 4.5cm 0cm},clip]{figs/SOGP_vs_MOGP_assessment1D_new_axis.png}  
      \IG[width={1.0\textwidth},trim={6cm 0cm 4.5cm 0cm},clip]{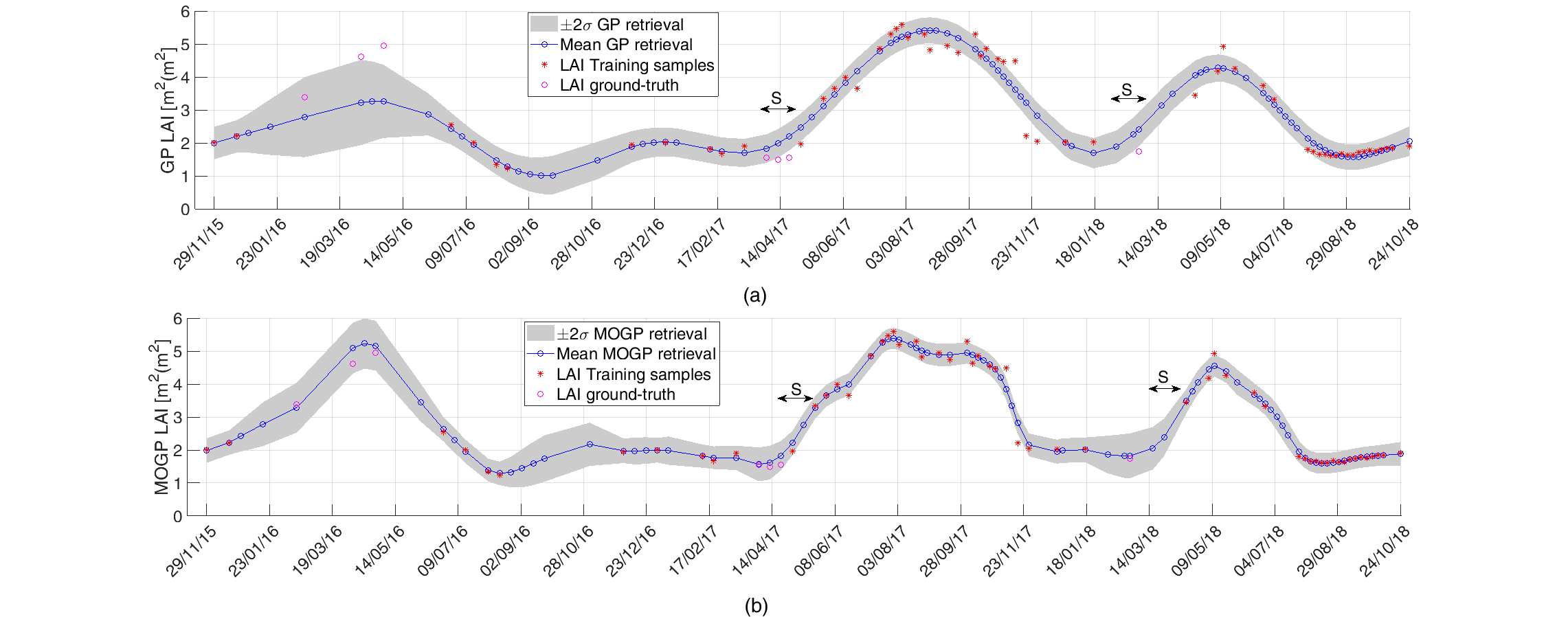}  
\end{tabular}}%\vspace{12pt}
\caption{Assessment of standard GPR(a) and MOGP(b) predictions (blue circles) for data gap filling on S2 cloud-free captures (magenta) eliminated from training information(red asterisks) and here used as reference. 
The $\pm$ 2$\sigma$ prediction uncertainty is represented by the boundary shade (grey) at any estimate. Intervals S in (a) and (b) \black{evince} the more realistic dynamics retrieved by MOGP during the stemming stages.}
 \label{Fig:MOGP_vs_SOGP_1D}
\end{figure}

\black{Here, we propose a quantitative assessment of MOGP predictions using the leave-one-image-out technique}. It consists in taking a group of LAI time samples out of the training collections and assessing their predictions using these extracted values as reference. For this purpose, \black{from the LAI time series in Figure \ref{Fig:InputTimeSeries1D}a} we eliminate 3 samples from February to April 2016, 3 samples from April to May 2017 and 1 sample in February 2018. Their elimination generates 3 \black{artificial} gaps ranging from 25 up to 90 days, which \black{exemplify realistic high-latitude S2 sampling conditions due to the prominent cloud cover}. \black{To highlight the benefits of the fusion approach, MOGP retrievals are also compared to estimations of standard GPR trained on the reduced collection of the LAI samples (Section \ref{SOGPR_theo})} \black{As shown in \ref{AppendixA}, GPR provided the best assessment performance among a set of conventional and advanced interpolation techniques. The result of GPR and MOGP retrievals are shown in Figure~\ref{Fig:MOGP_vs_SOGP_1D}. The predictions have been carried out only on the dates of S2 nominal captures to facilitate the identification of the samples useful for the assessment. Yet, it is worth stressing again that there is no limitation in the time sampling of the prediction output provided by the MOGP trained model.}

The lower capability of GPR to reconstruct LAI \black{over} long time data gaps can be easily observed for winter-spring 2016; the true phenological evolution of the crop (magenta circles) is completely lost because all missing values are significantly underestimated. Moreover, the predictions are characterized by \black{a} higher uncertaint\black{y} and, consequently, \black{a lower reliability}. The retrieval over the other \black{assessment} dates generally overestimates the LAI true value, with a consequent anticipation of the stemming stage \black{(indicated with an S in Figure \ref{Fig:MOGP_vs_SOGP_1D})}. Note that the true values fall often outside the 2$\sigma$ significance interval. 
Conversely, the simultaneous use of SAR-based information allows a more correct retrieval of crop phenological stages, within short and long data gaps. In most cases, the mean LAI estimation provided by MOGP almost coincides with the S2-based reference, causing no time shift or stretching in the description of the crop seasons. The quantitative assessment of the 7 predictions carried out by GPR and MOGP provides $R^2=0.90$ and $RMSE=0.95 m^2/m^2$ for the former, $R^2=0.99$ and $RMSE=0.24 m^2/m^2$ for the latter. 
\black{Another observation is that the LAI retrieval carried out over the longest time gap relies mainly on SAR data alone, while the contribution from optical data is marginal. Here, the meaning of passive-active data fusion is likely lost at local scale. Nonetheless, it is only thanks to the synergistic rules learned by MOGP over the fused active-passive complete time series that inferring the LAI behavior from RVI time evolution becomes feasible.} 

\subsection{Spatiotemporal gap filling} % \label{Results2D} % repeated label
The noteworthy results provided by MOGP over the selected parcel underpin its capability to merge \black{profitably} the active-passive information, \black{and reconstruct otherwise lost details of the vegetation phenology}. Yet, its performances are deeply related to the \black{time resemblance} of the training time series. \black{As the areas crop type often differs across multiple crop seasons}, it is interesting to analyze the behavior of MOGP when trained using long time series over time-heterogeneous areas, and at pixel instead of parcel level. 

\subsubsection{Spatial patterns of MOGP trained models}

\begin{figure}[!t]
 \centering
	\footnotesize
	\resizebox{1.0\textwidth}{!}{ 
	\begin{tabular}{ccc}
    \IG[width={0.35\textwidth},trim={2.5cm 0cm 2cm 0cm},clip]{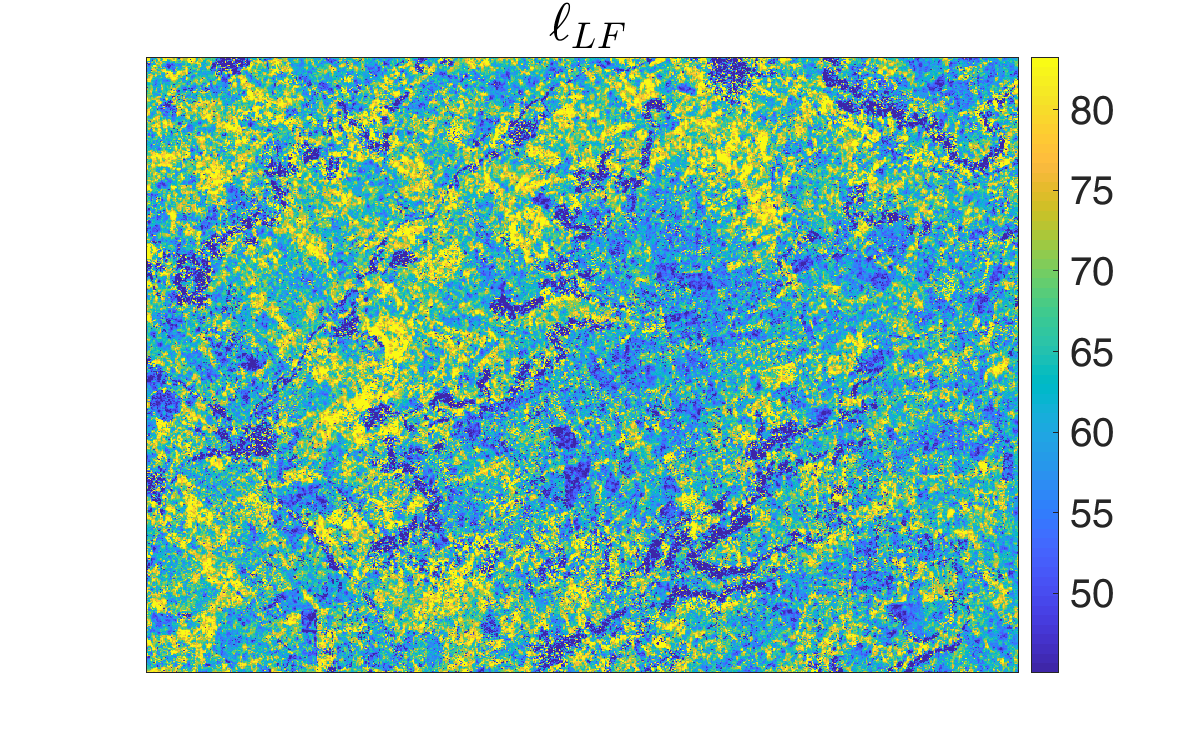}&
    \IG[width={0.35\textwidth},trim={2.5cm 0cm 2cm 0cm},clip]{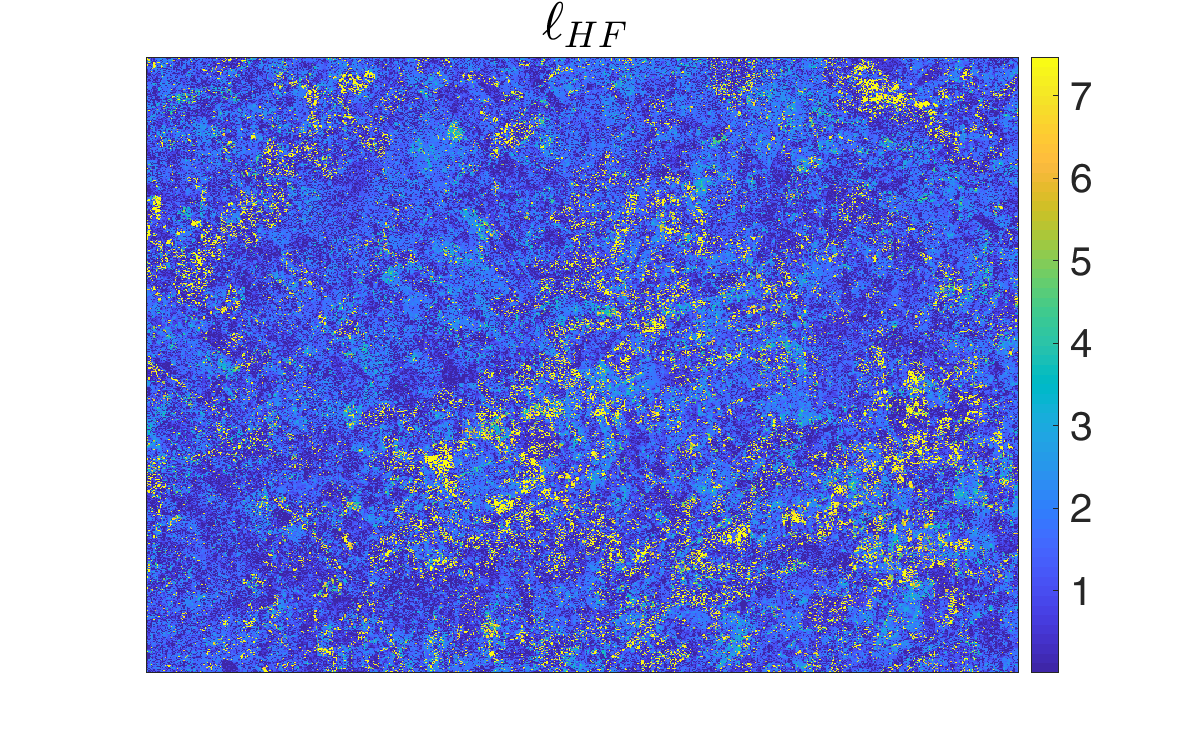}&  
    \IG[width={0.3\textwidth},trim={1.5cm 0cm 2cm 0cm},clip]{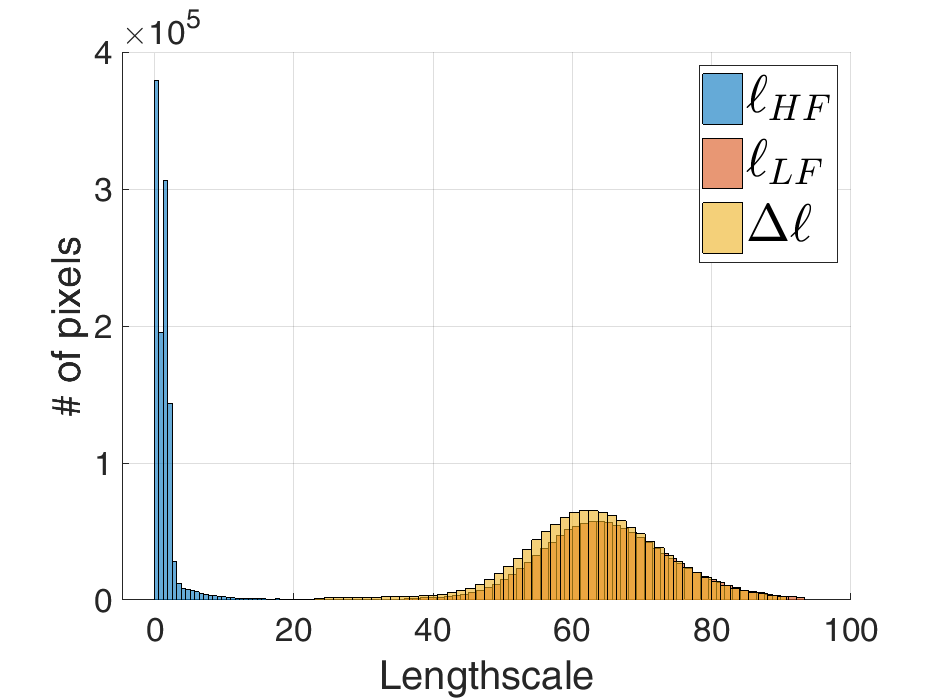}\\ 
    (a) & (b)&(c)\\
    \IG[width={0.35\textwidth},trim={3.5cm 0cm 1.5cm 0cm},clip]{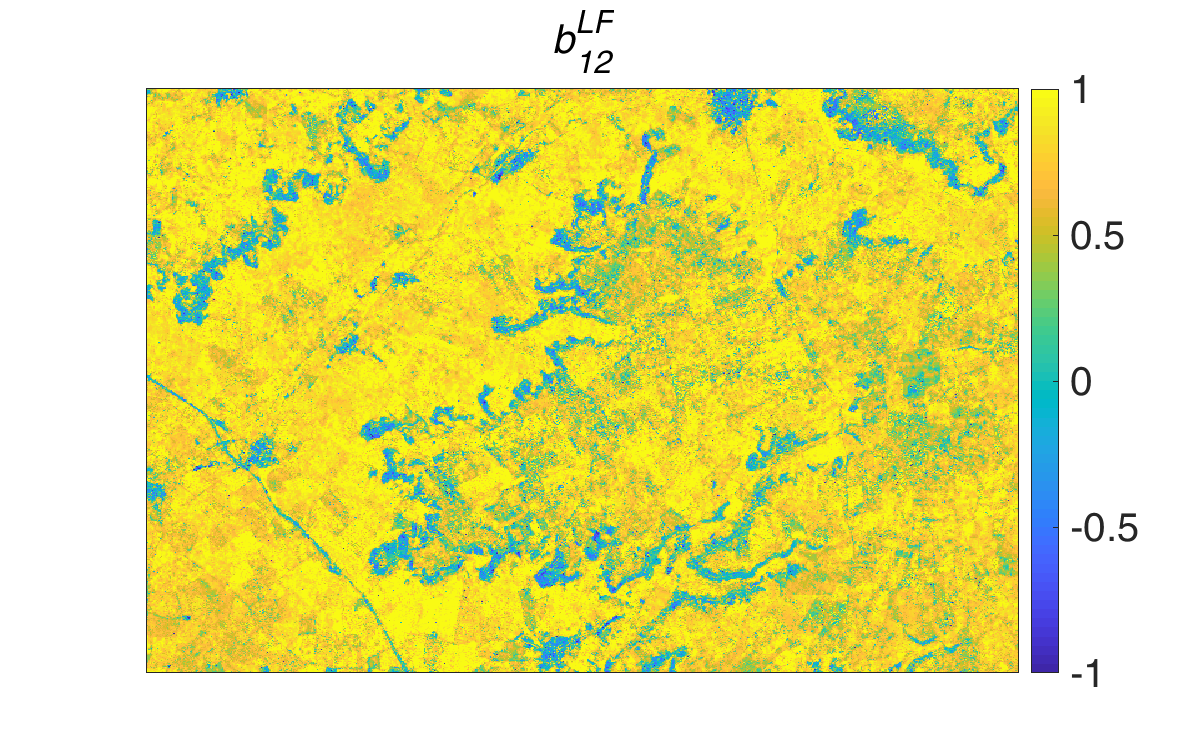}&  
    \IG[width={0.35\textwidth},trim={3.5cm 0cm 1.5cm 0cm},clip]{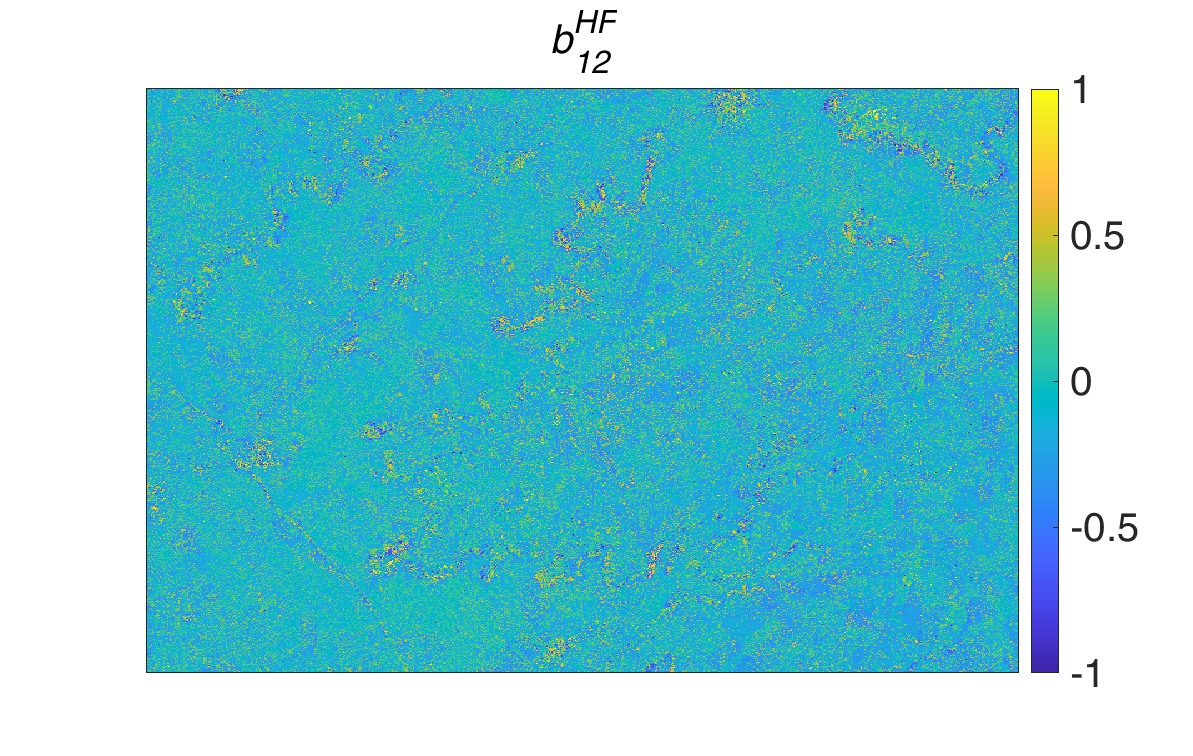}&
    \IG[width={0.36\textwidth},trim={1cm 0cm 1.5cm 0cm},clip]{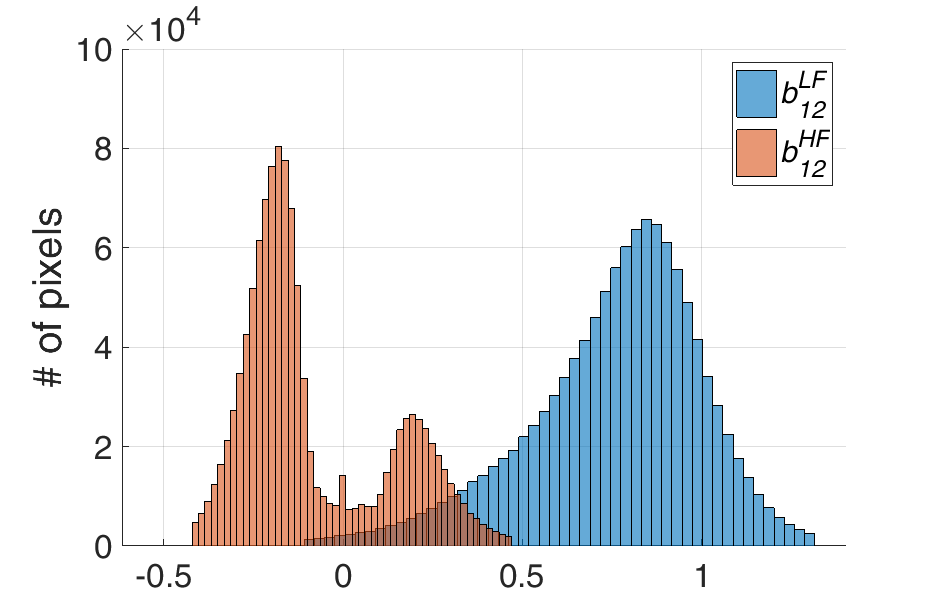}\\ 
    (d) & (e) & (f)\\
   \end{tabular}} \vspace{2pt}
\caption{Spatial distribution over the AOI and histograms over crop areas of $\ell$ parameter for LFGP and HFGP (a,b and c in upper \black{panel}), and of $b^{LF}_{12}$ and $b^{HF}_{12}$ (d,e,f in lower \black{panel}).}
 \label{Fig:MOGP_LengthSc}
\end{figure}

\black{Keeping in mind the interpretation of the MOGP model in Section \ref{UndstrMOGPmodel}, \black{here} we extend the analysis to the spatial domain. An independent model \black{is} trained for each pixel of the AOI using its corresponding LAI and RVI complete time series. Figure \ref{Fig:MOGP_LengthSc}a,b and c show the spatial distribution and the histograms of $\ell_{LF}$ and $\ell_{HF}$. The results over vegetated areas confirm MOGP identifies a LFGP accounting for the information shared between the two outputs, and a HFGP explaining their local differences along time. $\ell_{LF}$ varies from 40 to 80 days over croplands, being this range consistent with the time correlation expected for vegetation phenology descriptors. \black{Pasture lands and man-made surfaces such as road and urban areas, which present low $\ell_{LF}$ values in Figure \ref{Fig:MOGP_LengthSc}a, have not been included in any quantitative statistics \black{hereafter} presented}. Higher $\ell_{LF}$ values are generally associated with barley and wheat fields, although \black{the crop rotation applied during the three years covered by the training complicates the identification of time-homogeneous crop regions}. For each crop pixel, $\ell_{LF}$ provides the optimum time-scale of the covariance kernel to follow the vegetation dynamics throughout the seasons covered by the training; \black{the values of} $\ell_{HF}$ reflect the shorter time-scale of HFGP contributions to account for local behaviors.} The almost perfect overlapping between the histogram\black{s} of $\ell_{LF}$ and \black{ $\ell_{LF}-\ell_{HF}$} ($\Delta\ell$) in Figure \ref{Fig:MOGP_LengthSc}c demonstrates this high-low frequency dichotomy holds for almost all the pixels within the scene.

We can draw some additional considerations from the spatial distribution of $b^{LF}_{12}$ and $b^{HF}_{12}$, shown in Figures \ref{Fig:MOGP_LengthSc}d and \ref{Fig:MOGP_LengthSc}e. Their histograms are also given in \ref{Fig:MOGP_LengthSc}f to facilitate their comparison. High values of $b^{LF}_{12}$ characterize most of the crop polygons, with a slight lower distribution on the East part of the AOI. The results \black{confirm that} LFGP carries the shared information of the two outputs regarding the vegetation structure, which explains also the resemblance between $b^{LF}_{12}$ and $\rho_t(RVI,LAI)$ in Figure \ref{Fig:Pearson2DImgs}c. Following a similar reasoning, \black{we can state that} the two outputs are generally more independent with respect to HFGP contributions. 

\begin{figure}[t]
 \centering
	\footnotesize
	\resizebox{0.9\textwidth}{!}{ 
	\begin{tabular}{cc}
	\IG[width={0.50\textwidth},trim={3.5cm 1.8cm 1.5cm 0cm},clip]{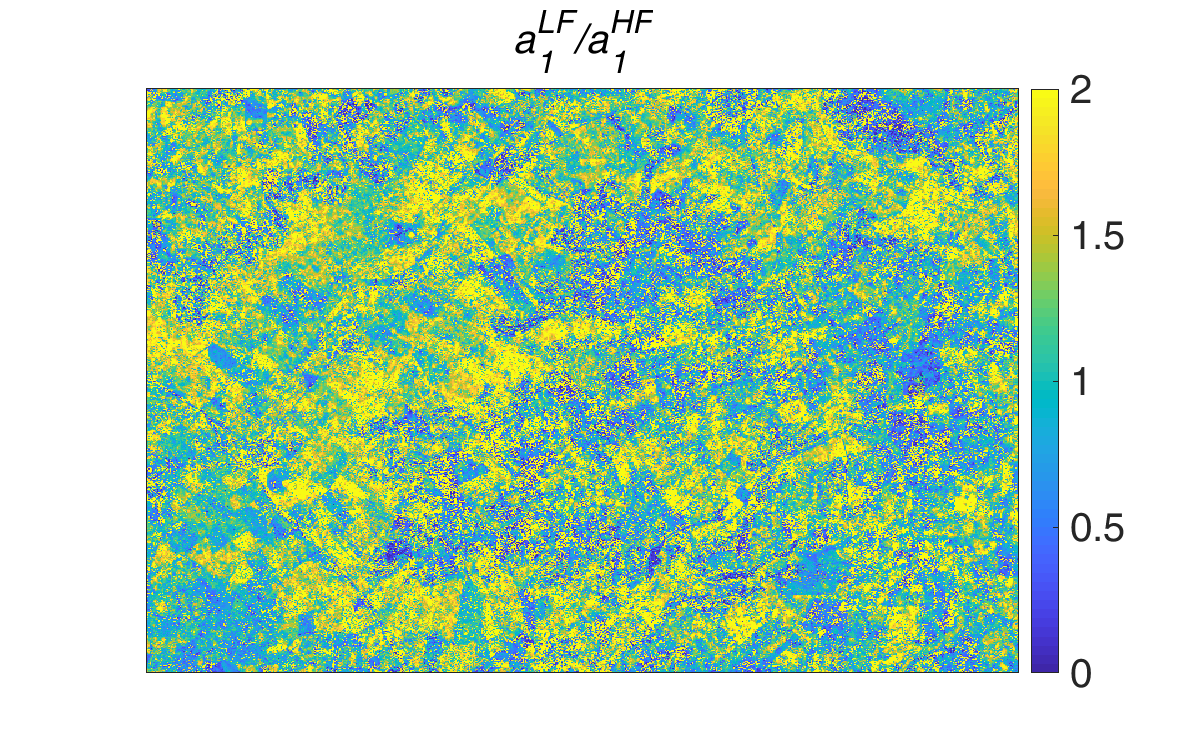} &
	\IG[width={0.50\textwidth},trim={3.5cm 1.8cm 1.5cm 0cm},clip]{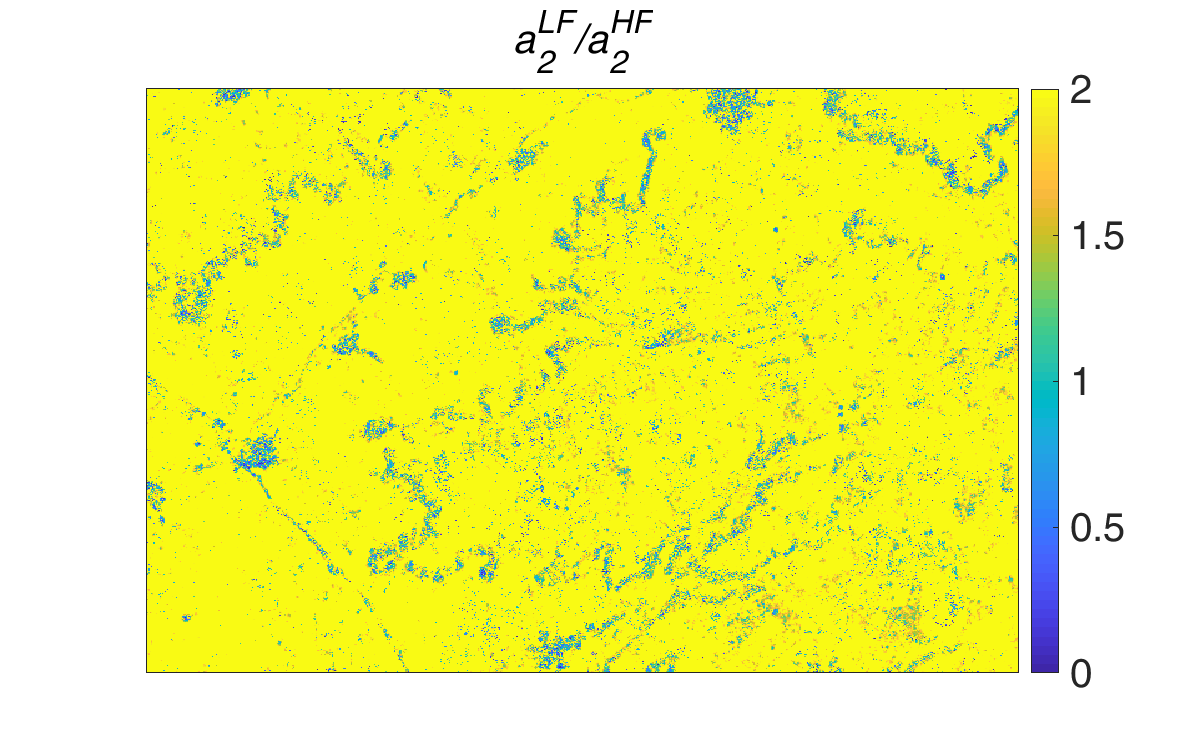}\\
	(a) &(b)\\
\end{tabular}}\vspace{3pt}
\caption{Spatial distribution of the ratios between high- (a) and low-frequency (b) coefficients in \eqref{Eqn:LAI_RVI_system}.}
 \label{Fig:MOGP_ModelParams_Ba}
\end{figure}

\black{A further step in the analysis is the study of the ratios between the\black{ir} coefficients in (\ref{Eqn:LAI_RVI_system}), i.e. $a_{1}^{LF}/a_{1}^{HF}$ and $a_{2}^{LF}/a_{2}^{HF}$.}
Their spatial maps are shown in images (a) and (b) of Figure \ref{Fig:MOGP_ModelParams_Ba}, respectively. $a^{LF}_2\gg a^{HF}_2$ \black{indicates that LFGP always dominates the linear combination providing RVI output, and indirectly that the SAR data time-smoothing preprocessing was effective (Section \ref{Timeseries_genS2}). An alternative interpretation is that RVI output information is fully contained in LAI output}. 

For LAI, the contribution balances in~\eqref{Eqn:LAI_RVI_system} becomes more complex because LFGP does not prevail everywhere. The spatial distribution of Figure \ref{Fig:MOGP_ModelParams_Ba}a reveals that 1) for about 60\% of pixels within the AOI holds the condition $a^{LF}_1>a^{HF}_1$, 2) only for 20\% of them it keeps above 1.5 and makes LFGP dominant. For the remaining 40\% pixels the contribution of HFGP is significant, and more than its half even dominant, i.e. $a_{2}^{LF}/a_{2}^{HF}<0.75$. For these latter pixels, LAI time series contains \black{non-negligible high-frequency} details \black{at higher frequency that are not sensed by RVI. As a consequence, RVI capability to facilitate LAI data gap filling becomes significantly lower}.

\iffalse
\black{Altogether, the ultimate purpose of MOGP is to generate outputs whose resemblance is maximized over periods where training samples of one of them are missing. In this reconstruction, the contributions related to HFGP are reduced to favor an evolution of LAI mainly based on LFGP. The relationship between the two LAI output's coefficients allows us creating two class labels accounting for higher and lower performances expected for the active-passive synergy. The first class contains areas where the mutual information between the active and passive time series controls the two outputs: lower prediction errors are here expected due to LFGP prevailing over HFGP. The second category gathers zones where active-passive non-shared information becomes important and MOGP data gap filling becomes less reliable.} 
\fi

\subsubsection{Spatial prediction analysis}\label{Results2D}

Following the leave-one-image-out assessment strategy proposed in Section \ref{Results1D}, the same \black{S2} acquisitions were left out for MOGP and GPR training at pixel level. For each assessment date, three LAI maps are \black{created}: the predictions of MOGP and GPR models and the map obtained from the original S2 capture \citep{amin2018}, which is used as reference for the assessment. 

\black{The quantitative comparison between reference and estimation is provided in terms of error histograms and scatter plot histograms, for two pixel selection criteria: 1) crop pixels fulfilling the condition $a^{LF}_1/a^{HF}_1>1.5$, assuring RVI contribution are dominant for the LAI \black{interpolation}; 2) all crop pixels available. For convenience, only the maps corresponding to longer gaps of 90 and 50 days are presented here, and assessed using the first pixel selection criterion. The results using both selection criteria for all the eliminated dates are presented in Section \ref{mapassessment}, along with a crop type analysis of MOGP and GPR prediction performance.}

\begin{figure}[!t]
 \centering
	\footnotesize
	\resizebox{1.1\textwidth}{!}{ 
	\begin{tabular}{ccc}
    \IG[width=3.5cm,trim={4.2cm 1.5cm 1cm 0cm},clip]{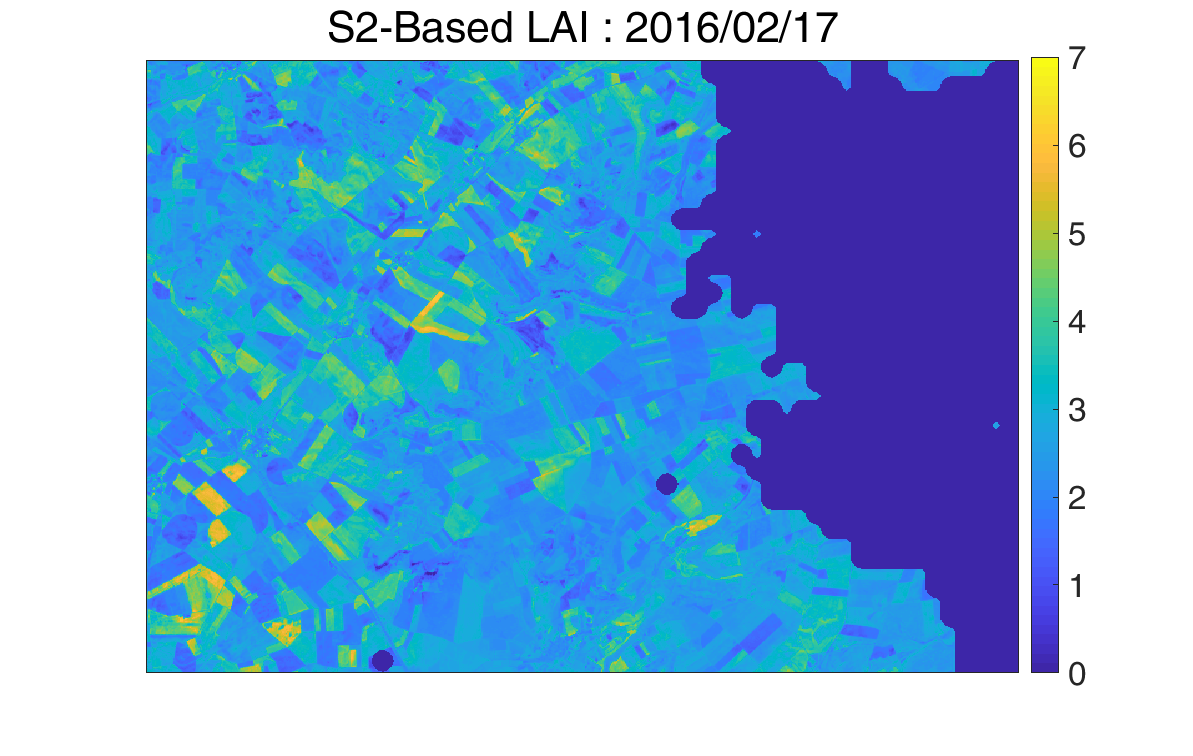}&  
    \IG[width=3.5cm,trim={4.2cm 1.5cm 1cm 0cm},clip]{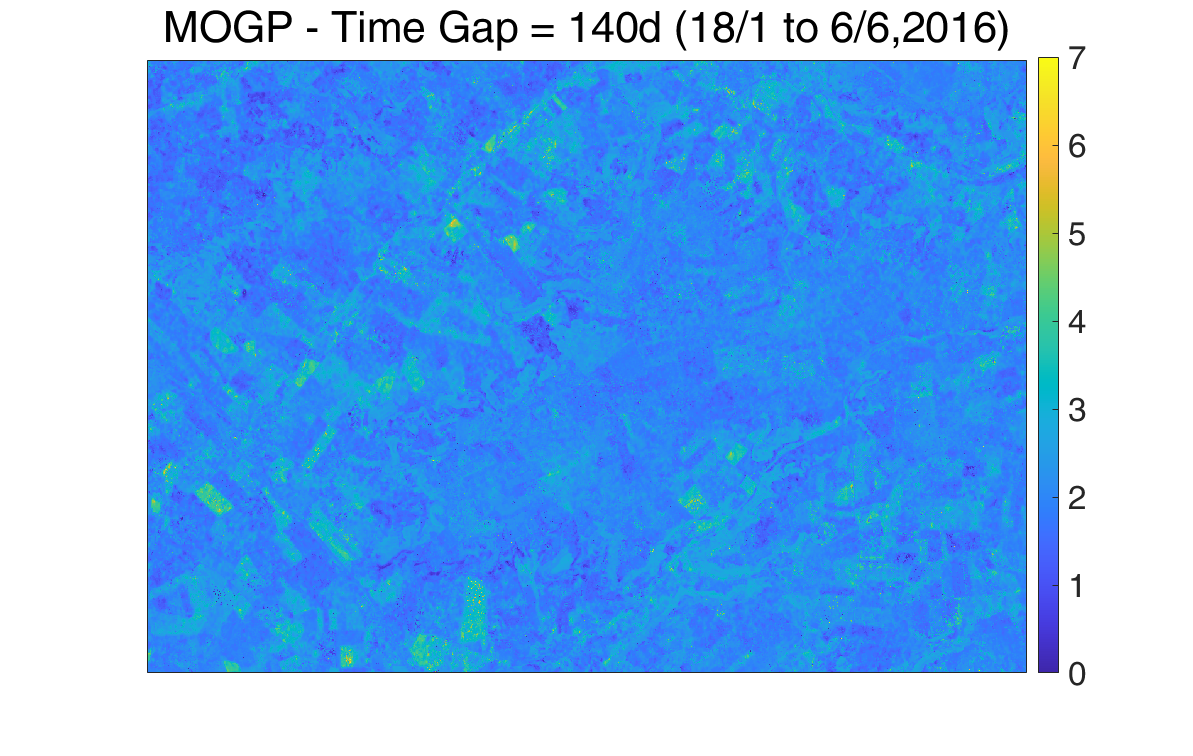}&
    \IG[width=3.5cm,trim={4.2cm 1.5cm 1cm 0cm},clip]{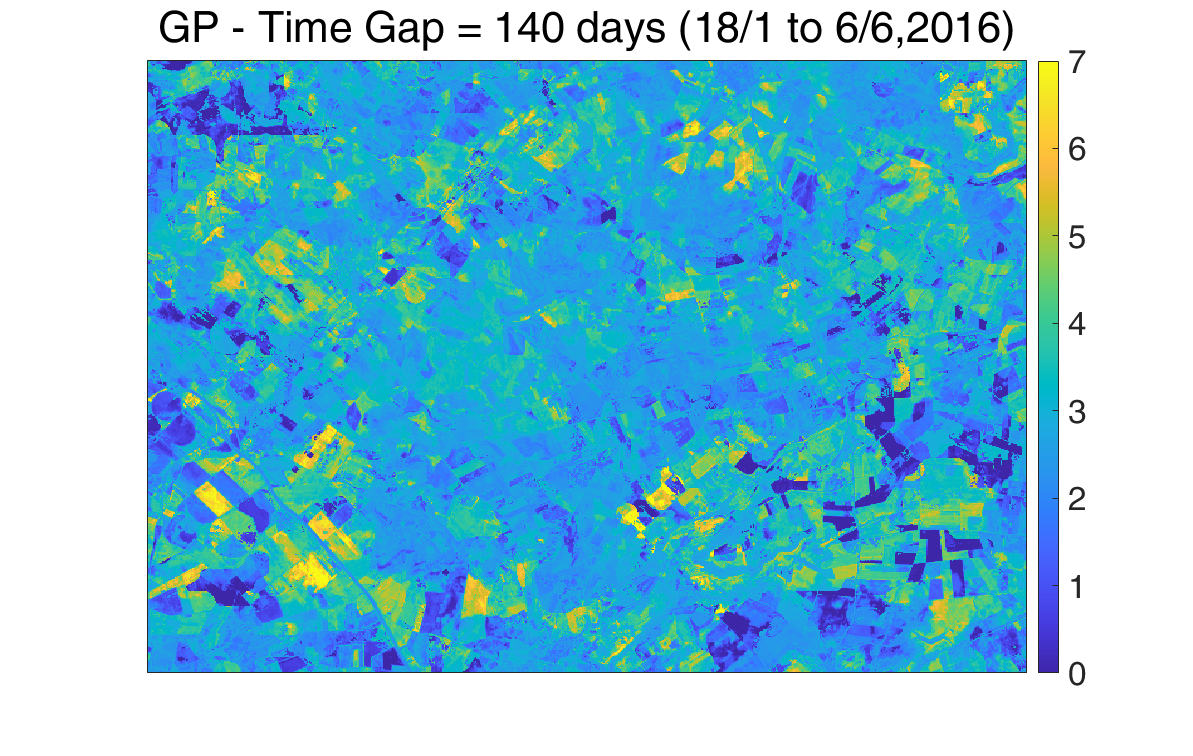}\\ 
    \IG[width=4.0cm,trim={0cm 0cm 0cm 0cm},clip]  {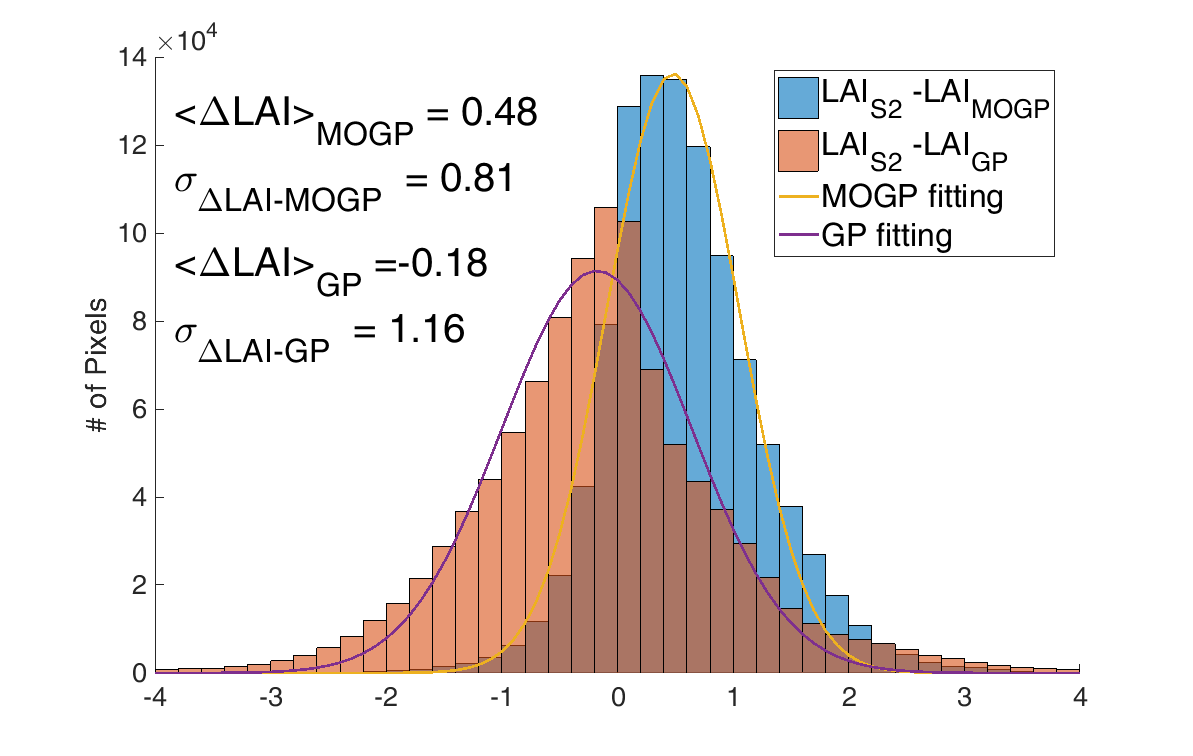}& 
    \IG[width=3.5cm,trim={0cm 0cm 0cm 0cm},clip]  {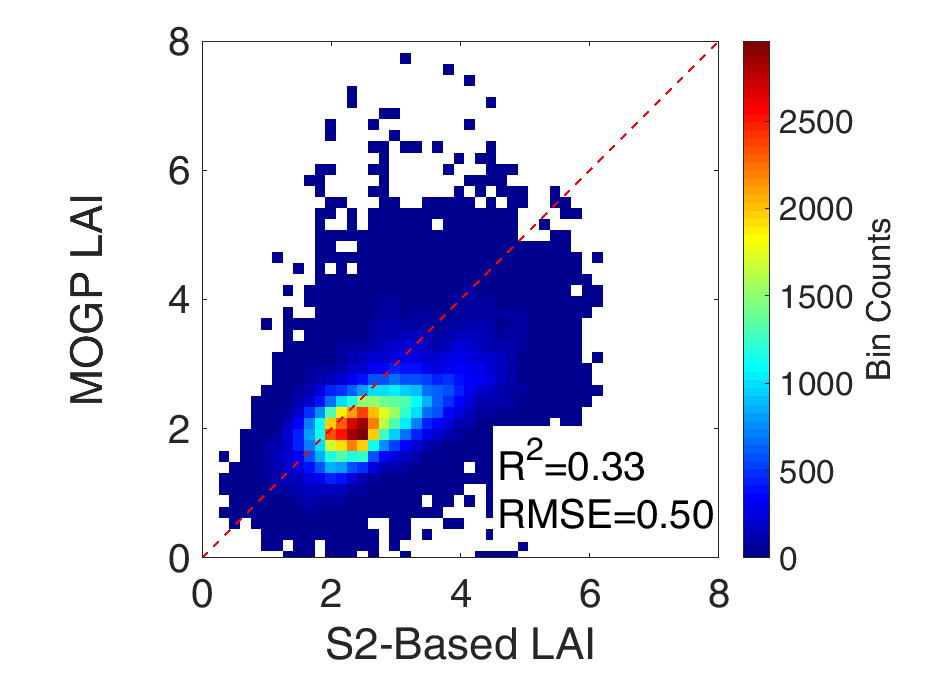}&
    \IG[width=3.5cm,trim={0cm 0cm 0cm 0cm},clip]  {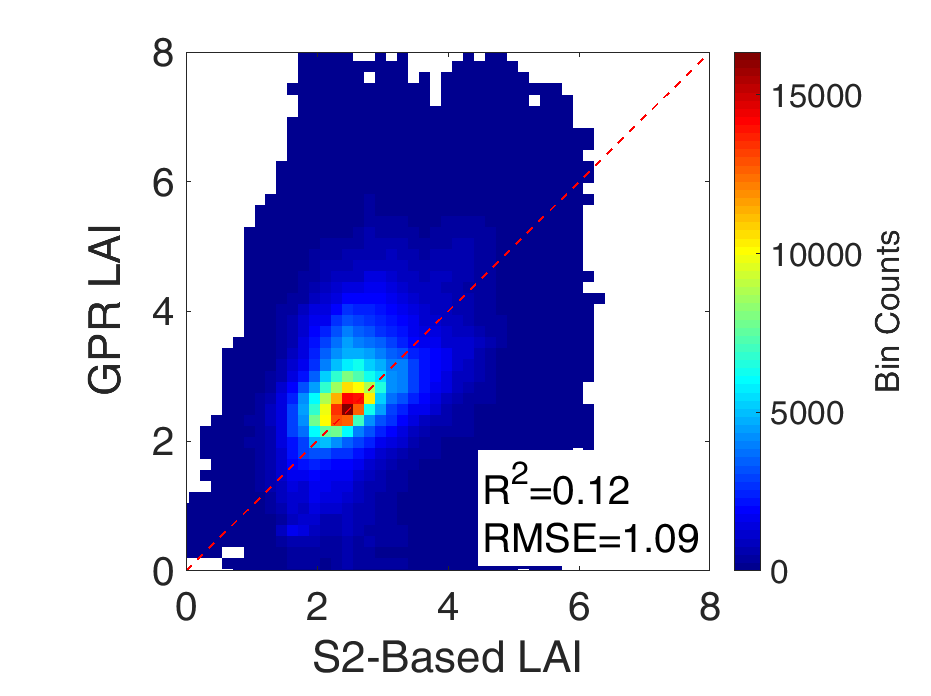}\\
    \IG[width=3.5cm,trim={4.2cm 1.5cm 1cm 0cm},clip]{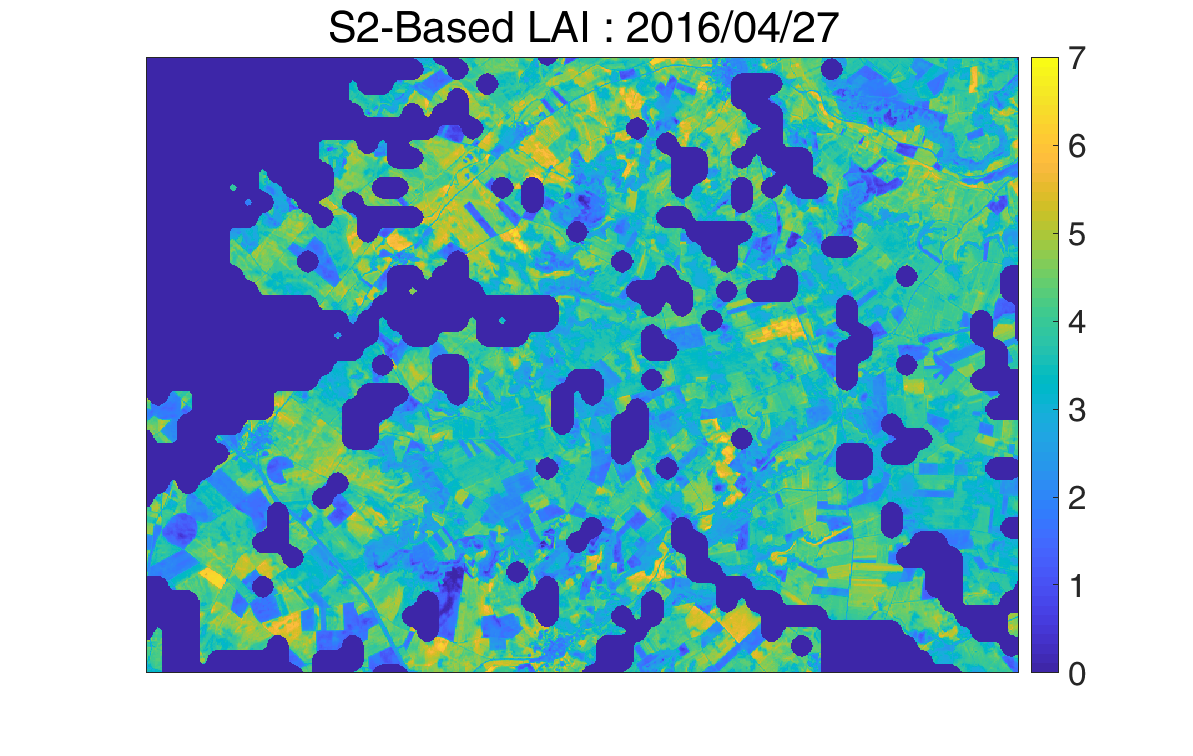}&  
    \IG[width=3.5cm,trim={4.2cm 1.5cm 1cm 0cm},clip]{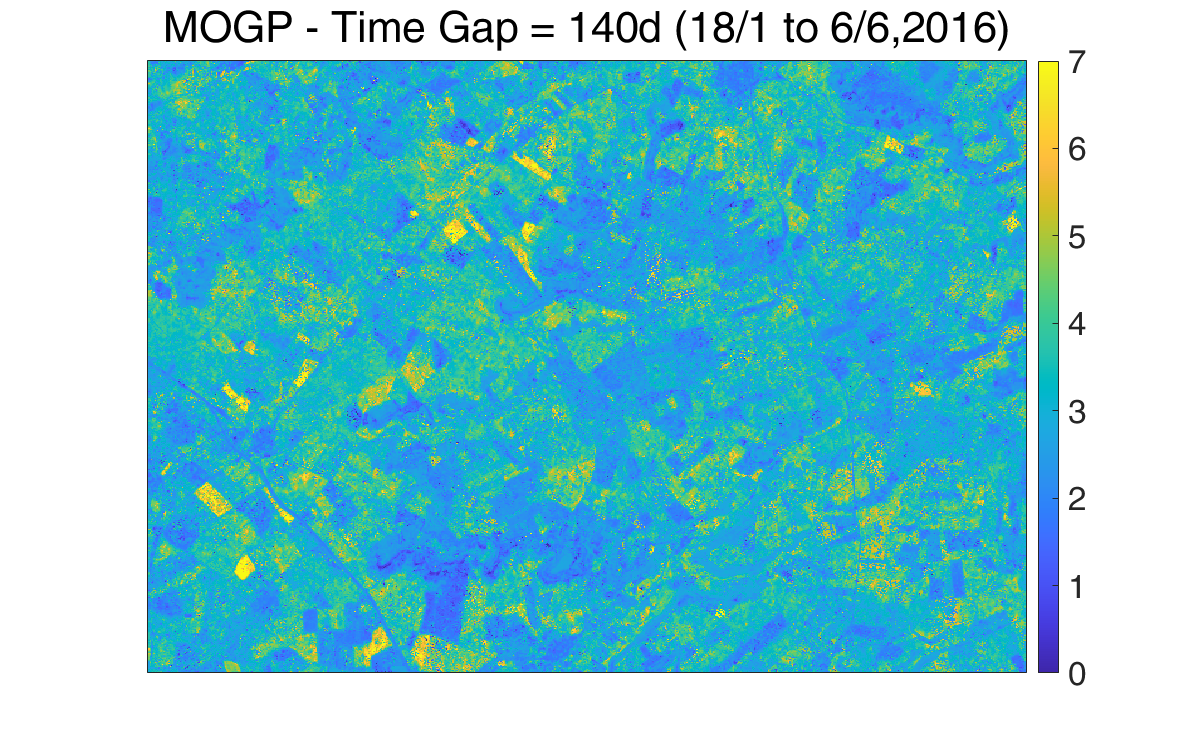}&
    \IG[width=3.5cm,trim={4.2cm 1.5cm 1cm 0cm},clip]{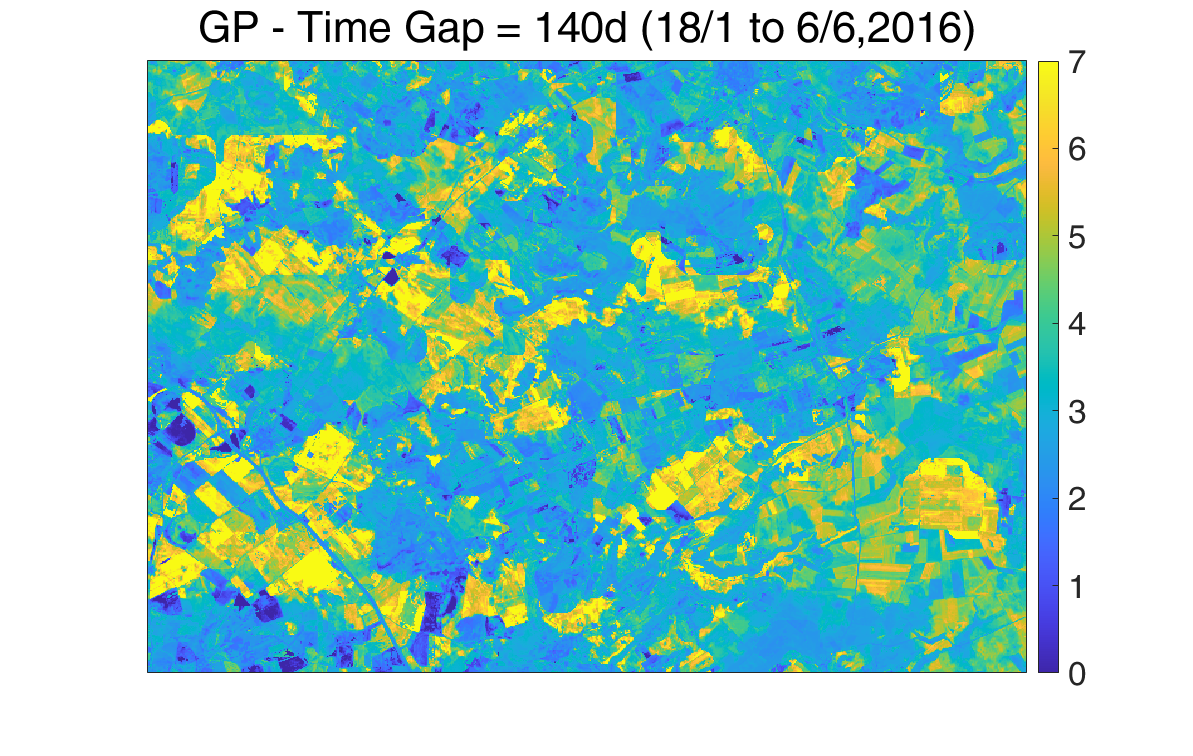}\\ 
    \IG[width=4.0cm,trim={0cm 0cm 0cm 0cm},clip]  {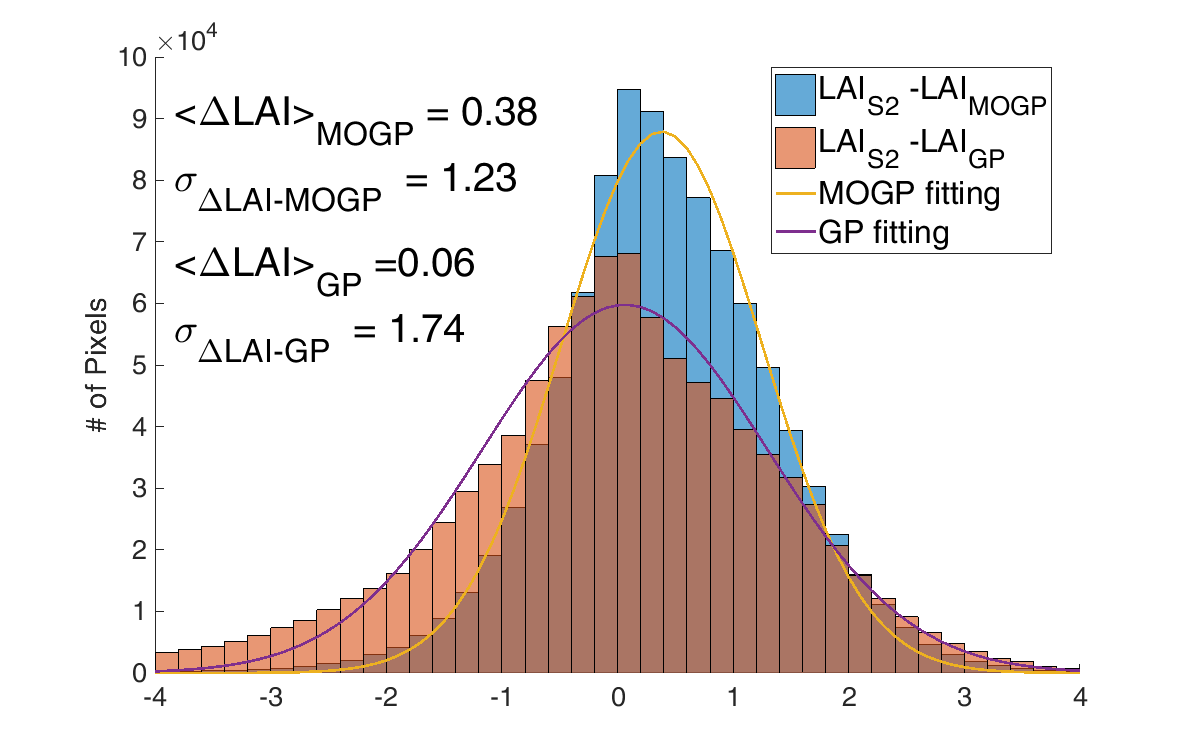}& 
    \IG[width=3.5cm,trim={0cm 0cm 0cm 0cm},clip]  {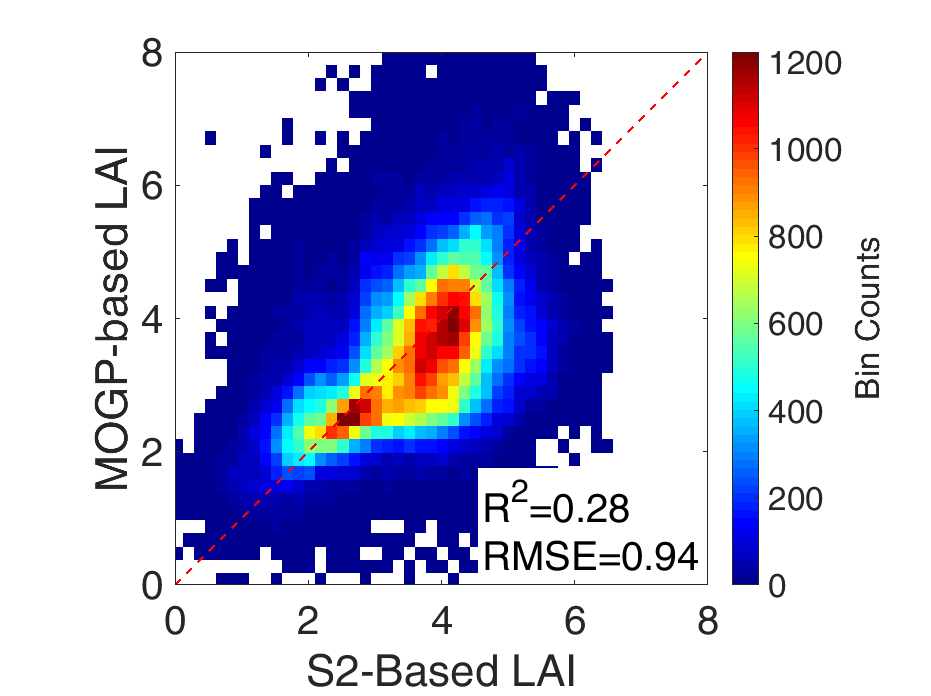}&
    \IG[width=3.5cm,trim={0cm 0cm 0cm 0cm},clip]  {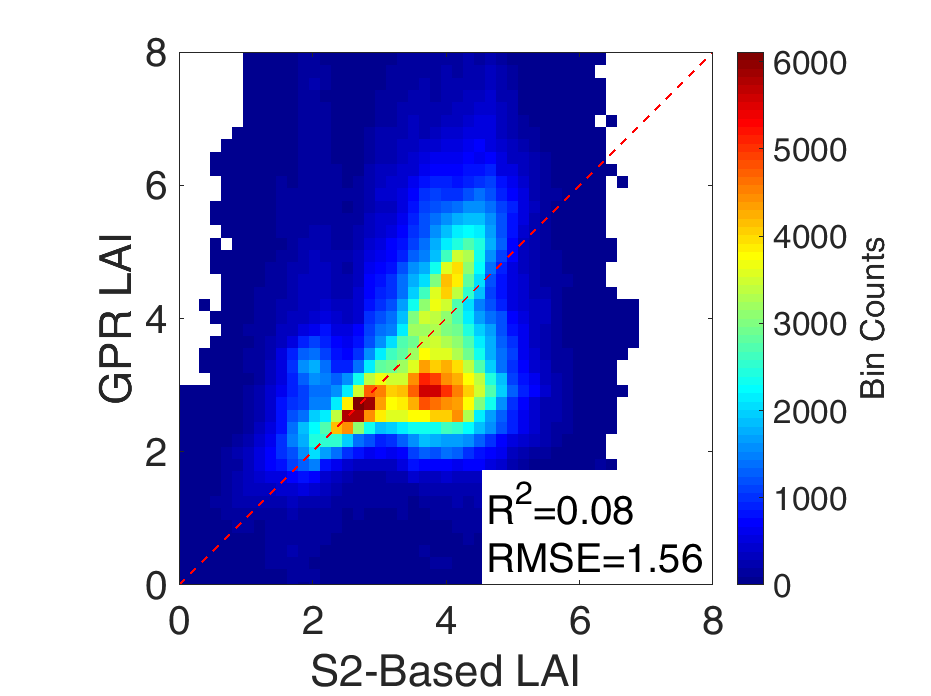}\\
\end{tabular}}%\vspace{12pt}
\caption{LAI maps from S2 image (reference), MOGP and standard GP, error histograms and scatter plots for two dates: 2016/2/17 and 2016/04/27 (90 days gap from 2016/1/18 to 2016/4/17).}
 \label{Fig:MOGPRetrieval2D_NewFigs_A}
\end{figure}

\paragraph{Scenario 1: long temporal gap}\label{longgap}

The first gap case being studied is the longest one: 90 days between 1/17 and 4/14 in 2016. The missing dates are 2/17, 4/7 and 4/27. Figure \ref{Fig:MOGPRetrieval2D_NewFigs_A} shows the results obtained for the first and the last dates. Qualitatively, the comparison between reference (1$^\text{st}$ column) and retrieved maps indicates that MOGP (2$^{nd}$ column) provides better LAI estimations than GPR (3$^{rd}$ column). GPR predictions tend to overestimate LAI, often providing unlikely values higher than 7. Conversely, MOGP prevents improbable predictions but slightly underestimates LAI values, as shown by the positive offset of the error histograms. The histograms indicate also a significantly lower dispersion of the errors given by MOGP, \black{confirmed by RMSE values}. A good match between reference and MOGP retrievals is obtained for LAI<3. Similar conclusions can be drawn by comparing the values of R$^2$, higher for MOGP even if quite low in absolute terms, and RMSE. 

\iffalse
Probably, R$^2$ is not the most appropriate assessment descriptor because as the validation is carried out using an (indirect) estimation of LAI from optical imagery as reference. On the one hand, R$^2$ is unable to reflect the lower 2D dispersion of scatter plot histograms given by MOGP with respect to the much broader response of GPR, to which RMSE is more sensitive. On the other hand, the smoothing process that both MOGP and GPR carry out, which aims to reduce the additive noise affecting the training samples, indirectly leads to lower R$^2$ (and higher RMSE). All the same, this penalizes the two approaches the same way. We can then assume R$^2$ and RMSE comparison is still meaningful, but with a relative and not absolute validity. 
\fi

\begin{figure}[!t]
 \centering
	\footnotesize
	\resizebox{1.1\textwidth}{!}{ 
	\begin{tabular}{ccc}
    \IG[width=3.5cm,trim={4.2cm 1.5cm 1cm 0cm},clip]{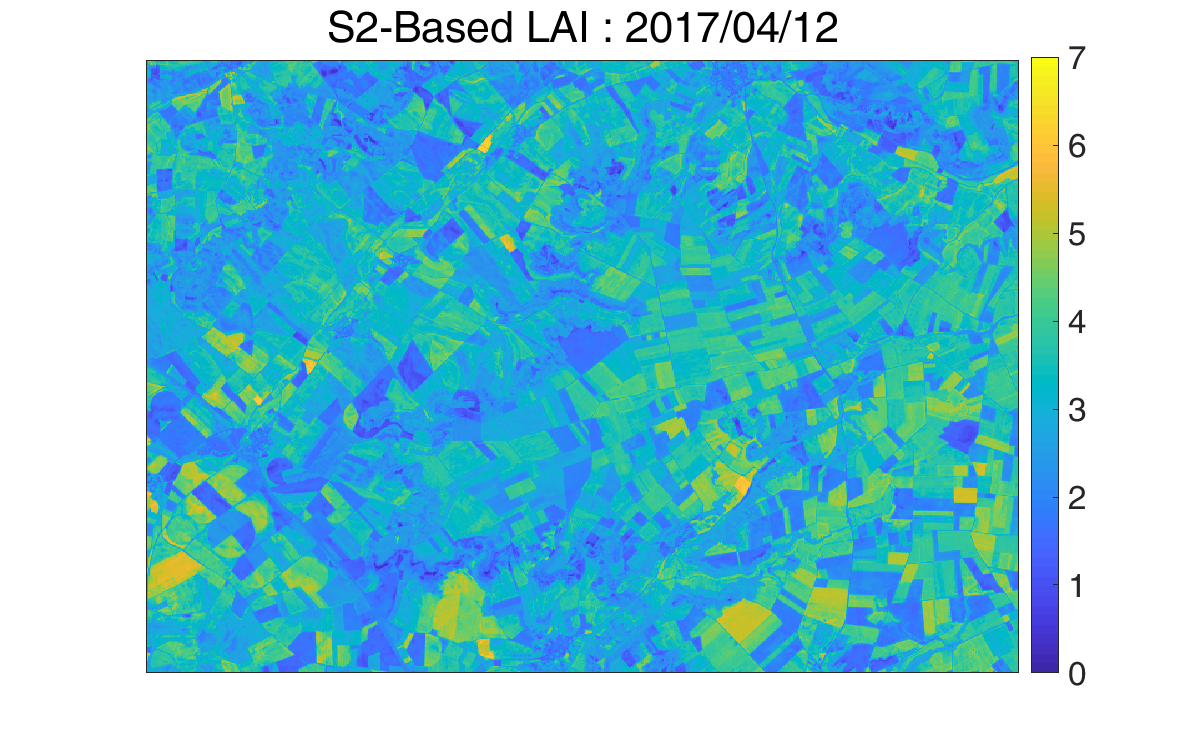}&  
    \IG[width=3.5cm,trim={4.2cm 1.5cm 1cm 0cm},clip]{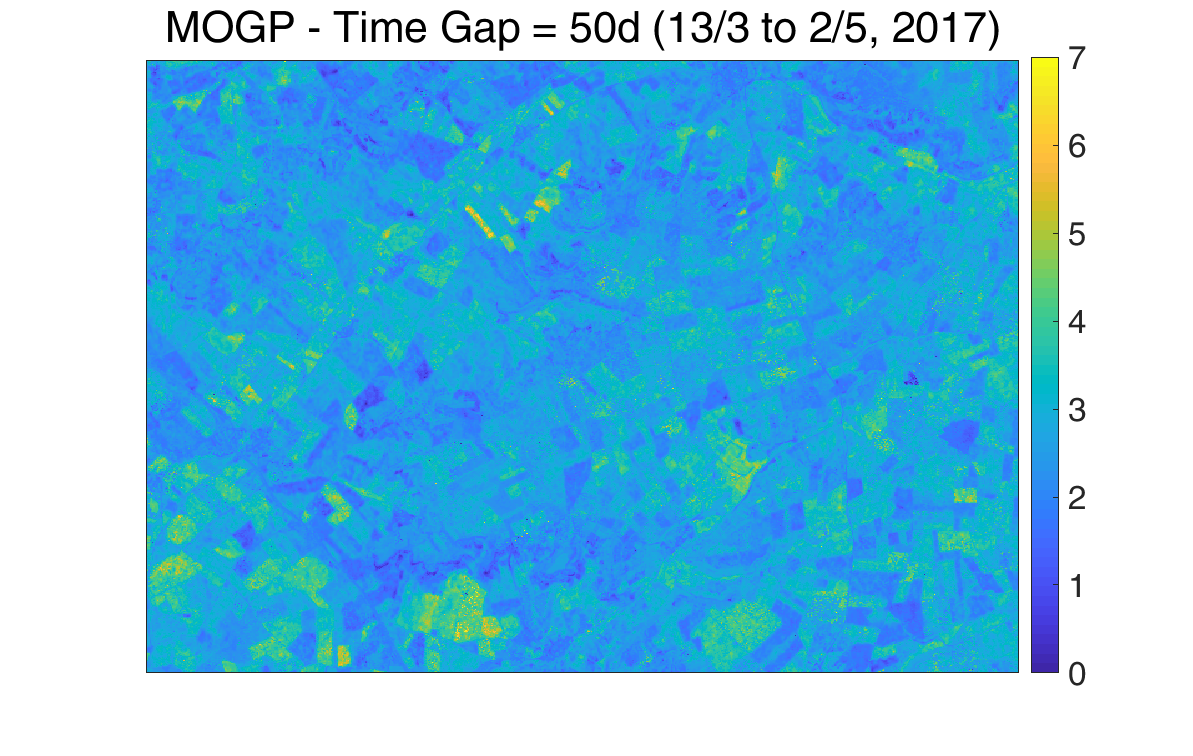}&
    \IG[width=3.5cm,trim={4.2cm 1.5cm 1cm 0cm},clip]{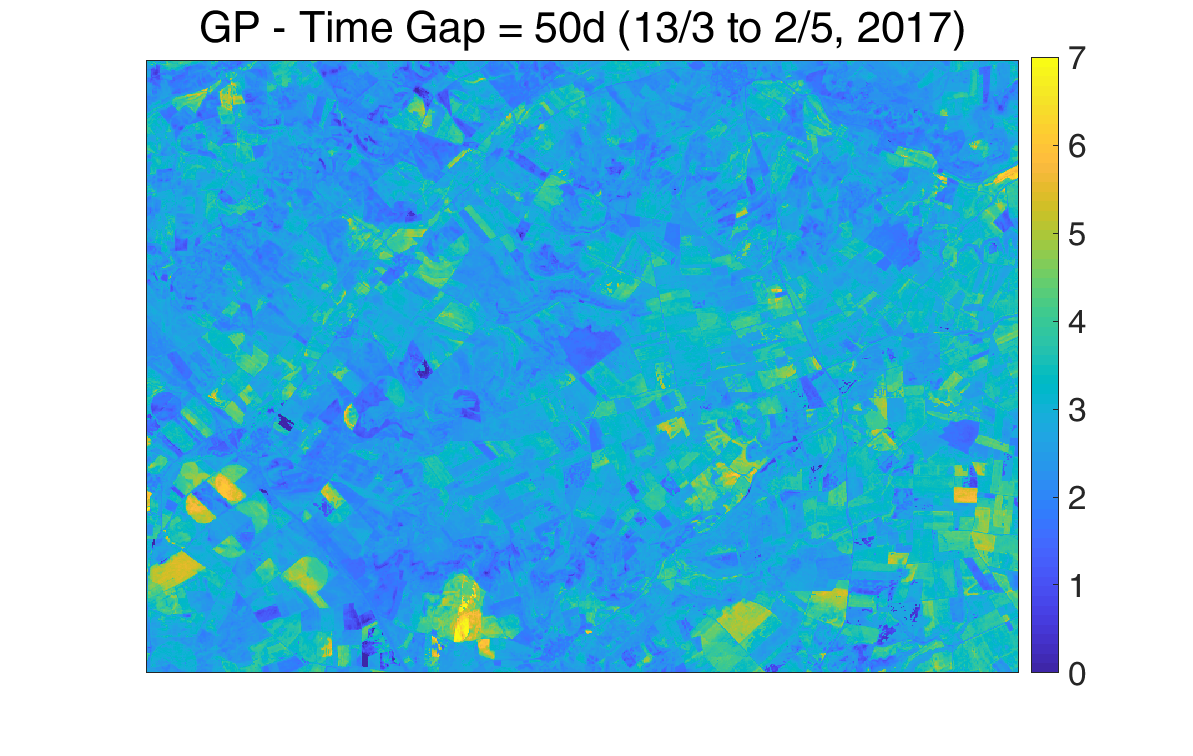}\\ 
    \IG[width=4.0cm,trim={0cm 0cm 0cm 0cm},clip]  {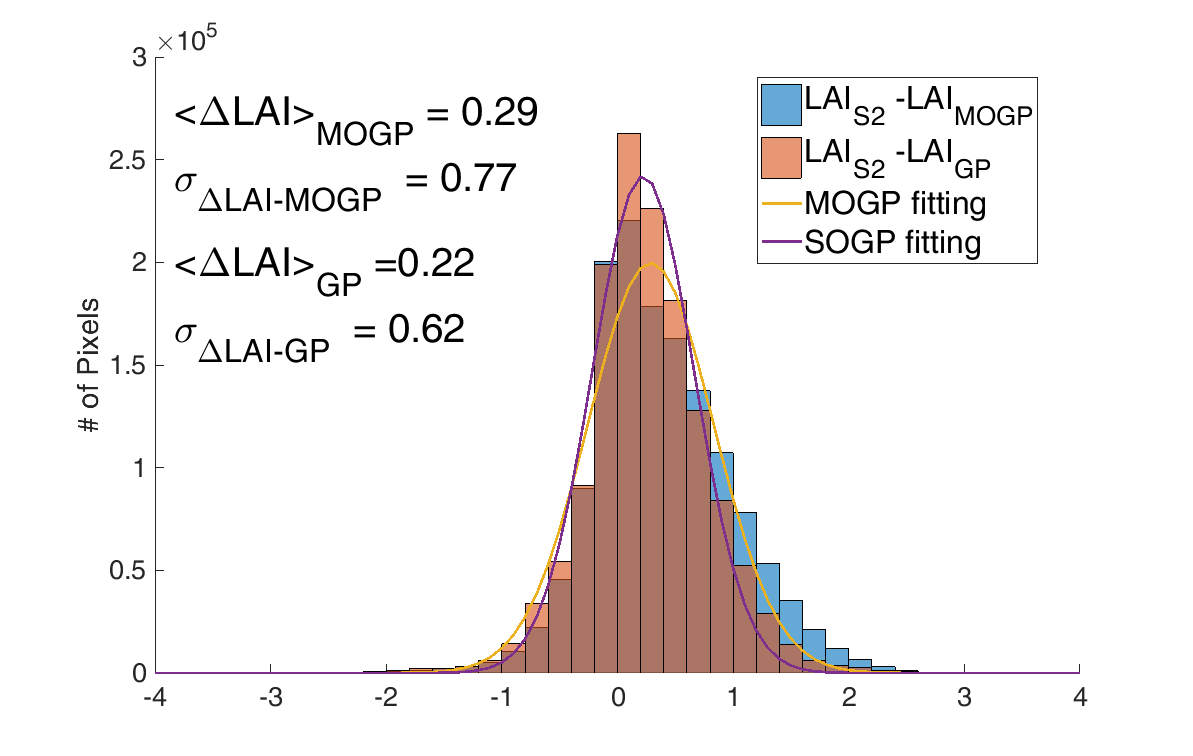}& 
    \IG[width=3.5cm,trim={0cm 0cm 0cm 0cm},clip]  {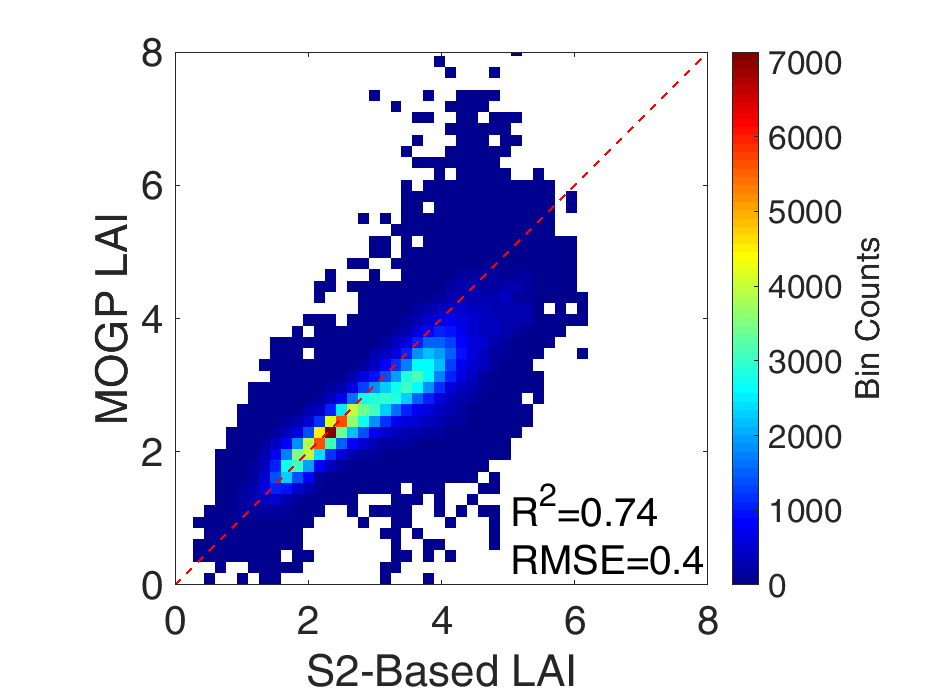}&
    \IG[width=3.5cm,trim={0cm 0cm 0cm 0cm},clip]  {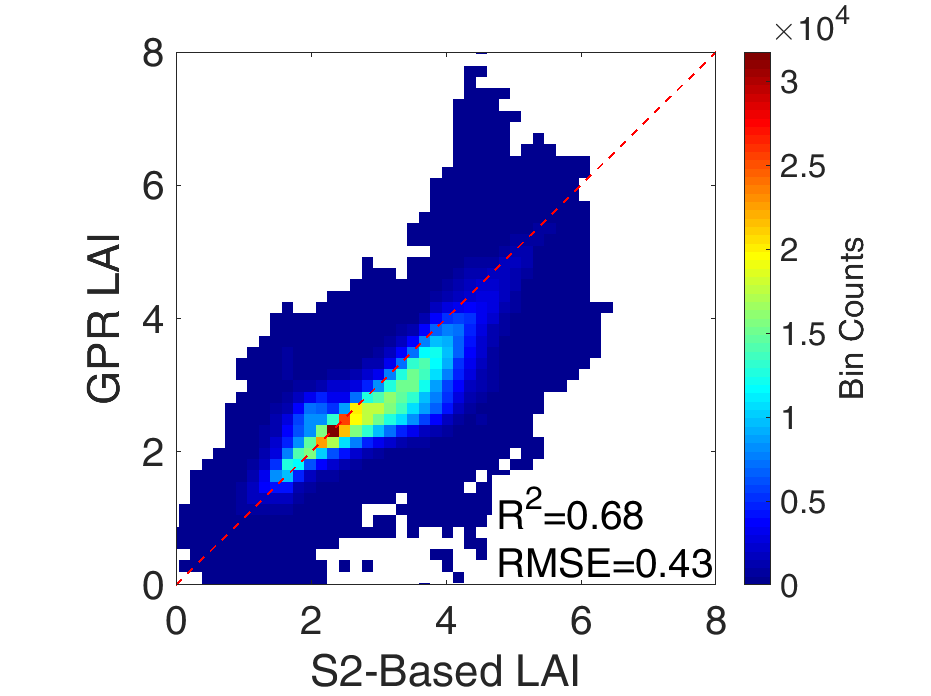}\\
\end{tabular}}
\caption{LAI maps from S2 image (reference), MOGP and standard GP, error histograms and scatter plots for 2017/4/12 (50 days gap from 2017/3/13 to 2017/5/2).} 
 \label{Fig:MOGPRetrieval2D_NewFigs_B}
\end{figure}

\begin{figure}[t]
%\begin{table}[t]
\centering
\resizebox{0.7\textwidth}{!}{
\begin{tabular}{rccc}
%\multicolumn{4}{c}{\textbf{Prediction vs S2-based LAI }}\\
%\cline{2-4} 
\multicolumn{1}{c}{\textbf{ }}  & \multicolumn{1}{c}{\textbf{2017/04/02}} & \multicolumn{1}{c}{\textbf{2017/04/12}} & \multicolumn{1}{c}{\textbf{2017/04/22}} \\ 
%\cline{1-4} 
%\multicolumn{1}{|r|}{\textbf{\rotatebox[origin=c]{90}{\raisebox{2 cm}{LAI$_{S2}$}}}}  & 
\multicolumn{1}{r}{\textbf{\rotatebox[origin=c]{90}{\parbox[c]{-3cm}{LAI$_{S2}$}}}}  & 

\multicolumn{1}{l}{\IG[width=3.5cm,trim={5cm 3cm 4cm 1.5cm},clip]{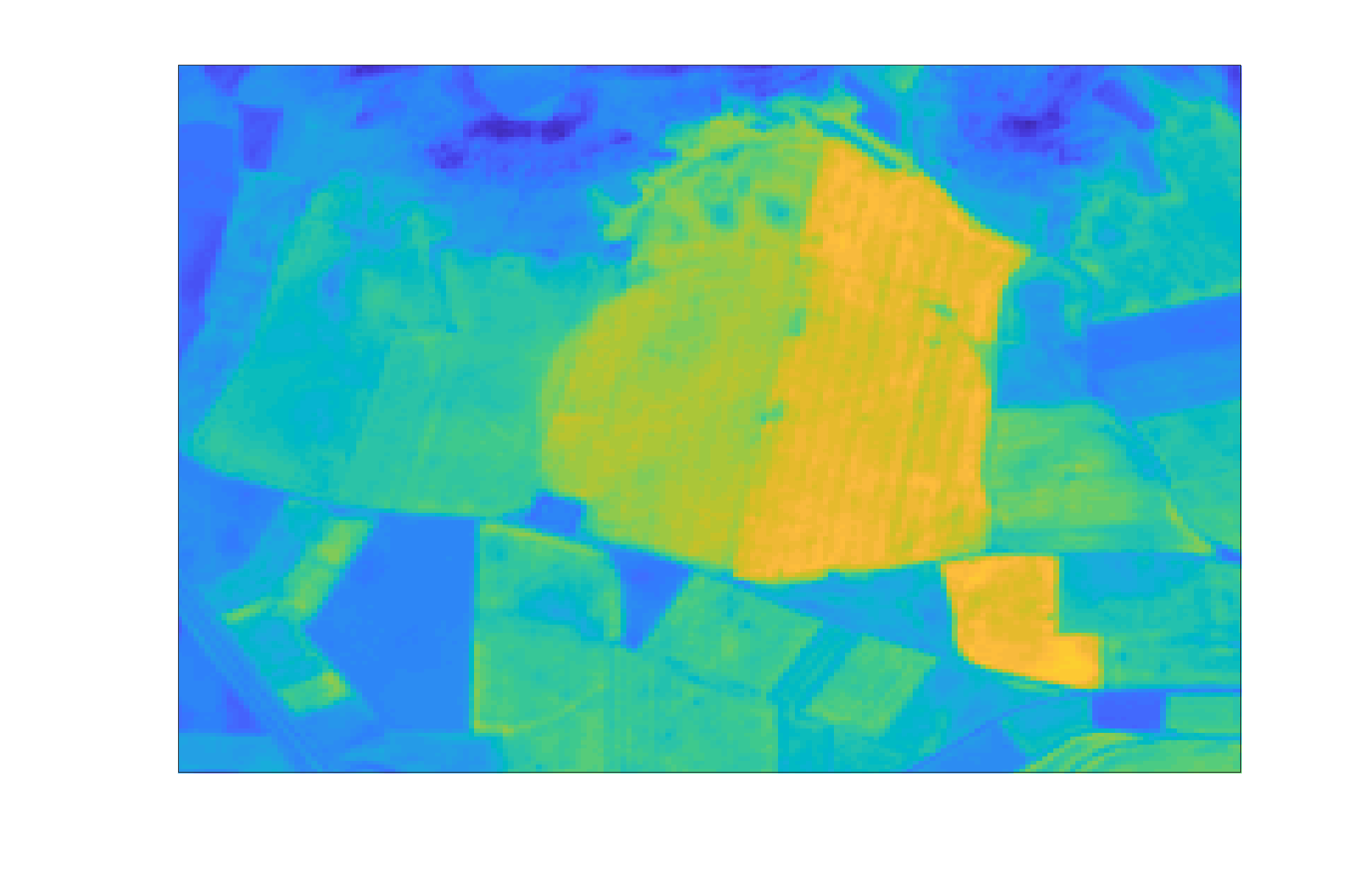} } & 
\multicolumn{1}{l}{\IG[width=3.5cm,trim={5cm 3cm 4cm 1.5cm},clip]{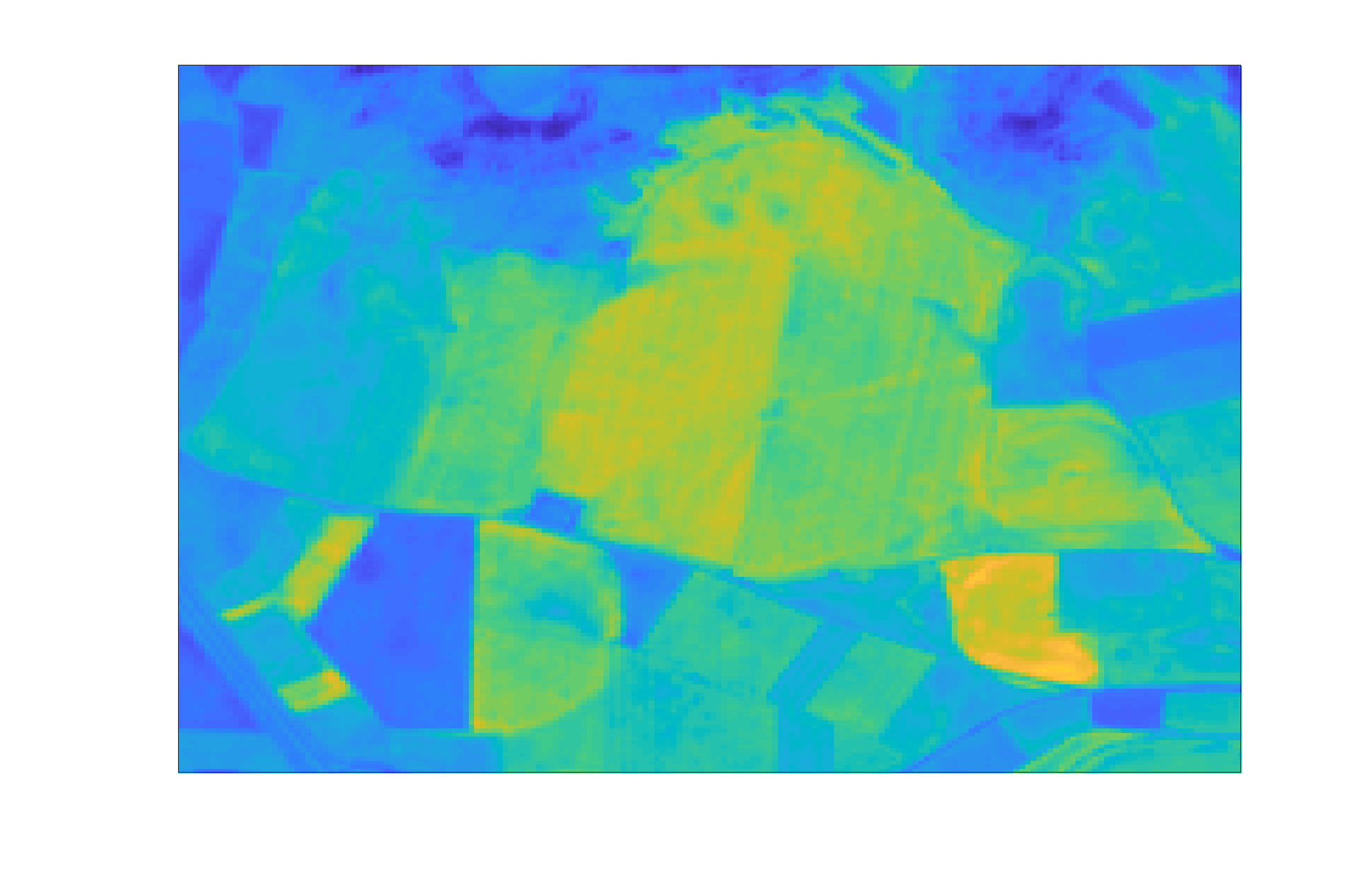} } & 
\multicolumn{1}{l}{\IG[width=3.5cm,trim={5cm 3cm 4cm 1.5cm},clip]{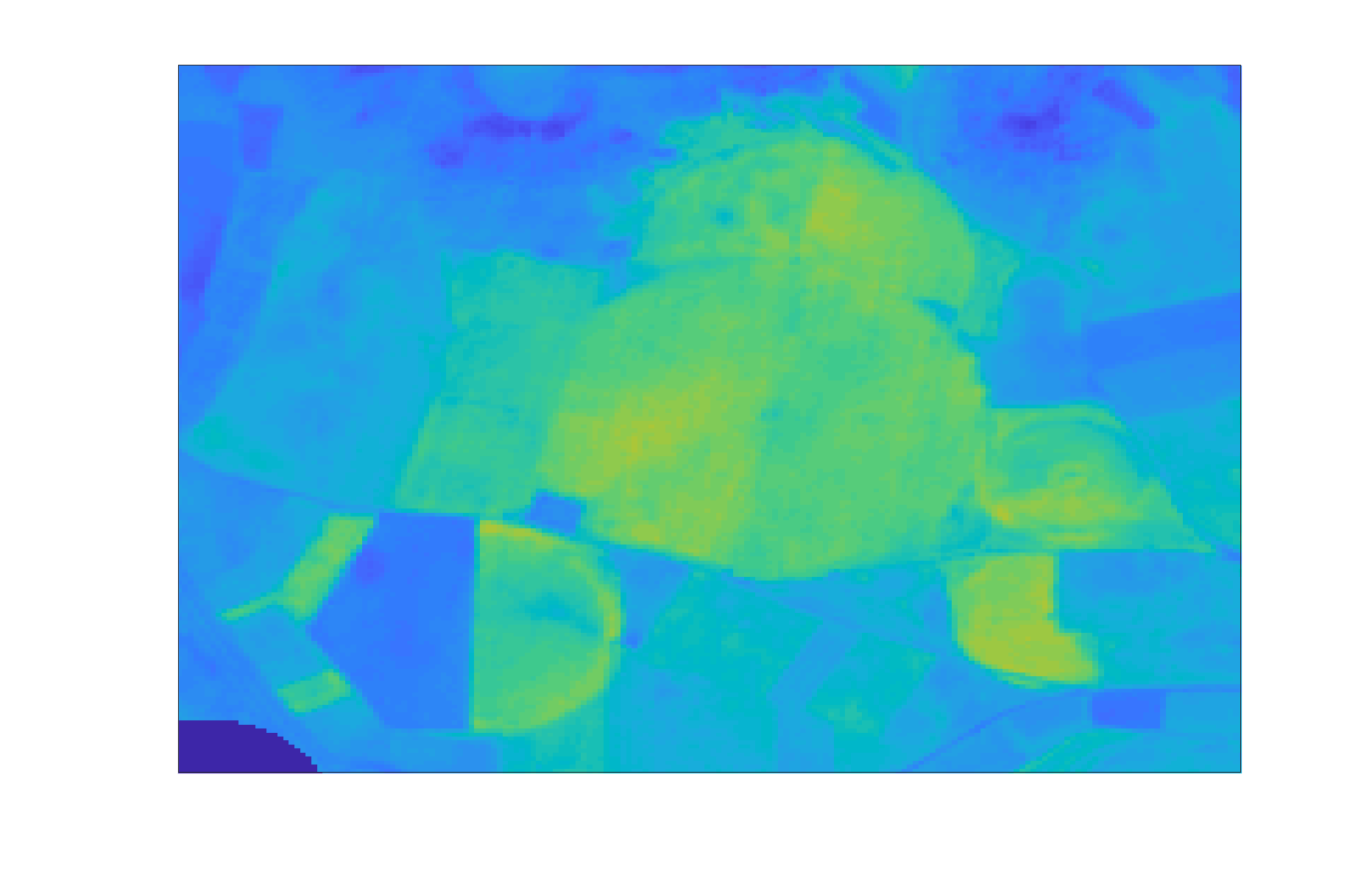} } \\
%\cline{1-4} 
\multicolumn{1}{r}{\textbf{\rotatebox[origin=c]{90}{\parbox[c]{-3cm}{MOGP}}}}  & 
%\multicolumn{1}{c}{\textbf{\rotatebox[origin=c]{90}{\raisebox{1 cm}MOGP}}}  & 
\multicolumn{1}{l}{\IG[width=3.5cm,trim={5cm 3cm 4cm 1.5cm},clip]{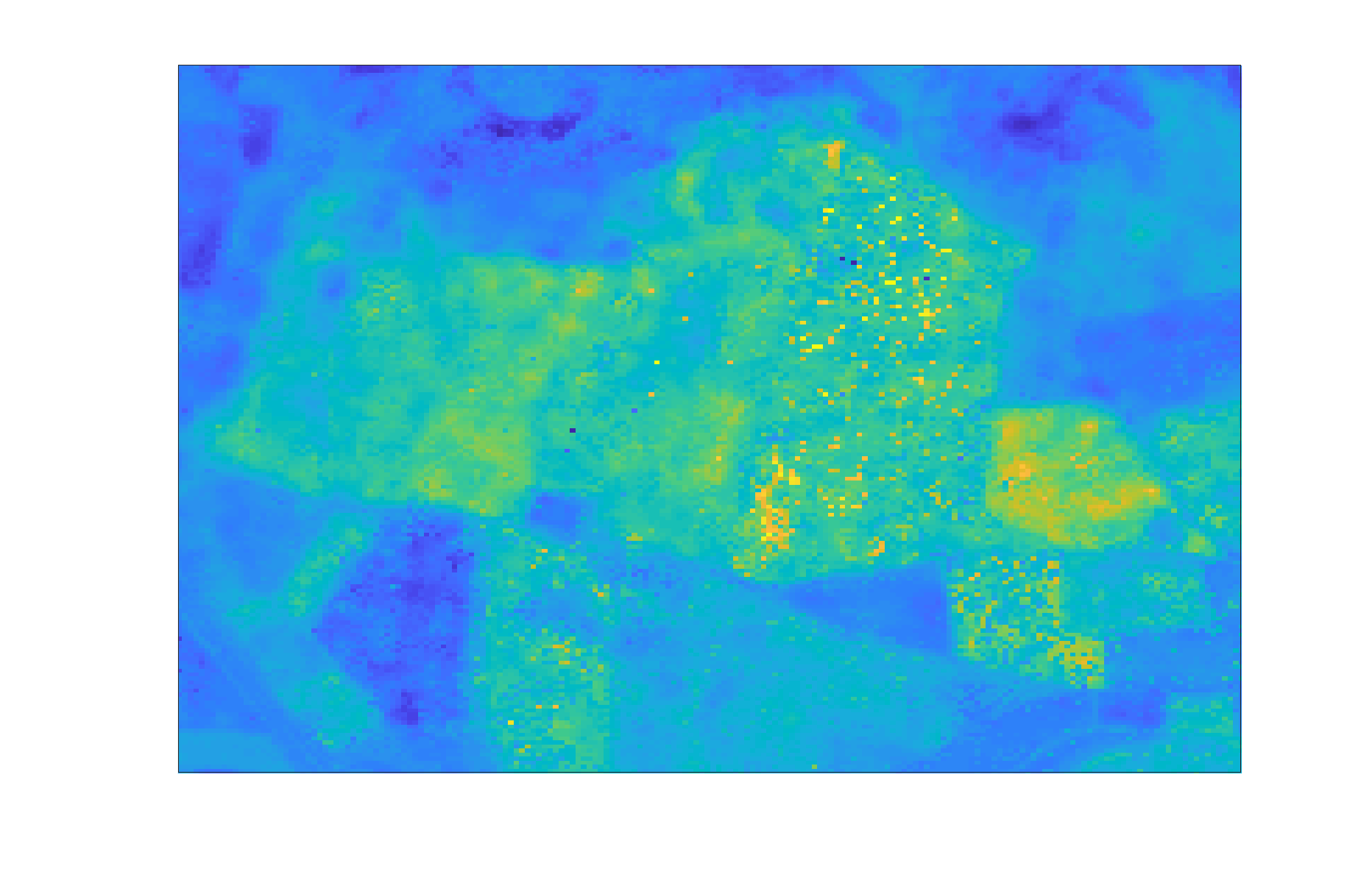} } & 
\multicolumn{1}{l}{\IG[width=3.5cm,trim={5cm 3cm 4cm 1.5cm},clip]{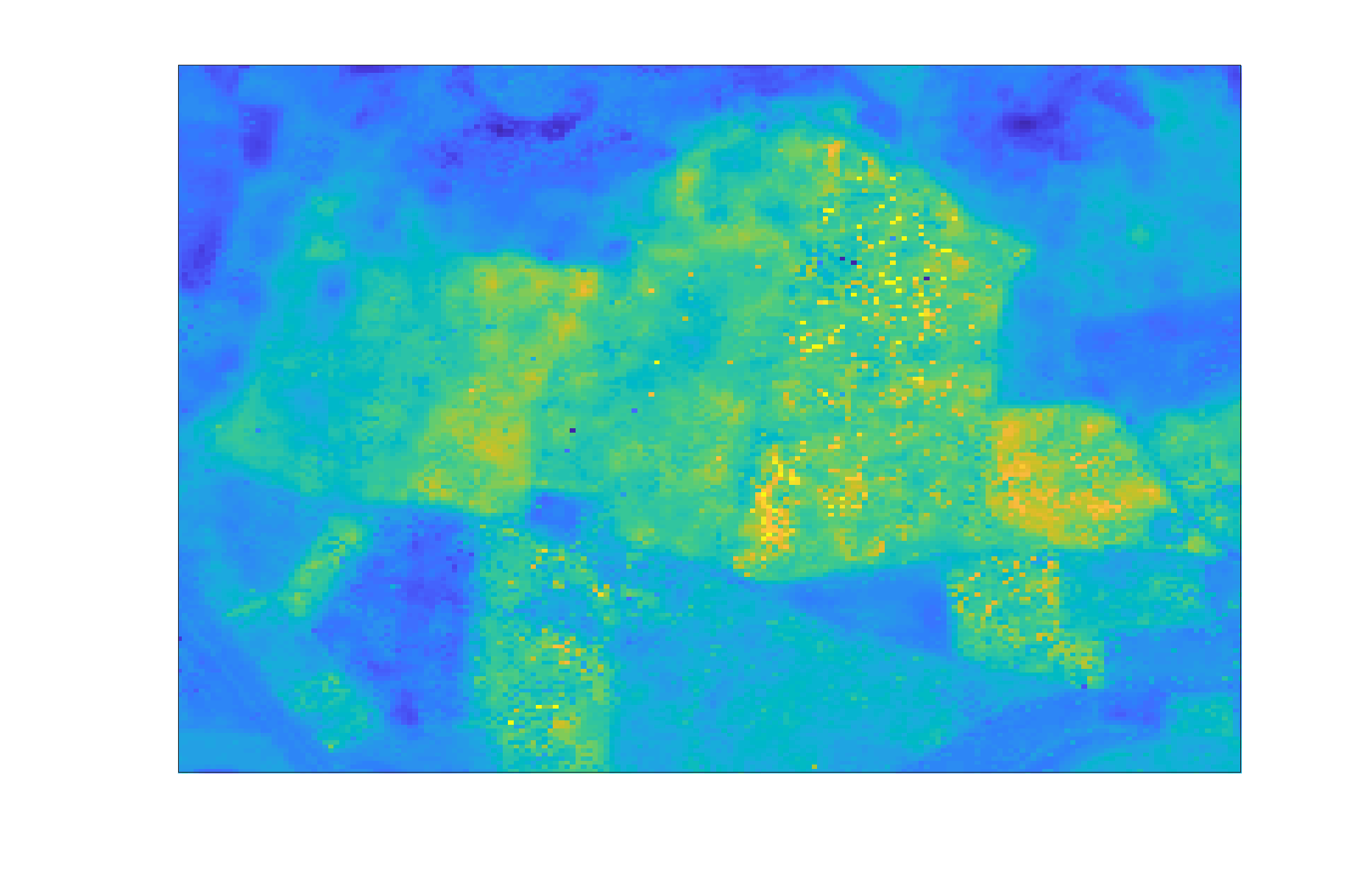} } & 
\multicolumn{1}{l}{\IG[width=3.5cm,trim={5cm 3cm 4cm 1.5cm},clip]{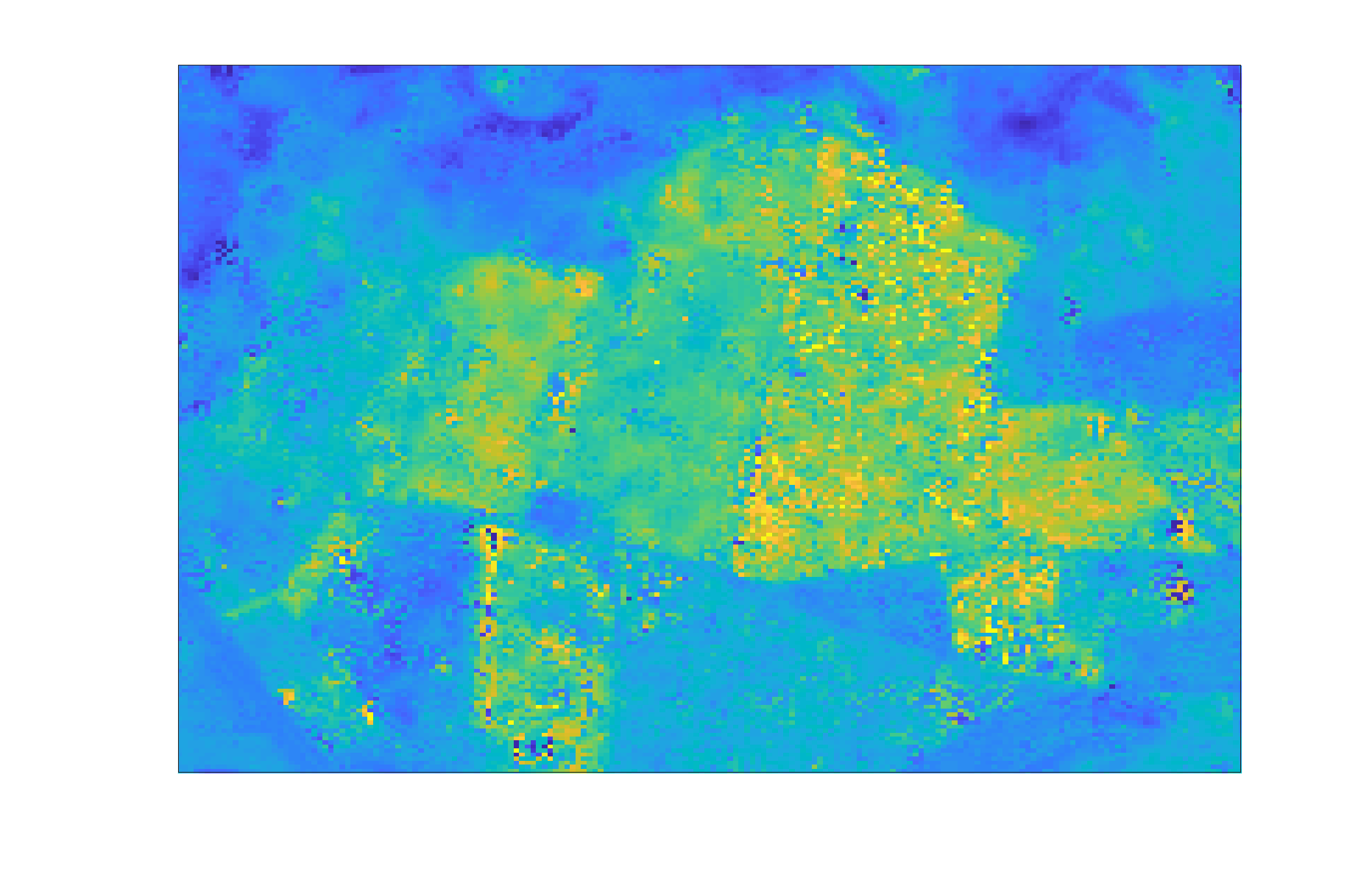} } \\
%\cline{1-4} 
\multicolumn{1}{r}{\textbf{\rotatebox[origin=c]{90}{\parbox[c]{-3cm}{GPR}}}}  & 
\multicolumn{1}{l}{\IG[width=3.5cm,trim={5cm 3cm 4cm 1.5cm},clip]{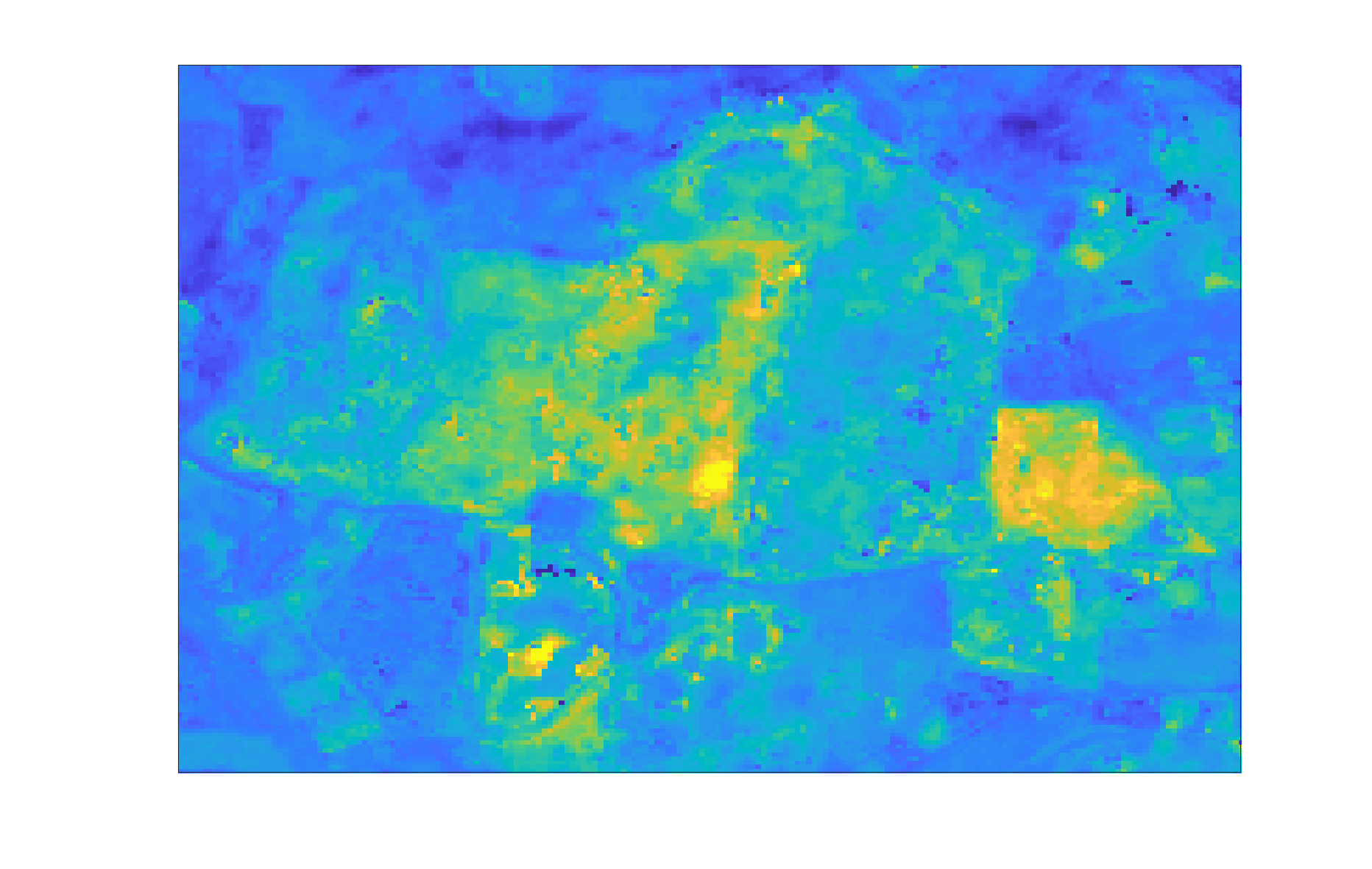} } & 
\multicolumn{1}{l}{\IG[width=3.5cm,trim={5cm 3cm 4cm 1.5cm},clip]{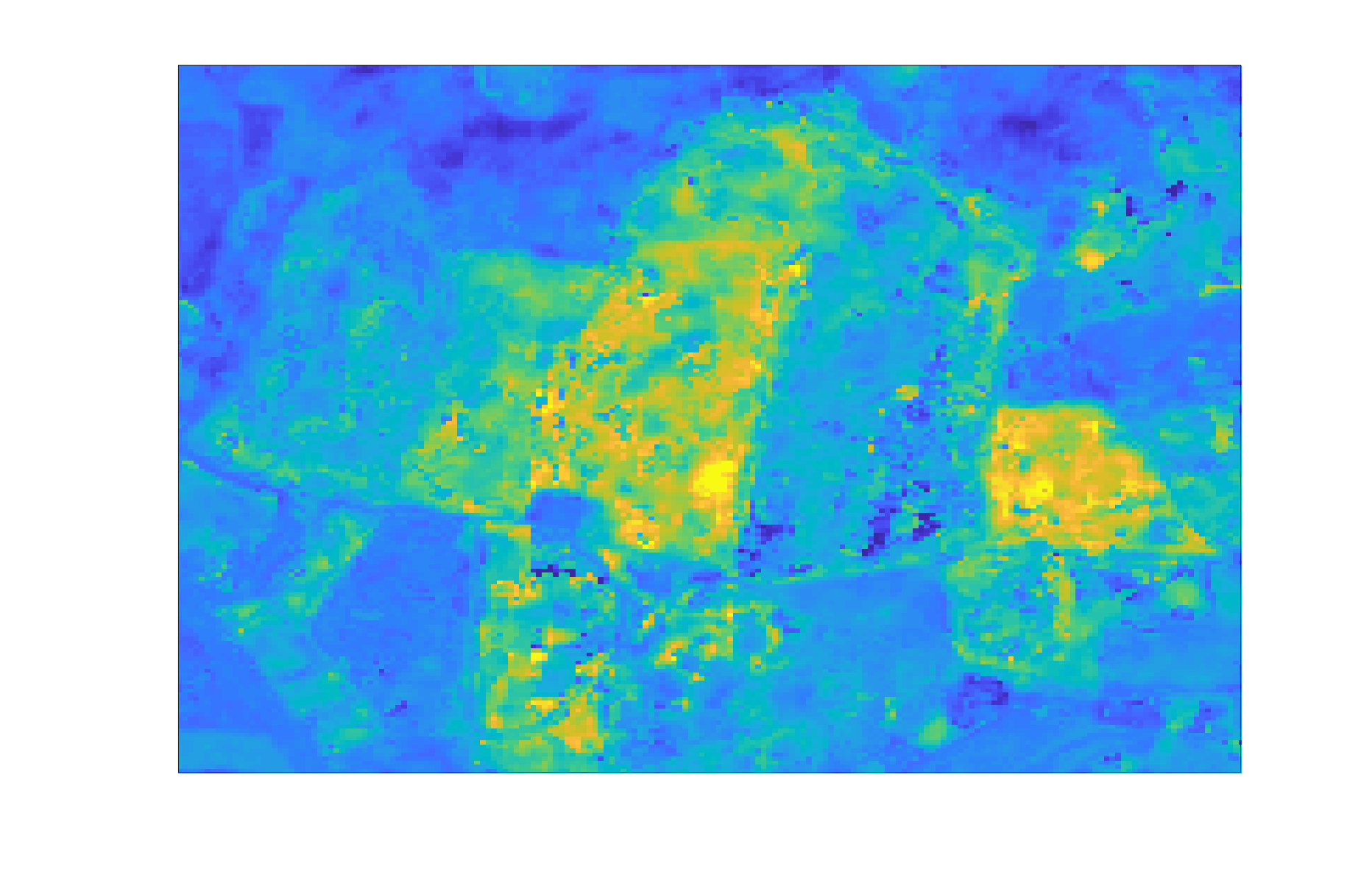} } & 
\multicolumn{1}{l}{\IG[width=3.5cm,trim={5cm 3cm 4cm 1.5cm},clip]{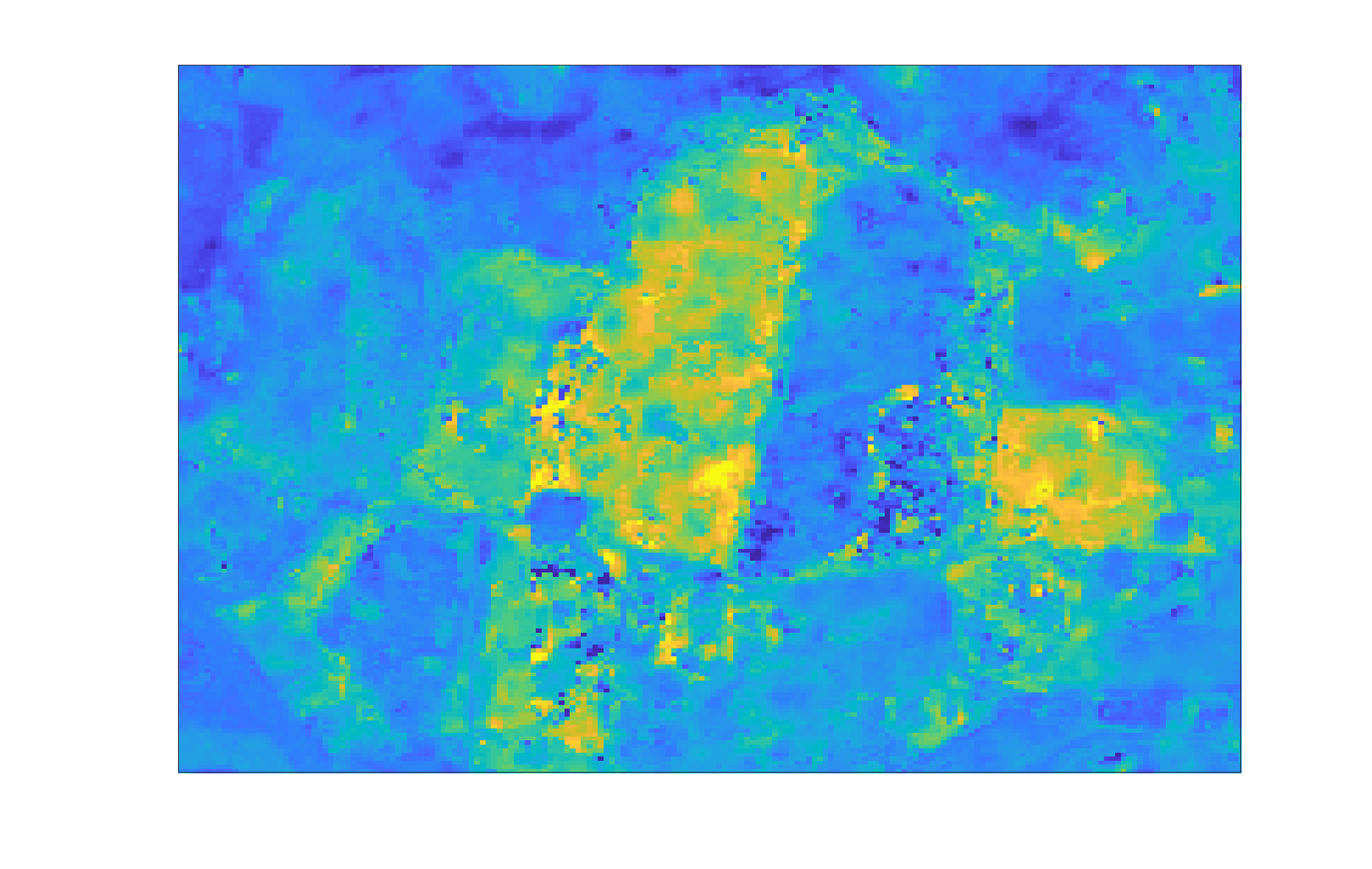} } \\
%\cline{1-4} 
\end{tabular}}
\caption{Comparison of LAI maps from S2 image (reference), MOGP and standard GP over a subset of AOI (yellow rectangle in Figure \ref{Fig:RGB_AOI}) corresponding to 4/2, 4/12 and 4/22 within the 50 days gap from 2017/3/13 to 2017/5/2. LAI evolution provided by GPR is more unlikely than MOGP's results.}
\label{Fig:Zooms}
%\end{table}
\end{figure}

\paragraph{Scenario 2: moderate temporal gap}\label{moderategap}

The retrievals shown in Figure~\ref{Fig:MOGPRetrieval2D_NewFigs_B} correspond to 4/12, the central date within the 50-day time gap from 3/13 to 5/2 in 2017. The predictions provided by the two approaches are alike. From the statistical point of view, the error dispersion is almost identical. The values of R$^2$ ranges around 70\% for both approaches, with MOGP slightly better than GPR; RMSE values area similar too. As already observed for the long temporal gap case, GPR tends to overestimate to unreliable values, whereas MOGP usually underestimates the highest values of LAI appearing in the reference maps. \black{The main reason is probably that the MOGP solution is tied to the RVI temporal dynamics, which prevents the sudden appearance of LAI peaks induced by noisy estimates from S2 data. As an example, note the unlikely LAI evolution given by GPR on the parcels in the down-left part of AOI (yellow rectangle in Figure \ref{Fig:RGB_AOI}), whose zooms are shown in \black{Figure}~\ref{Fig:Zooms}.} The overall values of LAI predicted by GPR range between 0 and 7, with local irregular changes too high to be plausible in less than one month (up to almost 7 and below 5 in 3 weeks). Over the same area, MOGP provides a gradually increase response along time, which can be considered more realistic.

\subsubsection{Assessment of retrieved maps}\label{mapassessment}

The performance of MOGP is strictly related to the existence of a favorable relationship between $a^{HF}_1$ and $a^{LF}_1$, which represents the core hypothesis for any active-passive synergy. The results summarized in Table \ref{Table:MOGPRetrieval2D_ReliabPixs} underlines exactly this point. Including all the pixels corresponding to crop areas in the assessment leads to R$^2$ values of MOGP predictions always lower than \black{those} obtained with the pixels fulfilling the condition $a^{LF}_1/a^{HF}_1>1.5$. Almost the same happens with RMSE, even if the magnitude of the change is lower. Conversely, the results for GPR are practically identical, confirming the samples considered for the two analyses \black{comprehend} all possible cases.

\begin{table}[t]
\resizebox{1.0\textwidth}{!}{
\begin{tabular}{lccccccc}
\multicolumn{8}{c}{\textbf{Prediction vs S2-based LAI }}\\
\cline{2-8} 
\multicolumn{1}{c|}{\textbf{ }}  &
\multicolumn{1}{c|}{\textbf{2016/02/17}} & \multicolumn{1}{c|}{\textbf{2016/04/07}} & \multicolumn{1}{c|}{\textbf{2016/04/27}} & \multicolumn{1}{c|}{\textbf{2017/04/02}} & \multicolumn{1}{c|}{\textbf{2017/04/12}} & \multicolumn{1}{c|}{\textbf{2017/04/22}} & \multicolumn{1}{c|}{\textbf{2018/02/26}} \\ 
\cline{1-8} 
\multicolumn{1}{|c|}{\textbf{ $R^2_{MOGP}$}}  & \multicolumn{1}{l|}{\textbf{\blue{0.29}{(0.33)}}} & \multicolumn{1}{l|}{\textbf{\blue{0.20}{(0.19)}}} & \multicolumn{1}{l|}{\textbf{\blue{0.29}(0.26)}} & \multicolumn{1}{l|}{\textbf{\blue{0.56}(0.70)}} & \multicolumn{1}{l|}{\textbf{\blue{0.58}(0.74)}} & \multicolumn{1}{l|}{\textbf{\blue{0.49}(0.71)}} & \multicolumn{1}{l|}{\textbf{\blue{0.25}(0.35)}} \\ \cline{1-8}
\multicolumn{1}{|c|}{\textbf{ $R^2_{GPR}$}}   & \multicolumn{1}{l|}{\textbf{\blue{0.12}{(0.12)}}} & \multicolumn{1}{l|}{\textbf{\blue{0.06}{(0.08)}}} & \multicolumn{1}{l|}{\textbf{\blue{0.08}(0.08)}} & \multicolumn{1}{l|}{\textbf{\blue{0.67}(0.67)}} & \multicolumn{1}{l|}{\textbf{\blue{0.68}(0.68)}} & \multicolumn{1}{l|}{\textbf{\blue{0.70}(0.70)}} & \multicolumn{1}{l|}{\textbf{\blue{0.40}(0.40)}} \\ \cline{1-8}
\multicolumn{1}{|c|}{\textbf{ $RMSE_{MOGP}$}} & \multicolumn{1}{l|}{\textbf{\blue{0.44}{(0.50)}}} & \multicolumn{1}{l|}{\textbf{\blue{0.80}{(0.90)}}} & \multicolumn{1}{l|}{\textbf{\blue{0.85}(0.94)}} & \multicolumn{1}{l|}{\textbf{\blue{0.42}(0.41)}} & \multicolumn{1}{l|}{\textbf{\blue{0.44}(0.40)}} & \multicolumn{1}{l|}{\textbf{\blue{0.55}(0.42)}} & \multicolumn{1}{l|}{\textbf{\blue{0.36}(0.44)}} \\ \cline{1-8}
\multicolumn{1}{|c|}{\textbf{ $RMSE_{GPR}$}}  & \multicolumn{1}{l|}{\textbf{\blue{1.12}{(1.09)}}} & \multicolumn{1}{l|}{\textbf{\blue{1.75}{(1.70)}}} & \multicolumn{1}{l|}{\textbf{\blue{1.62}(1.56)}} & \multicolumn{1}{l|}{\textbf{\blue{0.44}(0.43)}} & \multicolumn{1}{l|}{\textbf{\blue{0.46}(0.43)}} & \multicolumn{1}{l|}{\textbf{\blue{0.45}(0.41)}} & \multicolumn{1}{l|}{\textbf{\blue{0.50}(0.50)}} \\ \cline{1-8}
\end{tabular}}
\caption{R$^2$ and RMSE [$m^2/m^2]$ of LAI maps from S2 image (reference) vs MOGP and GPR over all the vegetated pixels (blue) and only pixels fulfilling $a^{LF}_1/a^{HF}_1>1.5$ (black).}
\label{Table:MOGPRetrieval2D_ReliabPixs}
\end{table}

MOGP retrievals present higher errors over low LAI areas, which typically corresponds to the beginning of the crop season (tillering) such as samples on 2016/2/17 and 2018/02/26. A possible explanation lies in the different properties of vegetation cover the two sensors are able to capture: its greenness for optical and its real volumetric structure, i.e. fresh biomass, for SAR. Albeit these two quantities are strictly related, depending on the land cover they may lead to slightly dissimilar estimation of LAI as the former usually anticipates the latter. From this point of view, the importance of the final findings summarized in Table \ref{Table:MOGPRetrieval2D_ReliabPixs} is twofold. On the one hand, they demonstrate the existence of a stochastic relationship between the time series of optical and radar vegetation descriptors among areas characterized by a dominant LFGP. On the other hand, they allow identifying crop areas where common long-time patterns between S2-LAI and S1-RVI time series are \black{barely detectable}.

\begin{table}[t]
\resizebox{1.0\textwidth}{!}{ 
\begin{tabular}{l|c|ll|ll|ll|c|ll|ll|ll|c|ll|}
\cline{2-18}
\multicolumn{1}{c|}{\multirow{2}{*}{\textbf{$R^2$}}}           & \multicolumn{3}{c|}{\textbf{2016/02/17}} & \multicolumn{2}{c|}{\textbf{2016/04/07}} & \multicolumn{2}{c|}{\textbf{2016/04/27}} & \multicolumn{3}{c|}{\textbf{2017/04/02}} & \multicolumn{2}{c|}{\textbf{2017/04/12}} & \multicolumn{2}{c|}{\textbf{2017/04/22}} & \multicolumn{3}{c|}{\textbf{2018/02/26}} \\ \cline{2-18} 
  & LCMap(\%) & MOGP & GPR  & MOGP  & GPR & MOGP & GPR & LCMap(\%)  & MOGP  & GPR  & MOGP   & GPR & MOGP  & GPR    & LCMap(\%)  & MOGP  & GPR  \\
   \hline
\multicolumn{1}{|l|}{\textbf{Wheat}}             & 30.2  & \cellcolor[RGB]{255.0 172.7 172.7}0.287 & \cellcolor[RGB]{ 255.0 172.7 172.7}0.148 & \cellcolor[RGB]{255.0 213.9 213.9}0.072 & \cellcolor[RGB]{ 255.0 213.9 213.9}0.015 & \cellcolor[RGB]{255.0 213.9 213.9}0.102 & \cellcolor[RGB]{ 255.0 213.9 213.9}0.036 & 25.7 & \cellcolor[RGB]{205.6 205.6 255.0}0.487 & \cellcolor[RGB]{ 205.6 205.6 255.0}0.578 & \cellcolor[RGB]{181.0 181.0 255.0}0.549 & \cellcolor[RGB]{ 181.0 181.0 255.0}0.694 & \cellcolor[RGB]{148.1 148.1 255.0}0.518 & \cellcolor[RGB]{ 148.1 148.1 255.0}0.728 & 29.5 & \cellcolor[RGB]{238.5 238.5 255.0}0.223 & \cellcolor[RGB]{ 238.5 238.5 255.0}0.251 \\
\multicolumn{1}{|l|}{\textbf{Maize}}             &  0.2  & \cellcolor[RGB]{0.0   0.0 255.0}0.121   & \cellcolor[RGB]{ 0.0   0.0 255.0}0.628   & \cellcolor[RGB]{238.5 238.5 255.0}0.355 & \cellcolor[RGB]{ 238.5 238.5 255.0}0.393 & \cellcolor[RGB]{255.0   0.0   0.0}0.811 & \cellcolor[RGB]{ 255.0   0.0   0.0}0.333 &  0.1 & \cellcolor[RGB]{255.0 230.3 230.3}0.854 & \cellcolor[RGB]{ 255.0 230.3 230.3}0.829 & \cellcolor[RGB]{255.0 230.3 230.3}0.88  & \cellcolor[RGB]{ 255.0 230.3 230.3}0.854 & \cellcolor[RGB]{255.0 222.1 222.1}0.928 & \cellcolor[RGB]{ 255.0 222.1 222.1}0.886 &  0.2 & \cellcolor[RGB]{213.9 213.9 255.0}0.812 & \cellcolor[RGB]{ 213.9 213.9 255.0}0.888 \\
\multicolumn{1}{|l|}{\textbf{Barley}}            & 32.2  & \cellcolor[RGB]{255.0 156.3 156.3}0.221 & \cellcolor[RGB]{ 255.0 156.3 156.3}0.04  & \cellcolor[RGB]{255.0 222.1 222.1}0.058 & \cellcolor[RGB]{ 255.0 222.1 222.1}0.009 & \cellcolor[RGB]{255.0 222.1 222.1}0.071 & \cellcolor[RGB]{ 255.0 222.1 222.1}0.015 & 37.4 & \cellcolor[RGB]{230.3 230.3 255.0}0.613 & \cellcolor[RGB]{ 230.3 230.3 255.0}0.664 & \cellcolor[RGB]{213.9 213.9 255.0}0.642 & \cellcolor[RGB]{ 213.9 213.9 255.0}0.72 & \cellcolor[RGB]{189.2 189.2 255.0}0.606  & \cellcolor[RGB]{ 189.2 189.2 255.0}0.737 & 31.6 & \cellcolor[RGB]{230.3 230.3 255.0}0.185 & \cellcolor[RGB]{ 230.3 230.3 255.0}0.24 \\
\multicolumn{1}{|l|}{\textbf{Rye} }              &  0.0  & \cellcolor[RGB]{255.0 246.8 246.8}0     & \cellcolor[RGB]{ 255.0 246.8 246.8}0     & \cellcolor[RGB]{255.0 246.8 246.8}0     & \cellcolor[RGB]{ 255.0 246.8 246.8}0     & \cellcolor[RGB]{255.0 246.8 246.8}0     & \cellcolor[RGB]{ 255.0 246.8 246.8}0     &  0.2 & \cellcolor[RGB]{238.5 238.5 255.0}0.781 & \cellcolor[RGB]{ 238.5 238.5 255.0}0.822 & \cellcolor[RGB]{213.9 213.9 255.0}0.766 & \cellcolor[RGB]{ 213.9 213.9 255.0}0.848 & \cellcolor[RGB]{181.0 181.0 255.0}0.719 & \cellcolor[RGB]{ 181.0 181.0 255.0}0.861 &  0.3 & \cellcolor[RGB]{131.6 131.6 255.0}0.232 & \cellcolor[RGB]{ 131.6 131.6 255.0}0.468 \\
\multicolumn{1}{|l|}{\textbf{Oats}}              &  0.2  & \cellcolor[RGB]{41.1  41.1 255.0}0.148  & \cellcolor[RGB]{ 41.1  41.1 255.0}0.564  & \cellcolor[RGB]{189.2 189.2 255.0}0.293 & \cellcolor[RGB]{ 189.2 189.2 255.0}0.424 & \cellcolor[RGB]{255.0  82.3  82.3}0.487 & \cellcolor[RGB]{ 255.0  82.3  82.3}0.163 &  0.0 & \cellcolor[RGB]{255.0  98.7  98.7}0.414 & \cellcolor[RGB]{ 255.0  98.7  98.7}0.121 & \cellcolor[RGB]{255.0  82.3  82.3}0.515 & \cellcolor[RGB]{ 255.0  82.3  82.3}0.195 & \cellcolor[RGB]{255.0  90.5  90.5}0.638 & \cellcolor[RGB]{ 255.0  90.5  90.5}0.339 &  0.4 & \cellcolor[RGB]{246.8 246.8 255.0}0.453 & \cellcolor[RGB]{ 246.8 246.8 255.0}0.472 \\
\multicolumn{1}{|l|}{\textbf{Sunflower}}         &  1.5  & \cellcolor[RGB]{255.0 156.3 156.3}0.253 & \cellcolor[RGB]{ 255.0 156.3 156.3}0.08  & \cellcolor[RGB]{255.0 189.2 189.2}0.154 & \cellcolor[RGB]{ 255.0 189.2 189.2}0.048 & \cellcolor[RGB]{255.0 213.9 213.9}0.145 & \cellcolor[RGB]{ 255.0 213.9 213.9}0.076 &  1.3 & \cellcolor[RGB]{255.0 181.0 181.0}0.522 & \cellcolor[RGB]{ 255.0 181.0 181.0}0.393 & \cellcolor[RGB]{255.0 164.5 164.5}0.596 & \cellcolor[RGB]{ 255.0 164.5 164.5}0.438 & \cellcolor[RGB]{255.0 164.5 164.5}0.545 & \cellcolor[RGB]{ 255.0 164.5 164.5}0.383 &  2.2 & \cellcolor[RGB]{164.5 164.5 255.0}0.349 & \cellcolor[RGB]{ 164.5 164.5 255.0}0.519 \\
\multicolumn{1}{|l|}{\textbf{Rapeseed}}          &  0.8  & \cellcolor[RGB]{246.8 246.8 255.0}0.276 & \cellcolor[RGB]{ 246.8 246.8 255.0}0.288 & \cellcolor[RGB]{106.9 106.9 255.0}0.075 & \cellcolor[RGB]{ 106.9 106.9 255.0}0.373 & \cellcolor[RGB]{255.0 222.1 222.1}0.061 & \cellcolor[RGB]{ 255.0 222.1 222.1}0.006 &  0.8 & \cellcolor[RGB]{238.5 238.5 255.0}0.493 & \cellcolor[RGB]{ 238.5 238.5 255.0}0.534 & \cellcolor[RGB]{189.2 189.2 255.0}0.199 & \cellcolor[RGB]{ 189.2 189.2 255.0}0.331 & \cellcolor[RGB]{230.3 230.3 255.0}0.295 & \cellcolor[RGB]{ 230.3 230.3 255.0}0.351 &  1.2 & \cellcolor[RGB]{255.0 238.5 238.5}0.579 & \cellcolor[RGB]{ 255.0 238.5 238.5}0.564 \\
\multicolumn{1}{|l|}{\textbf{Other leguminous}}  &  7.0  & \cellcolor[RGB]{255.0 230.3 230.3}0.041 & \cellcolor[RGB]{ 255.0 230.3 230.3}0.014 & \cellcolor[RGB]{255.0 230.3 230.3}0.045 & \cellcolor[RGB]{ 255.0 230.3 230.3}0.006 & \cellcolor[RGB]{255.0 222.1 222.1}0.06  & \cellcolor[RGB]{ 255.0 222.1 222.1}0.006 &  6.7 & \cellcolor[RGB]{255.0 123.4 123.4}0.399 & \cellcolor[RGB]{ 255.0 123.4 123.4}0.152 & \cellcolor[RGB]{255.0 115.2 115.2}0.5  & \cellcolor[RGB]{ 255.0 115.2 115.2}0.236 & \cellcolor[RGB]{255.0 164.5 164.5}0.544 & \cellcolor[RGB]{ 255.0 164.5 164.5}0.39   &  8.4 & \cellcolor[RGB]{230.3 230.3 255.0}0.028 & \cellcolor[RGB]{ 230.3 230.3 255.0}0.082 \\
\multicolumn{1}{|l|}{\textbf{Greenpeas}}         &  2.8  & \cellcolor[RGB]{255.0 213.9 213.9}0.069 & \cellcolor[RGB]{ 255.0 213.9 213.9}0     & \cellcolor[RGB]{255.0 123.4 123.4}0.315 & \cellcolor[RGB]{ 255.0 123.4 123.4}0.066 & \cellcolor[RGB]{255.0 106.9 106.9}0.381 & \cellcolor[RGB]{ 255.0 106.9 106.9}0.103 &  3.4 & \cellcolor[RGB]{255.0 197.4 197.4}0.624 & \cellcolor[RGB]{ 255.0 197.4 197.4}0.533 & \cellcolor[RGB]{255.0 213.9 213.9}0.686 & \cellcolor[RGB]{ 255.0 213.9 213.9}0.621 & \cellcolor[RGB]{230.3 230.3 255.0}0.644 & \cellcolor[RGB]{ 230.3 230.3 255.0}0.696 &  3.3 & \cellcolor[RGB]{181.0 181.0 255.0}0.047 & \cellcolor[RGB]{ 181.0 181.0 255.0}0.199 \\
\multicolumn{1}{|l|}{\textbf{Alfalfa} }          &  0.4  & \cellcolor[RGB]{164.5 164.5 255.0}0.417 & \cellcolor[RGB]{ 164.5 164.5 255.0}0.586 & \cellcolor[RGB]{255.0 230.3 230.3}0.281 & \cellcolor[RGB]{ 255.0 230.3 230.3}0.241 & \cellcolor[RGB]{255.0 189.2 189.2}0.339 & \cellcolor[RGB]{ 255.0 189.2 189.2}0.219 &  0.7 & \cellcolor[RGB]{255.0 222.1 222.1}0.603 & \cellcolor[RGB]{ 255.0 222.1 222.1}0.557 & \cellcolor[RGB]{255.0 205.6 205.6}0.661 & \cellcolor[RGB]{ 255.0 205.6 205.6}0.589 & \cellcolor[RGB]{255.0 246.8 246.8}0.62  & \cellcolor[RGB]{ 255.0 246.8 246.8}0.615 &  1.2 & \cellcolor[RGB]{197.4 197.4 255.0}0.497 & \cellcolor[RGB]{ 197.4 197.4 255.0}0.611 \\
\multicolumn{1}{|l|}{\textbf{Foragecrops}}       &  1.6  & \cellcolor[RGB]{255.0 115.2 115.2}0.315 & \cellcolor[RGB]{ 255.0 115.2 115.2}0.062 & \cellcolor[RGB]{255.0 222.1 222.1}0.127 & \cellcolor[RGB]{ 255.0 222.1 222.1}0.073 & \cellcolor[RGB]{255.0 189.2 189.2}0.115 & \cellcolor[RGB]{ 255.0 189.2 189.2}0.009 &  0.9 & \cellcolor[RGB]{255.0 238.5 238.5}0.566 & \cellcolor[RGB]{ 255.0 238.5 238.5}0.544 & \cellcolor[RGB]{246.8 246.8 255.0}0.599 & \cellcolor[RGB]{ 246.8 246.8 255.0}0.611 & \cellcolor[RGB]{181.0 181.0 255.0}0.546 & \cellcolor[RGB]{ 181.0 181.0 255.0}0.699 &  0.0 & \cellcolor[RGB]{189.2 189.2 255.0}0.671 & \cellcolor[RGB]{ 189.2 189.2 255.0}0.793 \\
\multicolumn{1}{|l|}{\textbf{Beet}}              &  1.4  & \cellcolor[RGB]{255.0 156.3 156.3}0.173 & \cellcolor[RGB]{ 255.0 156.3 156.3}0.003 & \cellcolor[RGB]{255.0 213.9 213.9}0.135 & \cellcolor[RGB]{ 255.0 213.9 213.9}0.07  & \cellcolor[RGB]{255.0 172.7 172.7}0.203 & \cellcolor[RGB]{ 255.0 172.7 172.7}0.065 &  1.4 & \cellcolor[RGB]{246.8 246.8 255.0}0.064 & \cellcolor[RGB]{ 246.8 246.8 255.0}0.081 & \cellcolor[RGB]{246.8 246.8 255.0}0.171 & \cellcolor[RGB]{ 246.8 246.8 255.0}0.186 & \cellcolor[RGB]{106.9 106.9 255.0}0.187 & \cellcolor[RGB]{ 106.9 106.9 255.0}0.477 &  1.2 & \cellcolor[RGB]{246.8 246.8 255.0}0.004 & \cellcolor[RGB]{ 246.8 246.8 255.0}0.017 \\
\multicolumn{1}{|l|}{\textbf{Potatoes}}          &  0.2  & \cellcolor[RGB]{255.0 106.9 106.9}0.524 & \cellcolor[RGB]{ 255.0 106.9 106.9}0.252 & \cellcolor[RGB]{230.3 230.3 255.0}0.002 & \cellcolor[RGB]{ 230.3 230.3 255.0}0.046 & \cellcolor[RGB]{115.2 115.2 255.0}0.427 & \cellcolor[RGB]{ 115.2 115.2 255.0}0.704 &  0.2 & \cellcolor[RGB]{222.1 222.1 255.0}0.087 & \cellcolor[RGB]{ 222.1 222.1 255.0}0.156 & \cellcolor[RGB]{172.7 172.7 255.0}0.024 & \cellcolor[RGB]{ 172.7 172.7 255.0}0.191 & \cellcolor[RGB]{255.0 172.7 172.7}0.367 & \cellcolor[RGB]{ 255.0 172.7 172.7}0.221 &  0.2 & \cellcolor[RGB]{148.1 148.1 255.0}0.559 & \cellcolor[RGB]{ 148.1 148.1 255.0}0.775  \\
\hline
\multicolumn{2}{c}{ }&\multicolumn{3}{r}{\textbf{$\Deltabf{R^2}=R^2_{MOGP}-R^2_{GPR}$:}} & \multicolumn{13}{l}{{\textbf{-0.5 }\IG[width={1\textwidth},trim={0cm 0.2cm 0cm 0cm},clip]{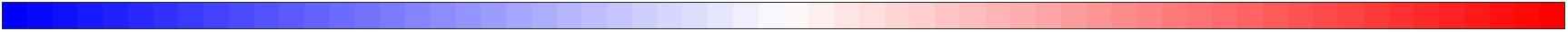}}\textbf{ 0.5 }} \\
\end{tabular}
}
\caption{Per-class R$^2$ calculated over pixels fulfilling the condition $a^{LF}_1/a^{HF}_1>1.5$. The cell background is proportional to MOGP and GPR performance difference in a blue-white-red colormap. Blueish tones indicate $R^2_{MOGP}$ is lower than $R^2_{GPR}$; reddish tones the other way around; white \black{color denotes negligible} differences.}
\label{Table:R2byclass}
\end{table}

\begin{table}[]
\resizebox{1.0\textwidth}{!}{ 
\begin{tabular}{l|c|ll|ll|ll|c|ll|ll|ll|c|ll|}
\cline{2-18}
\multicolumn{1}{c|}{\multirow{2}{*}{\textbf{RMSE}}}           & \multicolumn{3}{c|}{\textbf{2016/02/17}} & \multicolumn{2}{c|}{\textbf{2016/04/07}} & \multicolumn{2}{c|}{\textbf{2016/04/27}} & \multicolumn{3}{c|}{\textbf{2017/04/02}} & \multicolumn{2}{c|}{\textbf{2017/04/12}} & \multicolumn{2}{c|}{\textbf{2017/04/22}} & \multicolumn{3}{c|}{\textbf{2018/02/26}} \\ \cline{2-18} 
  & LCMap(\%) & MOGP & GPR  & MOGP  & GPR & MOGP & GPR & LCMap(\%)  & MOGP  & GPR  & MOGP   & GPR & MOGP  & GPR    & LCMap(\%)  & MOGP  & GPR    \\
   \hline
\multicolumn{1}{|l|}{\textbf{Wheat}}            & 30.2  & \cellcolor[RGB]{172.7 172.7 255.0}0.552 & \cellcolor[RGB]{ 172.7 172.7 255.0}1.059 & \cellcolor[RGB]{172.7 172.7 255.0}0.948 & \cellcolor[RGB]{ 172.7 172.7 255.0}1.45  & \cellcolor[RGB]{205.6 205.6 255.0}1.001 & \cellcolor[RGB]{ 205.6 205.6 255.0}1.313 & 25.7& \cellcolor[RGB]{255.0 230.3 230.3}0.442 & \cellcolor[RGB]{ 255.0 230.3 230.3}0.367 & \cellcolor[RGB]{255.0 230.3 230.3}0.422 & \cellcolor[RGB]{ 255.0 230.3 230.3}0.342 & \cellcolor[RGB]{255.0 230.3 230.3}0.449 & \cellcolor[RGB]{ 255.0 230.3 230.3}0.345 & 29.5& \cellcolor[RGB]{255.0 238.5 238.5}0.435 & \cellcolor[RGB]{ 255.0 238.5 238.5}0.398\\
\multicolumn{1}{|l|}{\textbf{Maize}}            &  0.2  & \cellcolor[RGB]{255.0 213.9 213.9}0.627 & \cellcolor[RGB]{ 255.0 213.9 213.9}0.436 & \cellcolor[RGB]{238.5 238.5 255.0}0.77  & \cellcolor[RGB]{ 238.5 238.5 255.0}0.889 & \cellcolor[RGB]{189.2 189.2 255.0}0.536 & \cellcolor[RGB]{ 189.2 189.2 255.0}0.906 &  0.1& \cellcolor[RGB]{255.0 246.8 246.8}0.351 & \cellcolor[RGB]{ 255.0 246.8 246.8}0.358 & \cellcolor[RGB]{246.8 246.8 255.0}0.407 & \cellcolor[RGB]{ 246.8 246.8 255.0}0.44  & \cellcolor[RGB]{238.5 238.5 255.0}0.361 & \cellcolor[RGB]{ 238.5 238.5 255.0}0.463 &  0.2& \cellcolor[RGB]{255.0 238.5 238.5}0.255 & \cellcolor[RGB]{ 255.0 238.5 238.5}0.194\\
\multicolumn{1}{|l|}{\textbf{Barley}}           & 32.2  & \cellcolor[RGB]{156.3 156.3 255.0}0.427 & \cellcolor[RGB]{ 156.3 156.3 255.0}1.002 & \cellcolor[RGB]{115.2 115.2 255.0}0.786 & \cellcolor[RGB]{ 115.2 115.2 255.0}1.606 & \cellcolor[RGB]{148.1 148.1 255.0}0.879 & \cellcolor[RGB]{ 148.1 148.1 255.0}1.517 & 37.4& \cellcolor[RGB]{255.0 246.8 246.8}0.398 & \cellcolor[RGB]{ 255.0 246.8 246.8}0.422 & \cellcolor[RGB]{255.0 246.8 246.8}0.387 & \cellcolor[RGB]{ 255.0 246.8 246.8}0.406 & \cellcolor[RGB]{255.0 238.5 238.5}0.42  & \cellcolor[RGB]{ 255.0 238.5 238.5}0.389 & 31.6& \cellcolor[RGB]{238.5 238.5 255.0}0.452 & \cellcolor[RGB]{ 238.5 238.5 255.0}0.548\\
\multicolumn{1}{|l|}{\textbf{Rye} }             &  0.0  & \cellcolor[RGB]{255.0 246.8 246.8}0     & \cellcolor[RGB]{ 255.0 246.8 246.8}0     & \cellcolor[RGB]{255.0 246.8 246.8}0     & \cellcolor[RGB]{ 255.0 246.8 246.8}0     & \cellcolor[RGB]{255.0 246.8 246.8}0     & \cellcolor[RGB]{ 255.0 246.8 246.8}0     &  0.2& \cellcolor[RGB]{255.0 246.8 246.8}0.343 & \cellcolor[RGB]{ 255.0 246.8 246.8}0.338 & \cellcolor[RGB]{255.0 238.5 238.5}0.338 & \cellcolor[RGB]{ 255.0 238.5 238.5}0.31  & \cellcolor[RGB]{255.0 238.5 238.5}0.34  & \cellcolor[RGB]{ 255.0 238.5 238.5}0.279 &  0.3& \cellcolor[RGB]{255.0 213.9 213.9}0.465 & \cellcolor[RGB]{ 255.0 213.9 213.9}0.286\\
\multicolumn{1}{|l|}{\textbf{Oats}}             &  0.2  & \cellcolor[RGB]{238.5 238.5 255.0}0.807 & \cellcolor[RGB]{ 238.5 238.5 255.0}0.883 & \cellcolor[RGB]{131.6 131.6 255.0}0.755 & \cellcolor[RGB]{ 131.6 131.6 255.0}1.478 & \cellcolor[RGB]{115.2 115.2 255.0}0.946 & \cellcolor[RGB]{ 115.2 115.2 255.0}1.756 &  0.0& \cellcolor[RGB]{213.9 213.9 255.0}0.536 & \cellcolor[RGB]{ 213.9 213.9 255.0}0.775 & \cellcolor[RGB]{222.1 222.1 255.0}0.589 & \cellcolor[RGB]{ 222.1 222.1 255.0}0.784 & \cellcolor[RGB]{230.3 230.3 255.0}0.548 & \cellcolor[RGB]{ 230.3 230.3 255.0}0.711 &  0.4& \cellcolor[RGB]{238.5 238.5 255.0}0.374 & \cellcolor[RGB]{ 238.5 238.5 255.0}0.466\\
\multicolumn{1}{|l|}{\textbf{Sunflower}}        &  1.5  & \cellcolor[RGB]{197.4 197.4 255.0}0.383 & \cellcolor[RGB]{ 197.4 197.4 255.0}0.744 & \cellcolor[RGB]{148.1 148.1 255.0}0.425 & \cellcolor[RGB]{ 148.1 148.1 255.0}1.061 & \cellcolor[RGB]{197.4 197.4 255.0}0.434 & \cellcolor[RGB]{ 197.4 197.4 255.0}0.763 &  1.3& \cellcolor[RGB]{255.0 246.8 246.8}0.303 & \cellcolor[RGB]{ 255.0 246.8 246.8}0.31  & \cellcolor[RGB]{246.8 246.8 255.0}0.305 & \cellcolor[RGB]{ 246.8 246.8 255.0}0.353 & \cellcolor[RGB]{246.8 246.8 255.0}0.362 & \cellcolor[RGB]{ 246.8 246.8 255.0}0.415 &  2.2& \cellcolor[RGB]{255.0 238.5 238.5}0.345 & \cellcolor[RGB]{ 255.0 238.5 238.5}0.298\\
\multicolumn{1}{|l|}{\textbf{Rapeseed}}         &  0.8  & \cellcolor[RGB]{131.6 131.6 255.0}0.613 & \cellcolor[RGB]{ 131.6 131.6 255.0}1.338 & \cellcolor[RGB]{123.4 123.4 255.0}0.764 & \cellcolor[RGB]{ 123.4 123.4 255.0}1.551 & \cellcolor[RGB]{106.9 106.9 255.0}1.255 & \cellcolor[RGB]{ 106.9 106.9 255.0}2.11  &  0.8& \cellcolor[RGB]{255.0 246.8 246.8}0.759 & \cellcolor[RGB]{ 255.0 246.8 246.8}0.739 & \cellcolor[RGB]{255.0 238.5 238.5}0.978 & \cellcolor[RGB]{ 255.0 238.5 238.5}0.91  & \cellcolor[RGB]{255.0 230.3 230.3}0.918 & \cellcolor[RGB]{ 255.0 230.3 230.3}0.84  &  1.2& \cellcolor[RGB]{255.0 230.3 230.3}0.636 & \cellcolor[RGB]{ 255.0 230.3 230.3}0.532\\
\multicolumn{1}{|l|}{\textbf{Other leguminous}} & 7.0    & \cellcolor[RGB]{123.4 123.4 255.0}0.474 & \cellcolor[RGB]{ 123.4 123.4 255.0}1.261 & \cellcolor[RGB]{0.0   0.0 255.0}0.718   & \cellcolor[RGB]{ 0.0   0.0 255.0}2.207   & \cellcolor[RGB]{90.5  90.5 255.0}0.824  & \cellcolor[RGB]{ 90.5  90.5 255.0}1.8    &  6.7& \cellcolor[RGB]{213.9 213.9 255.0}0.257 & \cellcolor[RGB]{ 213.9 213.9 255.0}0.49  & \cellcolor[RGB]{222.1 222.1 255.0}0.254 & \cellcolor[RGB]{ 222.1 222.1 255.0}0.46  & \cellcolor[RGB]{238.5 238.5 255.0}0.262 & \cellcolor[RGB]{ 238.5 238.5 255.0}0.376 &  8.4& \cellcolor[RGB]{255.0 238.5 238.5}0.434 & \cellcolor[RGB]{ 255.0 238.5 238.5}0.407\\
\multicolumn{1}{|l|}{\textbf{Greenpeas}}        &  2.8  & \cellcolor[RGB]{115.2 115.2 255.0}0.414 & \cellcolor[RGB]{ 115.2 115.2 255.0}1.232 & \cellcolor[RGB]{16.5  16.5 255.0}0.656  & \cellcolor[RGB]{ 16.5  16.5 255.0}2.057  & \cellcolor[RGB]{16.5  16.5 255.0}0.749  & \cellcolor[RGB]{ 16.5  16.5 255.0}2.129  &  3.4& \cellcolor[RGB]{246.8 246.8 255.0}0.273 & \cellcolor[RGB]{ 246.8 246.8 255.0}0.316 & \cellcolor[RGB]{246.8 246.8 255.0}0.281 & \cellcolor[RGB]{ 246.8 246.8 255.0}0.311 & \cellcolor[RGB]{255.0 238.5 238.5}0.322 & \cellcolor[RGB]{ 255.0 238.5 238.5}0.29  &  3.3& \cellcolor[RGB]{255.0 246.8 246.8}0.392 & \cellcolor[RGB]{ 255.0 246.8 246.8}0.395\\
\multicolumn{1}{|l|}{\textbf{Alfalfa} }         &  0.4  & \cellcolor[RGB]{255.0 246.8 246.8}0.953 & \cellcolor[RGB]{ 255.0 246.8 246.8}0.945 & \cellcolor[RGB]{41.1  41.1 255.0}1.102  & \cellcolor[RGB]{ 41.1  41.1 255.0}2.358  & \cellcolor[RGB]{82.3  82.3 255.0}1.193  & \cellcolor[RGB]{ 82.3  82.3 255.0}2.196  &  0.7& \cellcolor[RGB]{255.0 222.1 222.1}0.48 & \cellcolor[RGB]{ 255.0 222.1 222.1}0.356  & \cellcolor[RGB]{255.0 230.3 230.3}0.431 & \cellcolor[RGB]{ 255.0 230.3 230.3}0.357 & \cellcolor[RGB]{255.0 230.3 230.3}0.447 & \cellcolor[RGB]{ 255.0 230.3 230.3}0.359 &  1.2& \cellcolor[RGB]{255.0 230.3 230.3}0.47  & \cellcolor[RGB]{ 255.0 230.3 230.3}0.389\\
\multicolumn{1}{|l|}{\textbf{Foragecrops}}      &  1.6  & \cellcolor[RGB]{41.1  41.1 255.0}0.472  & \cellcolor[RGB]{ 41.1  41.1 255.0}1.724  & \cellcolor[RGB]{0.0   0.0 255.0}0.675   & \cellcolor[RGB]{ 0.0   0.0 255.0}2.916   & \cellcolor[RGB]{0.0   0.0 255.0}0.779   & \cellcolor[RGB]{ 0.0   0.0 255.0}3.026   &  0.9& \cellcolor[RGB]{246.8 246.8 255.0}0.254 & \cellcolor[RGB]{ 246.8 246.8 255.0}0.298 & \cellcolor[RGB]{246.8 246.8 255.0}0.262 & \cellcolor[RGB]{ 246.8 246.8 255.0}0.297 & \cellcolor[RGB]{255.0 246.8 246.8}0.291 & \cellcolor[RGB]{ 255.0 246.8 246.8}0.276 &  0.0& \cellcolor[RGB]{255.0 246.8 246.8}0.297 & \cellcolor[RGB]{ 255.0 246.8 246.8}0.278\\
\multicolumn{1}{|l|}{\textbf{Beet}}             &  1.4  & \cellcolor[RGB]{238.5 238.5 255.0}0.59  & \cellcolor[RGB]{ 238.5 238.5 255.0}0.707 & \cellcolor[RGB]{181.0 181.0 255.0}0.694 & \cellcolor[RGB]{ 181.0 181.0 255.0}1.128 & \cellcolor[RGB]{164.5 164.5 255.0}0.601 & \cellcolor[RGB]{ 164.5 164.5 255.0}1.141 &  1.4& \cellcolor[RGB]{255.0 238.5 238.5}0.494 & \cellcolor[RGB]{ 255.0 238.5 238.5}0.465 & \cellcolor[RGB]{255.0 246.8 246.8}0.512 & \cellcolor[RGB]{ 255.0 246.8 246.8}0.513 & \cellcolor[RGB]{255.0 222.1 222.1}0.595 & \cellcolor[RGB]{ 255.0 222.1 222.1}0.472 &  1.2& \cellcolor[RGB]{230.3 230.3 255.0}0.426 & \cellcolor[RGB]{ 230.3 230.3 255.0}0.572\\
\multicolumn{1}{|l|}{\textbf{Potatoes}}         &  0.2  & \cellcolor[RGB]{189.2 189.2 255.0}0.59  & \cellcolor[RGB]{ 189.2 189.2 255.0}0.973 & \cellcolor[RGB]{238.5 238.5 255.0}0.764 & \cellcolor[RGB]{ 238.5 238.5 255.0}0.85  & \cellcolor[RGB]{164.5 164.5 255.0}0.588 & \cellcolor[RGB]{ 164.5 164.5 255.0}1.137 &  0.2& \cellcolor[RGB]{222.1 222.1 255.0}0.266 & \cellcolor[RGB]{ 222.1 222.1 255.0}0.458 & \cellcolor[RGB]{222.1 222.1 255.0}0.288 & \cellcolor[RGB]{ 222.1 222.1 255.0}0.479 & \cellcolor[RGB]{230.3 230.3 255.0}0.29  & \cellcolor[RGB]{ 230.3 230.3 255.0}0.458 &  0.2& \cellcolor[RGB]{255.0 189.2 189.2}0.68  & \cellcolor[RGB]{ 255.0 189.2 189.2}0.357\\
\hline
\multicolumn{2}{c}{ }&\multicolumn{4}{r}{\textbf{$\Deltabf_{RMSE}=RMSE_{MOGP}-RMSE_{GPR}$:}} &\multicolumn{12}{l}{{\textbf{-1.5 }\IG[width={1\textwidth},trim={0cm 0.2cm 0cm 0cm},clip]{figs/Delta_RMSE_colorbar1.PNG}}\textbf{ 1.5 }} \\
\end{tabular}}
\caption{Per-class RMSE $[m^2\black{/}m^2]$ calculated over pixels fulfilling the condition $a^{LF}_1/a^{HF}_1>1.5$. The cell background is proportional to (MOGP,GPR) performance difference in a blue-white-red colormap. Blueish tones indicate \black{RMSE is lower for MOGP than for GPR}; reddish tones the other way around; white \black{color denotes negligible} differences.
}
\label{Table:RMSEbyclass}
\end{table}
To look further into this last issue, we analyze the performance of the synergy methodology per crop type. \black{A long} time series is required to optimize the free parameters of the model at pixel level, but the crop rotation may reduce MOGP performance in case of seasonal correlation changes. In order to workaround this issue, we consider only the pixels fulfilling the condition $a^{LF}_1/a^{HF}_1>1.5$, which guarantees the active-passive synergy is profi\black{t}able. For each assessment date, we used the corresponding land cover map to select pixels belonging to the same class, and calculated R$^2$ and RMSE. 

We identified 13 permanent classes along the three years, except for rye present for years 2017 and 2018. Information about the classes, their distribution in space and time, and the assessment results in terms of R$^2$ and RMSE are detailed in Tables \ref{Table:R2byclass} and \ref{Table:RMSEbyclass}, respectively. 

Taking into consideration R$^2$ in Table \ref{Table:R2byclass}, we observe a dominance of red colors over the long data gap (2016), meaning MOGP generally provides higher values than GPR. Higher absolute values are obtained for Oats, Greenpeas and Alfalfa; the highest performance is obtained over Maize, even if only for a fully developed crop at the end of April 2016. \black{As important as R$^2$ are also the values of RMSE, reported in Table \ref{Table:RMSEbyclass}}. The residuals \black{over the long time data gap are significantly higher for GPR than MOGP}, with the blueish color characterizing all the crop types. Concerning shorter data gaps, the two approaches present similar performances, as already \black{pointed out} in Section \ref{moderategap}. Still, some classes such as Oats, Greenpeas, Alfalfa and Maize, present a higher score for the synerg\black{istic approach}. Sunflowers deserves a specific mention: \black{all the artificial gaps lie from winter to spring time, when this crop type is at an early stage of development. Superior results from the active-passive synergy are expected over cloudy early-summer periods, when their development is more advanced}.

\section{Discussion}\label{Discussions}
\black{Establishing a physically interpretable link between active and passive time series over vegetated areas \black{becomes} feasible using MOGP modeling. \black{An absolute novelty of the solution we proposed is that the parameters of the trained model implicitly predict the meaningfulness of any fusion approach: they quantify the amount of information shared between the two time series and rule the interaction of low- and high-frequency GPs for output reconstruction}. The LAI data gap filling described in this work is just an example of MOGP possible applications. The strength of the two time series synergy strictly depends on the crop type, being some crops more prone to be well characterized by RVI (such as oats, wheats or maize), but also on crop-stage. All the same, the SAR information extracted by MOGP is key to reduce prediction uncertainties with respect to advanced single-output regression over long time data gaps.} 

Concerning the absolute values of R$^2$, higher performances are desirable: a best score of 74\% might be considered modest from an absolute point of view, except for the 81\% over maize within the long time data gap. Still, the estimation of vegetation parameters from single optical acquisition is notoriously affected by atmospheric correction residual errors and surface anisotropy \citep{jonsson2002}. These effects make the original time series unfit for any seasonality information extraction, unless smoothing procedures and outlier filtering are previously applied \citep{jonsson2004}. Then, our assessment based on direct LAI retrievals from S2 captures is likely to underestimate the real performance of MOGP (and GPR), due to the smoothing process performed by the covariance kernel. This means that differences between reference and validation do not always represent a worse estimation of true LAI. Having at disposal ground-truth information about season length, start-of-season and end-of-season for different crops for instance \citep{richardson2013}, a different assessment could be envisaged.

\black{Some limitations of the proposed approach must be also stressed, related to either its theoretical formulation or the specific implementation we carried out in this work.
The main theoretical limitation of MOGP is that it is unable to take into account possible time-shift between even highly correlated time series. If changes appear with a delay in the two sources of information, the efficiency of the reconstruction is affected in a way proportional with the magnitude of the time separation. Similarly, contraction or dilation effects of similar patterns cannot be exploited properly for the reconstruction, even if less probable.}

\black{Another weakness of the MOGP is its high computational cost, which makes it inefficient if the per-pixel training is applied at large scale. A more efficient usage of MOGP should exploit spatial homogeneities. Polygons from land cover maps, if available, enable to reduce the computational time by moving from a pixel-wise to an object-oriented approach, if one accepts to lose details about the spatial heterogeneity of each polygon.
Yet, the homogeneity of the same polygon must be checked when dealing with multiple-year time series, if we want to avoid the average of different crop types in space or time.}  The stationarity of the processes entering MOGP \black{is} not always fulfilled: LFGP stops being dominant and the identification of a statistical linking modeling covering the whole time series is not straightforward. Even if the crop-based assessment circumvented this problem by selected only those pixels showing a high active-passive correlation, the real performance of MOGP are likely underestimated. 
\black{A potential improvement is the estimation of crop-specific Coregionalization matrices via crop type space (multiple polygons) and time (multiple years) training}. This knowledge could be applied at pixels level in a transfer learning framework, if an yearly-updated land cover map is available. Using crop type \black{precalculated} $[\mathbf{B}]$ matrices, less samples are required for the estimation of the \black{remaining free parameters}, and processing single seasonalities for gap filling purposes becomes feasible. This \black{new} approach is expected to help deal with shorter time series and, to some extent, speed up the model training process. In fact, the main bottleneck represented by the inversion of $\sK$ is still present. Yet, all these follow-up studies go beyond the purpose of this work, whose main objectives have been 1) demonstrating the capability of MOGP to establish active-passive time series synergy through a statistically interpretable model, and 2) providing a practical example of MOGP for \black{tackling} a crucial problem of spaceborne vegetation monitoring, i.e. filling the long time-gaps of optical vegetation descriptors induced by clouds.

\section{Conclusions}\label{Conclusions}

The MOGP technique introduced in this work represents an innovative approach in the remote sensing community for establishing a fusion between optical and SAR information. Unlike other black-box machine learning solutions, the interpretation of the MOGP model is easy: the two imageries are modeled as combinations of independent GPs and their influence in the final prediction is weighted by the corresponding \emph{coregionalization} matrix. 
The approximations held in this study constitutes a first demonstration of MOGP capability to extract information shared between collections of LAI and RVI acquisitions. The results demonstrate the capability of MOGP for retrieving missing LAI values, improving the prediction of standard GP over short-time gaps (R$^2$ of 74\% vs 68\%) and especially over long-time gaps (R$^2$ of 33\% vs 12\%, RMSE of 0.5 vs 1.09). The method we proposed always guarantees that the LAI time series obtained as output follows RVI time evolution if no optical sample is available over long periods, keeping the solution tied to real measurements. 
\black{The crop-specific analysis reve\black{a}l\black{s} the synergy is profitable for most crops, even if MOGP performance is crop type as well as crop stage dependent. Specially maize, but also wheat, oats, rye and barley, can profit from the LAI-RVI synergy. For beet or potatoes results are less encouraging, but alternative active descriptors to RVI should be explored. In conclusion, active-passive sensor fusion with MOGP represents a novel and promising approach to cope with crop monitoring over \black{cloud-dominated} areas.} 

Further studies are now required to advance towards a more efficient usage of MOGP. For instance, spatial homogeneities can be exploit to move from a pixel-wise to an object-oriented approach, accepting to lose the possible spatial heterogeneity within the polygon. Other two aspects to be examined are including multiple SAR descriptors to MOGP model over specific crop types, and exploring model training over single-seasonality. However, shorter observation periods imply having fewer time samples to train a more complex model. All these aspects must be carefully studied and motivate future analysis. 

\section{Acknowledgement}\label{Acknowledgement}

The research leading to these results has received funding from the European Union's Horizon 2020 Research and Innovation Programme, under Grant Agreement no 730074. Luca Pipia and Jochem Verrelst were supported by the European Research Council (ERC) under the ERC-2017-STG SENTIFLEX project (grant agreement 755617). Gustau Camps-Valls was supported by ERC under the ERC-COG-2014 SEDAL (grant agreement 647423). GCV and JMM acknowledge the funding received from the Spanish Ministry of Economy and Competitiveness (MINECO) and FEDER co-funding through the projects TIN2012-38102-C03-01 and TIN2015-64210-R.

% References should be produced using the bibtex program from suitable
% BiBTeX files (here: strings, refs, manuals). The IEEEbib.bst bibliography
% style file from IEEE produces unsorted bibliography list.
% -------------------------------------------------------------------------
\bibliographystyle{elsarticle-harv}
\bibliography{MOGP}
\newpage

\appendix
\section{Single- vs multi-output regression performance}\label{AppendixA}
\begin{table}[h]
\resizebox{0.75\textwidth}{!}{
\begin{tabular}{|llll|}
\hline
\rowcolor[HTML]{C0C0C0} 
\textbf{Interpolation Method}        & \textbf{R$^2$} & \textbf{$\Sigmabf$Res} & \textbf{RMSE} \\ %& \textbf{Reference} \\
\hline
\textbf{Fourier analysis: Offset + Harmonic analysis}         & 0.5307      & 5.3803             & 0.9896                            \\
\textbf{Polynomial curve fitting}                             & 0.8329      & 16.0240            & 2.3874                            \\
\textbf{Double logistic curve}                                & 0.3127      & 8.8674             & 1.6912                            \\
\textbf{Linear interpolation}                                 & 0.3006      & 7.9026             & 1.3926                            \\
\textbf{Nearest neighbor interpolation}                       & 0.9734      & 7.1444             & 1.3061                            \\
\textbf{Next neighbor interpolation}                          & 0.0645      & 8.2933             & 1.4231                            \\
\textbf{Previous neighbor interpolation}                      & 0.8433      & 7.6838             & 1.4651                            \\
\textbf{Spline interpolation using not-a-knot end conditions} & 0.6620      & 5.0872             & 0.9798                            \\
\textbf{Shape-preserving piecewise cubic interpolation}       & 0.6963      & 7.3767             & 1.3220                            \\
\textbf{Bagging trees}                                        & 0.2120      & 10.1841            & 1.7216                            \\
\textbf{Adaptive Regression Splines}                          & 0.7446      & 12.2503            & 1.9048                            \\
\textbf{Boosting random trees}                                & 0.0014      & 8.9694             & 1.5074                            \\
\textbf{Boosting trees}                                       & 0.2859      & 8.2312             & 1.5876                            \\
\textbf{k-nearest neighbours regression}                      & 0.1452      & 8.2350             & 1.5764                            \\
\textbf{Gaussian Process Regression}                          & {\textbf{\blue{0.9081}}}      & {\textbf{\blue{5.7816}}}             & {\textbf{\blue{0.9527}}}     \\
\textbf{Neural networks}                                      & 0.7481      & 7.5766             & 1.3005                            \\
\textbf{Random forests}                                       & 0.9734      & 7.1444             & 1.3061                            \\
\hline
\rowcolor[HTML]{C0C0C0} 
%Multi-Output Interpolation Method                             & \textbf{R2} & \textbf{Residuals} & \textbf{RMSE} \\%& \textbf{Reference} \\
\textbf{Multi-Output Gaussian Process}                        & \multicolumn{1}{l}{\textbf{\blue{0.9900}}}  & {\textbf{\blue{1.3035}}}  & {\textbf{\blue{0.2377}}} \\
\hline
\end{tabular}}
\caption{Performance of the set of 19 single-output regression methods applied to the LAI time series in Section \ref{Assessment1D} vs the proposed multi-output regression technique. In blue, the method providing the best score as a trade-off between R$^2$ ans $RMSE$ for the two cases. A general description of the methods along with specific references to each of them can be found in \cite{verrelst2019}.}
\label{SOvsMO_1DAssess}
\end{table}

\newpage

%\listoffigures
%{\renewcommand*\numberline[1]{Figure.\,#1:\space}
%\makeatletter
%\renewcommand*\l@figure[2]{\noindent#1\par}
%\makeatother
%\listoffigures}
\newlength{\fig}
\settowidth{\fig}{Figure\,99:~}
{\renewcommand*\numberline[1]{\llap{\makebox[\fig][l]{Figure\,#1:~}}}
\makeatletter
\renewcommand*\l@figure[2]{\leftskip\fig\noindent#1\par}
\makeatother
\listoffigures}

%\listoftables

\end{document}